\documentclass[]{aastex}
\usepackage{emulateapj5,onecolfloat,graphicx}
\usepackage[hyperindex,breaklinks]{hyperref}

\def\cf{{\it cf.}}
\def\eg{{\it e.g.}}
\def\etal{{\it et al.}}
\def\etc{{\it etc.}}
\def\ie{{\it i.e.}}

\def\DF{{\small DF}}

\def\GCS{{\small GCS}}
\def\Hipp{{\small HIPPARCOS}}
\def\LSR{{\small LSR}}
\def\BH{{\small BH}}
\def\CR{{\small CR}}
\def\ILR{{\small ILR}}
\def\OLR{{\small OLR}}
\def\LON{{\small LoN}}
\def\LMC{{\small LMC}}
\def\ML{{\small M/L}}
\def\MW{{\small MW}}
\def\WASER{{\small WASER}}
\def\muB{\mu_{\rm B}}
\def\Rd{R_{\rm d}}
\def\BTii{BT08}
\def\LCDM{{\small$\Lambda$CDM}}
\def\be{\vspace{0.1cm}\begin{equation}}
\def\ee{\vspace{0.05cm}\end{equation}}
\def\ssection{\vspace{0.2cm}\section}

\long\def\Ignore#1{\relax}

\def\bk{\;\pmb{\mit k}}

\def\bv{\;\pmb{\mit v}}
\def\bx{\;\pmb{\mit x}}
\def\cross{\;\pmb{\mit \times}}

\def\pmb#1{\setbox0=\hbox{$#1$}%
  \kern-0.25em\copy0\kern-\wd0
  \kern.05em\copy0\kern-\wd0
  \kern-0.025em\raise.0433em\box0}
\def\spmb#1{\setbox1=\hbox{${\scriptstyle #1}$}%
  \kern-0.25em\copy1\kern-\wd1
  \kern.05em\copy1\kern-\wd1
  \kern-0.025em\raise.0433em\box1}

\slugcomment{Chapter for {\it Planets, Stars and Stellar Systems\/} v5.
 Includes minor changes made in proofs.}

\begin{document}

\twocolumn[
\title{Dynamics of Disks and Warps}
\author{J. A. Sellwood}
\affil{Rutgers University, Department of Physics \& Astronomy, \\
       136 Frelinghuysen Road, Piscataway, NJ 08854-8019 \\
       {\it sellwood@physics.rutgers.edu}}

\vspace{0.25cm}

\begin{abstract}
This chapter reviews theoretical work on the stellar dynamics of
galaxy disks.  All the known collective global instabilities are
identified, and their mechanisms described in terms of local wave
mechanics.  A detailed discussion of warps and other bending waves is
also given.  The structure of bars in galaxies, and their effect on
galaxy evolution, is now reasonably well understood, but there is
still no convincing explanation for their origin and frequency.
Spiral patterns have long presented a special challenge, and ideas and
recent developments are reviewed.  Other topics include scattering of
disk stars and the survival of thin disks.
\end{abstract}


\vspace{0.25cm}
]

\ssection{Introduction}
A significant fraction of the stars in the universe resides in the
rotationally supported disks of galaxies.  Disks are mostly thin and
flat, but the disk is often warped away from its principal plane in
the outer parts.  Disk galaxies usually manifest spiral patterns,
and rather more than half host bars.  Most, but not all, disk galaxies
have a central bulge, perhaps also a thick stellar disk, and generally
a small fraction of the stars resides in a quasi-spherical stellar
halo, while the central attraction at large distances from the center
is dominated by a dark halo.  The material in most disks
overwhelmingly orbits in a single sense, although a small fraction of
galaxies have been found to host substantial counter-rotating
components.

This chapter is primarily concerned with the dynamics of rotationally
supported disks of stars.  Stellar disks are believed to have formed
over time from gas that had previously settled into centrifugal
balance in the gravitational well of the galaxy, and the process of
star formation continues to the present day in most disk galaxies.
While stars are the dominant dynamical component today, the small gas
fraction (usually $\la 10\%$ by mass) can still play an important
dynamical role in some contexts.

Disk dynamics is a rich topic for two principal reasons: (a) the
organized orbital motion facilitates gravitationally-driven collective
behavior and (b) outward transfer of angular momentum extracts energy
from the potential well.  Space limitations preclude a detailed
development and this review will mostly be confined to a summary of
the principal results and open issues.  The derivations of the
principal formulae can be found in the excellent textbook
by \citet[][hereafter \BTii]{BT08}.  Furthermore, no attempt is made
to cite every paper that relates to a topic.

The distribution of gas within the Milky Way, the properties of the
Galactic bulge, and other closely related topics are described
elsewhere in this volume.  The distributions of light and mass within
galaxies and our current understanding of the processes that lead to
the formation of galaxies are described in volume 6.

\ssection{Preliminaries}
\subsection{Relaxation Rate}
\label{relax}
Because stellar disks contain many stars, the attraction from
individual nearby stars is negligible in comparison with the
aggregated gravitational field of distant matter.  The Appendix
explains how the usual rough calculation to support this assertion,
which was derived with quasi-spherical systems supported by random
motion in mind (\eg\ \BTii\ pp.~35-38), must be revised for disks.
Three factors all conspire to reduce the relaxation time in disks by
several orders of magnitude below the traditional estimate
(eq.~\ref{trelax}), although it remains many dynamical times.

\begin{figure}[t]
\begin{center}
\includegraphics[width=.8\hsize]{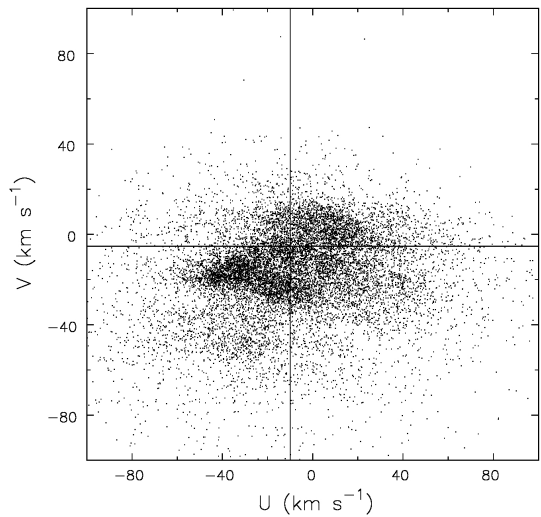}
\end{center}
\caption{The velocity distribution, in Galactic coordinates, of stars
  near the Sun, as given in the \GCS.  $U$ is the velocity of the star
  toward the Galactic center and $V$ is the component in the direction
  of Galactic rotation, both measured relative to the Sun.  The
  intersection of the vertical and horizontal lines shows the local
  standard of rest estimated by \cite{AB09}.}
\label{GCSsimp}
\end{figure}

Fig.~\ref{GCSsimp} shows the non-smooth distribution of stellar
velocities of $>14\,000$ F \& G dwarf stars in the vicinity of the
Sun, as found in the Geneva-Copenhagen Survey \citep[][hereafter
  \GCS]{Nord04,HNA9}.  The determination of the radial velocities has
confirmed the substructure that was first identified by \cite{Dehn98}
from a clever analysis of the \Hipp\ data without the radial
velocities (The implications of this Figure are discussed more fully
in \S\ref{scatt}.)  As collisional relaxation would erase
substructure, this distribution provides a direct illustration of the
collisionless nature of the solar neighborhood \citep[unless the
  substructure is being recreated rapidly, \eg][]{DeSi04}.

The usual first approximation that stars move in a smooth
gravitational potential well therefore seems adequate.

\subsection{Mathematical Formulation}
This approximation immediately removes the need to distinguish stars
by their masses.  A stellar system can therefore be described by a
{\bf distribution function} (\DF), $f(\bx,\bv,t)$ that specifies the
stellar density in a 6D phase space of position $\bx$ and velocity
$\bv$ at a particular time $t$.  Since masses are unimportant, it is
simplest conceptually to think of the stars being broken into
infinitesimal fragments so that discreteness is never an issue.

With this definition, the mass density at any point is
\be
\rho(\bx,t) = \int f d^3\bv,
\ee
which in turn is related to the gravitational potential, $\Phi$, through
{\bf Poisson's equation}
\be
\nabla^2\Phi(\bx,t) = 4\pi G \rho(\bx,t).
\ee
This is, of course, just the potential from the stellar component
described in $f$; the total potential includes contributions from dark
matter, gas, external perturbations, \etc \ Finally, the evolution of
the \DF\ is governed by the {\bf collisionless Boltzmann equation}
(\BTii\ eq.~4.11):
\be
{\partial f \over \partial t} + \bv \cdot {\partial f \over \partial
  \bx} + \dot{\bv} \cdot {\partial f \over \partial \bv} = 0,
\ee
where the acceleration is the negative gradient of the\break smooth
total potential: $\dot{\bv} = -\nabla\Phi_{\rm tot}$.  The time
evolution of a stellar system is completely described by the solution
to these three coupled equations.  Note that collisionless systems
have no equation of state that relates the density to quantities such
as pressure.

The most successful way to obtain global solutions to these coupled
equations is through {\bf $N$-body simulation}.  The particles in a
simulation are a representative sample of points in phase space whose
motion is advanced in time in the gravitational field.  At each step,
the field is determined from a smoothed estimate of the density
distribution derived from the instantaneous positions of the particles
themselves.\footnote{\ie\ a simulation solves eqs.~(1)--(3) by the
  method of characteristics.}  This rough and ready approach is
powerful, but simulations have limitations caused by {\bf noise} from
the finite number of particles, and {\bf bias} caused by the smoothed
density and approximate solution for the field, and other possible
artifacts.

Understanding the results from simulations, or even\break knowing when
they can be trusted, requires dynamical insight that can be obtained
only from analytic treatments.  This chapter therefore stresses how
the basic theory of stellar disks inter-relates with well designed,
idealized simulations to advance our understanding of these complex
systems.

\begin{figure}[t]
\begin{center}
\includegraphics[width=.6\hsize]{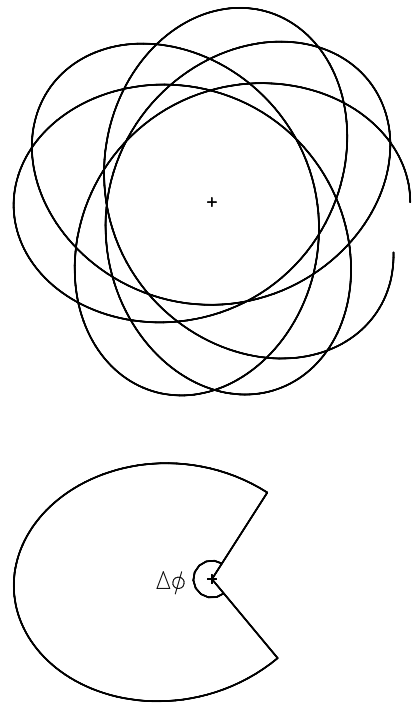}
\end{center}
\caption{An orbit of a star is generally a non-closing rosette (upper
  panel).  The lower panel shows one radial oscillation, from
  pericenter to pericenter say, during which time the star moves in
  azimuth through the angle $\Delta\phi$.}
\label{orbit}
\end{figure}

\subsection{Orbits}
When orbital deflections by mass clumps can be neglected
(\S\ref{relax}), the stars in a disk move in a smooth gravitational
potential.  The effects of mass clumps and other non-uniformities are
considered in \S10.  It is simplest to first discuss motion of
stars in the disk mid-plane, before considering full 3D motion.

The orbit of a star of {\bf specific energy} $E$ and {\bf specific
  angular momentum} $L_z$ in the mid-plane of an axisymmetric
potential is, in general, a non-closing rosette, as shown in
Fig.~\ref{orbit}.  The motion can be viewed as a {\bf retrograde
  epicycle} about a {\bf guiding center} that itself moves at a
constant rate around a circle of radius $R_g$, which is the radius of
a circular orbit having the same $L_z$.  In one complete radial
period, $\tau_R$ the star advances in azimuth through an angle
$\Delta\phi$, as drawn in the lower panel.  In fact, $\pi \leq
\Delta\phi \leq 2\pi$ in most reasonable gravitational potentials;
specifically $\Delta\phi = \surd2 \pi$ for small epicycles in a flat
rotation curve.

These periods can be used to define two angular frequencies for the
orbit: $\Omega_\phi = \Delta\phi / \tau_R$, which is the angular rate
of motion of the guiding center, and $\Omega_R = 2\pi / \tau_R$.  In
the limit of the radial oscillation amplitude, $a \rightarrow 0$,
these frequencies tend to $\Omega_\phi \rightarrow \Omega$, the
angular frequency of a circular orbit at $R=R_g$, and
$\Omega_R \rightarrow \kappa$, the {\bf epicylcic frequency}, defined
through
\be
\kappa^2(R_g) = \left( R{d\Omega^2 \over dR} + 4\Omega^2 \right)_{R_g}.
\ee

\subsection{Resonances}
\label{resonances}
If the potential includes an infinitesimal non-axisym\-metric
perturbation having $m$-fold rotational symmetry and which turns at
the angular rate $\Omega_p$, the {\bf pattern speed}, stars in the
disk encounter wave crests at the Doppler-shifted frequency
$m|\Omega_p-\Omega_\phi|$.  Resonances arise when
\be
m(\Omega_p-\Omega_\phi) = l\Omega_R,
\label{rescon}
\ee
where simple orbit perturbation theory breaks down for steady
potential perturbations.  At {\bf corotation} (\CR),\break where
$l=0$, the guiding center of the star's orbit has the same angular
frequency as the wave.  At the {\bf inner Lindblad resonance} (\ILR),
where $l=-1$, or at the {\bf outer Lindblad resonance} (\OLR), where
$l=+1$, the guiding center respectively overtakes, or is overtaken by,
the wave at the star's unforced radial frequency.  Other resonances
arise for larger $|l|$, but these three are the most important.  Note
that resonant orbits close after $m$ radial oscillations and $l$ turns
about the galactic center in a frame that rotates at angular rate
$\Omega_p$.

\begin{figure}[t]
\includegraphics[width=\hsize]{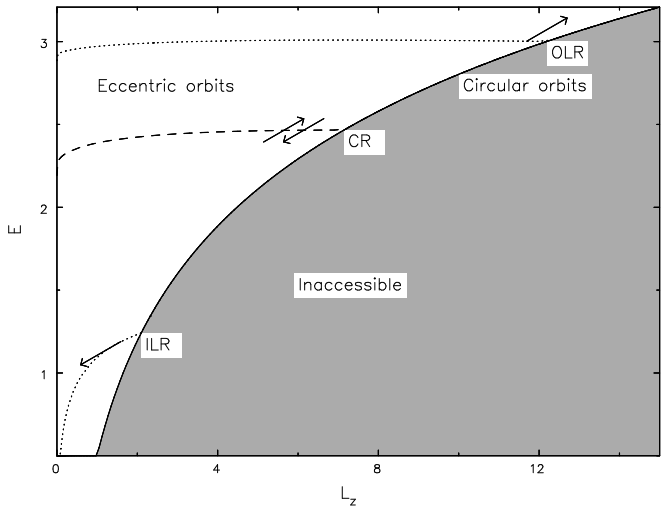}
\caption{The Lindblad diagram for a disk galaxy model.  Circular
  orbits lie along the full-drawn curve and eccentric orbits fill the
  region above it.  Angular momentum and energy exchanges between a
  steadily rotating disturbance and particles move them along lines of
  slope $\Omega_p$ as shown.  The dotted and dashed lines are the
  loci of resonances for an $m=2$ perturbation of arbitrary pattern
  speed.}
\label{lindblad}
\end{figure}

The solid curve in Fig.~\ref{lindblad}, the so-called Lindblad
diagram, marks the locus of circular orbits for stars in an
axisymmetric disk having a flat rotation curve.  Stars having more
energy for their angular momentum pursue non-circular orbits.  The
broken curves show the loci of the principal resonances for some
bi-symmetric wave having an arbitrarily chosen pattern speed; thus
resonant orbits, that close in the rotating frame, persist even to
high eccentricities.

\subsection{Wave-Particle Scattering}
The arrows in Figure~\ref{lindblad} indicate how these classical
integrals are changed for stars that are scattered by a steadily
rotating mild potential perturbation.  Since {\bf Jacobi's invariant},
\be
I_J \equiv E - \Omega_pL_z
\label{jacobi}
\ee
(\BTii\ eq.~3.112), is conserved in axes that rotate with the
perturbation, the slope of all scattering vectors in
Fig.~\ref{lindblad} is $\Delta E / \Delta L_z = \Omega_p$.

\cite{LBK} showed that a steadily rotating potential perturbation
causes lasting changes to the integrals only at resonances.  Since
$\Omega_c = \Omega_p$ at \CR, scattering vectors are parallel to the
circular orbit curve at this resonance, where angular momentum changes
do not alter the energy of random motion.  Outward transfer of angular
momentum involving exchanges at the Lindblad resonances, on the other
hand, does move stars away from the circular orbit curve, and enables
energy to be extracted from the potential well and converted to
increased random motion.

Notice also from Fig.~\ref{lindblad} that the direction of the
scattering vector closely follows the resonant locus (dotted curve) at
the \ILR\ only.  Thus, when stars are scattered at this resonance,
they stay on resonance as they gain random energy, allowing strong
scatterings to occur.  The opposite case arises at the \OLR, where a
small gain of angular momentum moves a star off resonance.

\subsection{Vertical Motion}
\label{vres}
Stars also oscillate vertically about the disk mid-plane.  If the
vertical excursions are not large, and the in-plane motion is nearly
circular, the vertical motion is harmonic with the characteristic
frequency $\nu = (\partial^2 \Phi / \partial z^2)^{1/2}$, which
principally depends on the density of matter in the disk mid-plane at
$R_g$.  As $\nu$ is significantly larger than $\kappa$, the frequency
of radial oscillation, the two motions can be considered to be
decoupled for small amplitude excursions in both coordinates.  In
practice, the vertical excursions of most stars are large enough to
take them to distances from the mid-plane where the potential departs
significantly from harmonic \citep[\ie\ for $|z| \ga 200\;$pc in the
  solar neighborhood,][]{HF04}.

The principal {\bf vertical resonances} with a rotating density wave
occur where $m(\Omega_p-\Omega_\phi) = \pm\nu$, which are at radii
farther from corotation than are the in-plane Lindblad resonances.
Since spiral density waves are believed to have substantial amplitude
only between the Lindblad resonances, the vertical resonances do not
affect density waves, and anyway occur where the forcing amplitude
will be small.

The situation is more complicated when the star's radial amplitude is
large enough that $\nu$ varies significantly between peri- and
apo-galacticon.  Also, the vertical periods of stars lengthen for
larger vertical excursions, allowing vertical resonances to become
important (see \S\ref{bentsheet}).

\ssection{Local Theory of Density Waves}
\label{local}
\subsection{Plane Waves in a Thin Mass Sheet}
Since solutions of Poisson's equation can be obtained analytically
only for the simplest mass distributions, theoretical treatments
necessarily make quite drastic simplifying assumptions.  Local theory,
summarized in this section, is built around a WKB-type potential
solution for density variations in the form of short wavelength plane
waves in a locally uniform, razor thin disk.  In this case, a
plane-wave disturbance $\Sigma_1e^{ikx}$ to the surface density in the
$(x,y)$-plane at some instant gives rise to the potential
\be
\Phi_1(\bx) = -{2\pi G \over |k|}\Sigma_1 e^{ikx} e^{-|kz|},
\label{planew}
\ee
(\cf\ eq.~6.31 of \BTii).  This relation strictly applies only to
straight waves of infinite extent.  However, the sinusoidal density
variations cause the field of distant waves to cancel quickly, and the
formula is reasonably accurate near the center of a short wave packet.
Note that eq.~(\ref{planew}) does not depend on the inclination of the
wavefronts to the radial direction, but the curvature of the
wavefronts must also be negligible, which requires $|kR| \gg 1$, with
$R$ being the distance from the disk center.

\subsection{Dispersion Relations}
\label{WKB}
A dispersion relation gives the relationship between the wavenumber
$k$ and frequency $\omega$ of self-consistent waves.  Assuming (a)
the above relation between density and potential for a razor-thin disk
and (b) the wave is of small amplitude, so that second order terms are
negligible, the local dispersion relation for short axisymmetric
density waves may be written
\be
\omega^2 = \kappa^2 - 2\pi G\Sigma |k| {\cal F}(s,\chi)
\label{LSKa}
\ee
\citep{Kaln65}, where $\Sigma$ is the local undisturbed surface
density and the complicated {\bf reduction factor} $\cal F$, with two
dimensionless arguments, is explained below.  \cite{LS66} independently
derived a generalized relation for {\it tightly-wrapped\/} spiral
waves, which required two additional assumptions: (c) that the wave
vector $\bk$ is approximately radial, and (d) that the disturbance is
not close to any of the principal resonances.  Their better-known
relation simply replaces the frequency of the wave with the
Doppler-shifted forcing frequency experienced by the orbit guiding
center:
\be
(\omega - m\Omega)^2 = \kappa^2 - 2\pi G\Sigma |k| {\cal F}(s,\chi)
\label{LSK}
\ee
(see also \BTii\ \S6.2.2), where now $\omega \equiv m\Omega_p$.

The reduction factor, ${\cal F}$, is always positive and is unity when
the disk has no random motion.  Since all the factors in the
self-gravity term on the RHS are intrinsically positive,
eqs.~(\ref{LSKa}) \& (\ref{LSK}) say that self-gravity enables a
supporting response from the stars at a frequency that is lower than
their free epicycle frequency $\kappa$.

When the disk has random motion, the vigor of the stellar response
depends on two factors: the ratio of the forcing frequency experienced
by a star to its natural frequency, $s \equiv |\omega -
m\Omega|/\kappa$, and the ratio of the typical sizes of the stellar
epicycles ($\propto \langle v_R^2 \rangle^{1/2}/\kappa$) to the
wavelength of the wave ($\propto |k|^{-1}$).  Thus for a Gaussian
distribution of radial velocities, $\cal F$ is a function of both $s$
and $\chi \equiv \sigma_R^2 k^2 / \kappa^2$.  Clearly, when $\chi$ is
large, the unforced epicyclic amplitude of most stars is larger than
the radial wavelength, and the weakened supporting response arises
mainly from the small fraction of stars near the center of the
velocity distribution.

\begin{figure}[t]
\includegraphics[width=\hsize]{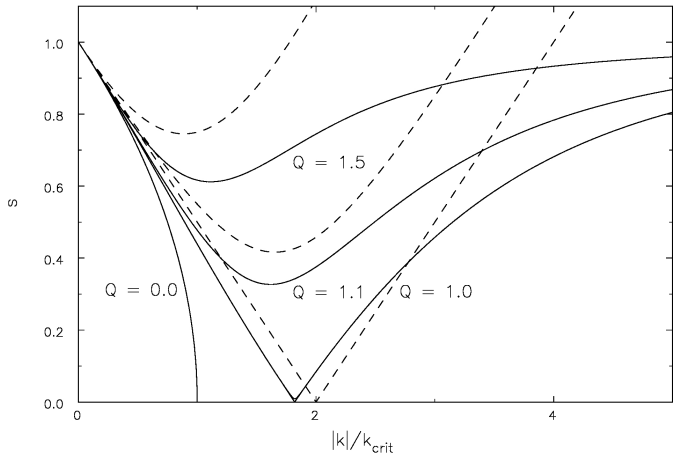}
\caption{The solid lines show the dispersion relation (\ref{LSK}),
  while the dashed lines are from (\ref{gasDR}), each for several
  values of $Q$.  The ordinate $s = |\omega - m\Omega|/\kappa$.  Since
  the signs of both the abscissa and the ordinate are suppressed, the
  graph is reflection symmetric about both axes.  Note that $s=0$ at
  \CR\ and $s=1$ at the Lindblad resonances.}
\label{disprel}
\end{figure}

The dispersion relation for barotropic fluid (gaseous)
disks\footnote{\cite{GLB1} derived the vertically-integrated equations
  for ``2D pressure'' in a gas disk.} is similar (\BTii\ eq.~6.55)
\be
(\omega - m\Omega)^2 = \kappa^2 - 2\pi G\Sigma |k| + v_s^2k^2,
\label{gasDR}
\ee
where $v_s$ is the sound speed in the gas.  Both dispersion relations
are shown graphically in Fig.~\ref{disprel}, below.

\subsection{Axisymmetric Stability}
\cite{Toom64} determined the condition for marginal stability of short
axisymmetric waves by solving for $\omega^2=0$ in
eq.~(\ref{LSKa}).  Since all the factors in the self-gravity term are
intrinsically positive, $\omega^2$ could be negative for large $|k|$,
implying instability.  Without random motion, ${\cal F} = 1$ and short
waves with $k > k_{\rm crit}$ or $\lambda < \lambda_{\rm crit} =
4\pi^2G\Sigma/\kappa^2$, will be unstable.  Unlike for the Jeans
instability in a stationary medium, longer waves are stabilized by
rotation, embodied in the $\kappa^2$ term.

For disks with random motion, the reduction factor ${\cal F}
\rightarrow 0$ for large $|k|$, irrespective of the frequency, for the
reason given above.  Thus random motion stabilizes short waves, as for
the Jeans instability.  \cite{Toom64} showed for a Gaussian
distribution of radial velocities that the RHS of eq.~(\ref{LSKa}) is
$\geq 0$ for all $|k|$ when $\sigma_R \geq \sigma_{R,{\rm crit}}$,
where
\be
\sigma_{R,{\rm crit}} \simeq {3.36G\Sigma \over \kappa}.
\ee
Thus {\bf Toomre's local axisymmetric stability criterion} is $Q
\equiv \sigma_R/ \sigma_{R,{\rm crit}} \geq 1$.  When reasonably
constant with radius, the locally estimated $Q$ value is a good
indicator of global axisymmetric stability.  For example, the local
values in the equilibrium models proposed by \cite{Kaln76} are in
reasonable agreement with those he derived from global axisymmetric
modes (his Fig.~2).  It should be noted that an axisymmetrically
stable disk, \ie\ for which $Q \geq 1$ everywhere, could still be
unstable to non-axisymmetric modes.

The local axisymmetric stability criterion for rotating gas disks is
similar.  The longest unstable wavelength, $\lambda_{\rm crit}$ in a
cold ($v_s=0$) disk is the same as for a stellar disk.  The minimum of
the quadratic expression in $k$ on the RHS of eq.~(\ref{gasDR}) occurs
for $k=\pi G\Sigma/v_s^2$, and the condition of marginal stability
($\omega^2=0$) at this minimum is readily solved, yielding $v_{s,\rm
  crit} = \pi G\Sigma/\kappa$.

\cite{Vand70}, \cite{Rome92}, and others have extended the discussion
of axisymmetric stability to disks of finite thickness.  The principal
difference is the dilution of the self-gravity term caused by
spreading the disk matter in the vertical direction, resulting in a
somewhat more stable disk.  The precise correction depends on the
assumed vertical structure of both the disk and the wave, but is minor
when the characteristic disk thickness $z_0 \ll \lambda_{\rm crit}$,
which is usually the case.

More realistic composite disks consist of multiple stellar populations
having a range of velocity dispersions, as well as a cool gas
component.  \cite{JS92} considered the stability of two
gravitationally coupled gas disks having different sound speeds, but
\cite{Rafi01} improved their analysis to include multiple stellar
components.  These analyses show that while the gas component may
contain a small fraction (typically $\la 10\%$) of the total mass, its
low sound speed causes it to have a disproportionate effect on the
overall stability.

\subsection{Self-consistent Density Waves}
\label{scwaves}
Fig.~\ref{disprel} gives, for four different values of $Q$, a
graphical representation of the stellar (solid curves),\footnote{The
  reduction factor used assumes a Gaussian distribution of radial
  velocities} and gaseous (dashed curves) dispersion relations
eqs.~(\ref{LSK}) \& (\ref{gasDR}) respectively.  The curves do not
differ for $Q=0$ for which $s$ is imaginary when $k>k_{\rm crit}$.
The dashed and solid curves have similar forms for moderate $k$ but
differ substantially for large $k$, where frequencies in the stellar
sheet cannot exceed $|\omega-m\Omega|=\kappa$.  Thus waves in a
stellar sheet extend only from \CR, where $s=0$, to the Lindblad
resonance on either side, where $s=1$.\footnote{Additional
  Bernstein-type waves exist near integer values of $s>1$, but such
  solutions seem to be of little dynamical importance.}

A gas layer, on the other hand, can support gravitationally modified
sound waves of arbitrarily high frequency.  Note that this important
difference from the stellar case can lead to artificial modes when
hydrodynamic equations are used as a simplifying approximation for a
stellar disk; while meaningful results can be obtained using this
approximation, it is important to check that the derived modes do not
cross Lindblad resonances.

Since the dispersion relation (\ref{LSK}) does not depend on the sign
of $k$, curves for negative $k$, or leading waves, are simple
reflections about the $k=0$ axis.  Furthermore, the relation gives
the square of $(\omega-m\Omega)$, implying that solutions when this
frequency is negative, which arise inside \CR, are simple reflections
about the $s=0$ axis.  Thus were the signs not suppressed,
Fig.~\ref{disprel} shows only the panel for trailing waves outside \CR,
while the other quadrants would be reflected images.

In the quadrant shown, there are either two values of $k$ for each
value of $s$, known as the {\bf short-} and {\bf long-wave branches},
or there are none.  Only for the marginally stable case of $Q=1$ do
solutions exist for all frequencies $0 \leq s \leq 1$ and all $|k|$.
When $Q>1$, a {\bf forbidden region} opens up in the vicinity of
\CR\ where waves are evanescent.

Recall that the relation (\ref{LSK}) was derived assuming both $|kR|
\gg 1$ and that the waves are tightly wrapped -- \ie\ that the wave
vector is directed radially.  These approximations may not be too bad
on the short-wave branch (large $k$) but long-wave branch solutions
may not exist at all, except in a very low-mass disks where $k_{\rm
  crit}$ is large, and {\it must\/} fail as $|k| \rightarrow 0$.
\cite{LBK} gave an improved dispersion relation (their eq.~A11) that
relaxed the requirement that the wave vector be radial, but still used
the plane wave potential (\ref{planew}) that requires $|kR| \gg 1$.
Their relation is no different for tightly-wrapped waves, but has no
long-wave branch or forbidden region for open waves when $Q \ga 1$.

\subsection{Group Velocity}
\label{vgroup}
The waves in a disk described by the local dispersion relations
(\ref{LSK}) \& (\ref{gasDR}) have a phase speed, equal to the pattern
speed, in the azimuthal direction.  However, \cite{Toom69} pointed out
that these dispersive waves also have a group velocity that is
directed radially.  Since $v_g = \partial\omega/\partial k$, the group
velocity is proportional to the slope of the lines in
Fig.~\ref{disprel}.  Each portion of a wavetrain, or wave packet, is
carried radially at the group velocity, retaining its pattern speed
$\omega$ whilst gradually adjusting its wavelength, and also
transporting {\bf wave action} or angular momentum.

For trailing waves outside \CR, the situation illustrated in
Fig.~\ref{disprel}, the group velocity is positive on the short-wave
branch, and the waves carry angular momentum outwards towards
the \OLR.  The curves have the opposite slope inside \CR, where
density waves are disturbances of negative angular
momentum \citep[stars there lose angular momentum as the wave is set
  up,][]{LBK}, and therefore the inward group velocity on the
short-wave branch leads once more to outward angular momentum
transport.  Outward angular momentum transport is in the sense
expected from the gravitational stresses of a trailing spiral
\citep{LBK}.

However, the sign of the slope $\partial\omega/\partial k$ on the
dubious long-wave branch is opposite to that of short waves,
indicating that angular momentum in that regime is transported in the
direction opposite to the gravity torque!  This apparent contradiction
is resolved by the phenomenon of {\bf lorry transport}, a term coined
by \cite{LBK} to describe an advective Reynolds stress (see also
Appendix J of \BTii).  Stars in their epicyclic oscillations gain
angular momentum from the wave near apo-center and lose it back to the
wave near peri-center; no star suffers a net change, but angular
momentum is still carried inward at a sufficient rate to overwhelm the
gravity torque.

Thus where eq.~(\ref{LSK}) holds in a disk with approximately uniform
$Q \ga 1$, a {\it tightly wrapped\/} trailing wave packet originating
on the long-wave branch, will travel towards \CR\ until it reaches the
edge of the forbidden zone where it ``refracts'' into a short wave
that carries it back towards the Lindblad resonance.  The details of
the turning point at the forbidden zone, which requires an evanescent
wave propagating into the forbidden region, were described by
\cite{Mark76} and by \cite{GT78}.

The fate of the short-wave propagating towards the\break Lindblad
resonance also requires more sophisticated analysis because the
dispersion relation (\ref{LSK}) does not hold near resonances.
\cite{Mark74} carried the analysis to second order to show that the
wave is absorbed there through the wave-particle coupling described
by \cite{LBK}.

\subsection{Swing Amplification}
\label{swing}
\cite{GLB2}, \cite{JT66} and \cite{Toom81} extended the above analysis
for tightly-wrapped waves to waves of arbitrary inclination.  They
again adopted the approximate potential of a short-wave\-length plane
wave (eq.~\ref{planew}), and focused on a local patch of the disk
whose dimensions were assumed small compared with $R_0$, the radial
distance of the patch center, which orbits the disk center at the
local circular speed $V=R_0\Omega_0$.

Following \cite{Hill}, they then introduced Cartesian-like
coordinates, $x \equiv R-R_0$ and $y \equiv R_0(\phi-\phi_0)$, where
$\phi_0$ is the azimuth of the orbiting center of the patch.  The
vertical coordinate, $z$, is unchanged from the usual cylindrical
polar coordinate.  Since the angular rotation rate decreases outwards
in galaxies, the motion of stars in the patch is that of a shear flow,
with a speed that varies as $\dot y = -2Ax$, together with Coriolis
forces arising from the rotation of the patch at angular rate
$\Omega_0$.  The Oort ``constant'' $A = -{1\over2}R d\Omega/dR$.  The
undisturbed surface density of matter, $\Sigma$, is assumed constant
over the patch.  This set of approximations is described as the {\bf
sheared sheet} (see \BTii\ pp.~678-681).

If $\alpha$ is the inclination angle of a plane density wave to the
azimuthal direction, with $180^\circ > \alpha > 90^\circ$ for leading
waves and $90^\circ > \alpha > 0^\circ$ for trailing waves, then the
shear causes the pitch angle to change with time as ${\rm cot}\,\alpha
= 2At$, with $t=0$ when $\alpha=90^\circ$, \ie\ when the wave is
purely radial.  Both \cite{GLB2} for a gaseous sheet and \cite{JT66}
for a stellar sheet found waves amplify strongly, but transiently, as
they shear from leading to trailing.

The unremitting shear flow ultimately tears apart any transient
disturbance, and indeed \cite{JT66} asserted that the stellar sheet is
locally stable at small-amplitudes when $Q \geq 1$, but they did not
give the proof.

\begin{figure}[t]
\includegraphics[width=\hsize]{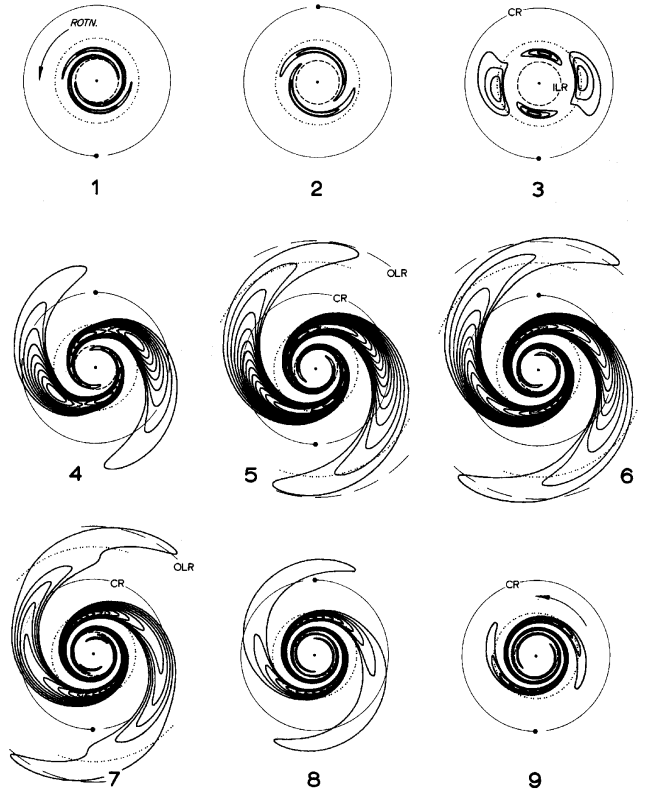}
\caption{The time evolution of an input leading wave packet in the
  half-mass Mestel disk.  The unit of time is half a circular orbit
  period at the radius marked \CR. Reproduced from \cite{Toom81}.}
\label{dusttoashes}
\end{figure}

Fig.~\ref{dusttoashes}, from \cite{Toom81}, illustrates the dramatic
transient trailing spiral that results from a small input leading
disturbance.  This calculation illustrates that even a global,
small-amplitude disturbance does not persist, but decays after its
transient flourish.  In the late stages, the ``wave action''
propagates at the group velocity (\S\ref{vgroup}) away from
corotation, where it is absorbed at the Lindblad resonances.

Notice also that the wave in this $Q=1.5$ disk extends across the
forbidden region near \CR.  The absence of solutions near $s=0$ when
$Q>1$ in Fig.~\ref{disprel} is a consequence of assuming a steady,
tightly-wrapped wave; shearing waves do not respect this restriction.

The responsiveness of a disk to input leading signal depends both on
the degree of random motion -- it decreases rapidly over the range $1
< Q < 2$ -- and upon the azimuthal wavelength, which is $\lambda_y =
2\pi R/m$ for an $m$-armed pattern.  \cite{JT66} defined the parameter
\be
X \equiv {\lambda_y \over \lambda_{\rm crit}} = {R k_{\rm crit} \over m}.
\label{Xdef}
\ee
In a disk having a flat rotation curve, \cite{Toom81} showed that the
strongest amplification occurs when $1 \la X \la 2.5$, or about {\it
  twice\/} the scale of the longest unstable axisymmetric wave.  The
peak amplification shifts to larger $X$ values for declining rotation
curves and {\it vice versa\/} for rising ones.  Naturally, the
phenomenon of swing amplification disappears entirely in uniformly
rotating disks.

By making the assumption that spiral patterns in galaxies are
amplified solely by this mechanism, \cite{ABP87} argued that the
multiplicity of the observed spiral arm pattern can be used to place
bounds on the mass density in the disk.  The circular orbit speed in a
nearly axisymmetric galaxy affords a direct estimate of the central
attraction, but it is hard to determine how much of the total
attraction is contributed by the disk, and how much by other
components, such as the dark matter halo.  Because $\lambda_{\rm
  crit}$ is proportional to the disk surface density, for a given
rotation curve, the most vigorous amplification shifts to higher
sectoral harmonics in lower mass disks.  Generally, disks that
contribute most of the central attraction, \ie\ close to {\bf maximum
  disks}, would prefer bi-symmetric spirals, while higher arm
multiplicities would indicate significantly sub-maximal disks
\citep[see also][]{SC84}.

\begin{figure}[t]
\includegraphics[width=\hsize]{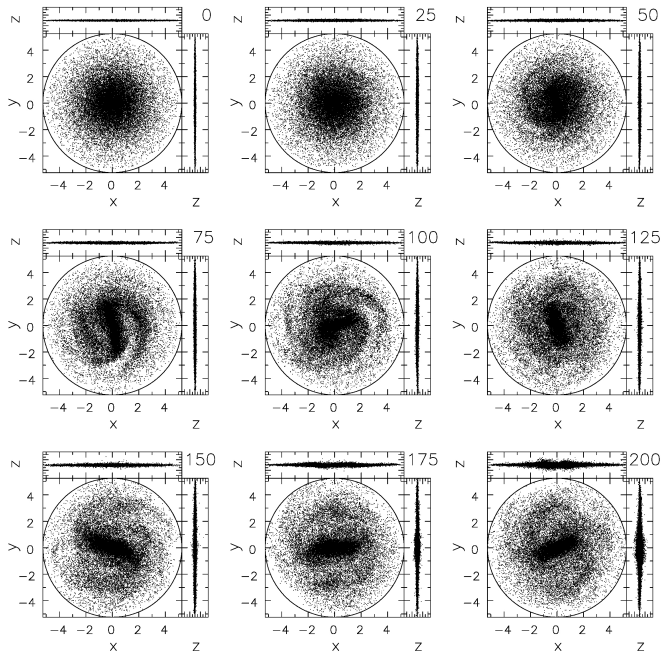}
\caption{The formation of a bar in an (unpublished) $N$-body
  simulation.  The three orthogonal projections show only particles in
  the initially exponential disk, halo particles are omitted.  The
  disk, which started with $Q=1.2$, has unit mass and unit scale
  length and $G=1$.  One orbit at $R=2$ takes 15 time units, where the
  central attractions of the disk and halo are nearly equal.}
\label{barform}
\end{figure}

\ssection{Bar Instability}
\label{global}
The puzzle of how galaxy disks could be stable presented a major
obstacle to the development of our understanding of disk dynamics for
many years.  Superficially reasonable models of disk galaxies were
found repeatedly, both in simulations
\citep[\eg][]{MPQ70,Hohl71,ZH78,CS81,Sell81,AS86,KJKJ,DBS09} and in
linear stability analyses
\citep{Kaln78,ANI79,Toom81,Sawa88,VD96,PC97,Korc05,Poly05,Jala07} to
possess vigorous, global and disruptive bar-forming instabilities.
Fig.~\ref{barform} illustrates the global, disruptive nature of this
instability.  While it is premature to claim that this problem has
been completely solved, it now seems that the stability of a massive
disk galaxy requires only a dense bulge-like mass component near the
center, and owes little to the inner density of a dark matter halo.
Galaxies lacking a central mass concentration, however, are still
believed to require significant inner halo mass for global stability.

\subsection{Mechanism for the Bar Mode}
\label{barmech}
\cite{Toom81} provided the most important step forward by elucidating
the mechanism of the bar instability (see also \BTii\ \S6.3).  Linear
bar-forming modes are standing waves in a cavity, akin to the familiar
modes of organ pipes and guitar strings.  Reflections in galaxies take
place at the center and at the corotation radius, except that outgoing
leading spiral waves incident on the corotation circle are
super-reflected into amplified ingoing trailing waves (\ie\ swing
amplification), while also exciting an outgoing transmitted trailing
wave.  The feedback loop is closed by the ingoing trailing wave
reflecting off the disk center into a leading wave, which propagates
outwards because the group velocity of leading waves has the opposite
sign to that of the corresponding trailing wave.  The amplitude of the
continuous wave-train at any point in the loop rises exponentially,
because the circuit includes positive feedback.

Toomre supported this explanation with linear stability studies of two
disks.  The Gaussian disk, which has a low central density, manifests
a set of modes in which the more slowly-growing, higher ``overtones''
display the kind of standing wave pattern to be expected from the
superposition of ingoing trailing and outgoing leading waves.  (The
eigenfrequencies of these modes are shown below in Fig.~\ref{modeD}.)

The other linear stability study he presented was for the
inappropriately-named ``Mestel'' disk,\footnote{\cite{Mest63} solved
the far greater challenge of a disk of finite radius that has an
exactly flat rotation curve.} whose unusual stability properties were
first described by Toomre's student \cite{Zang76} in an unpublished
thesis, and later by \cite{ER98}.  This disk has the scale-free
surface density profile $\Sigma = V_c^2/(2\pi GR)$, with $V_c$ being
the circular orbital speed that is independent of $R$.  Zang had
carried through a global, linear stability analysis of this disk, with
random motions given by a smooth distribution function.  In order to
break the scale-free nature of the disk, Zang introduced a central
cutout, and later an outer taper, in the active mass density,
replacing the removed mass by rigid components in order that the
central attraction remained unchanged at all radii.  The dominant
linear instabilities he derived for the tapered disk were confirmed in
$N$-body simulations by \cite{SE01}.

\cite{Zang76} showed that a full-mass Mestel disk is stable to
bi-symmetric modes, even if $Q<1$, provided the tapers are gentle
enough.  \cite{ER98} extended this important result to other power-law
disks, finding they have similar properties.  Such disks are not
globally stable, however, because they suffer from lop-sided
instabilities (see \S\ref{lopsided}).  By halving the active disk
mass, with rigid mass preserving the overall central attraction,
\cite{Toom81} was well able to eliminate the lop-sided mode from the
Mestel disk.  In fact, he claimed that when $Q=1.5$, the half-mass
Mestel disk was globally stable, making it the only known model of a
non-uniformly rotating disk that is stable to all small amplitude
perturbations.

Cutting the feedback loop really does stabilize a disk (see next
section), which seems to confirm Toomre's cavity mode mechanism.
Despite this, \cite{Poly04} pointed out that the strong emphasis on
phenomena at corotation places the blame for the instability on a
radius well outside where the resulting bar has its peak amplitude.
He therefore proposed an alternative mechanism for the bar instability
based upon orbit alignment \citep[][see also \S\ref{barorig}]{LB79}.
Even though the mechanism was originally envisaged as a slow trapping
process, \cite{Poly04} argued it may also operate on a dynamical
timescale, and he devised \citep{Poly05} an approximate technique
to compute global modes that embodied this idea.  While his method
should yield the same mode spectrum as other techniques, his
alternative characterization of the eigenvalue problem may shed
further light on the bar-forming mechanism.

\subsection{Predicted Stability}
The fact that small-amplitude, bi-symmetric instabilities are so
easily avoided in the Mestel disk, together with his understanding of
bar-forming modes in other models, led \cite{Toom81} to propose that
stability to bar-formation merely required the feedback cycle through
the center to be cut.  He clearly hoped that a dense bulge-like mass
component, which would cause an \ILR\ to exist for most reasonable
pattern speeds, might alone be enough to stabilize a cool, massive
disk.

Unfortunately, this prediction appeared to be contradicted almost
immediately by the findings of \cite{ELN82}, whose $N$-body
simulations formed similar bars on short time\-scales irrespective of
the density of the central bulge component!  They seemed to confirm
previous conclusions \citep{OP73} that only significantly sub-maximal
disks can avoid disruptive bar-forming instabilities.

However, \cite{Sell89a} found that Toomre's prediction does not apply
to noisy simulations because of non-linear orbit trapping.
Simulations in which shot noise was suppressed by quiet start
techniques did indeed manifest the tendency towards global stability
as the bulge was made more dense, as Toomre's linear theory predicted.
Shot noise from randomly distributed particles, on the other hand,
produced a large enough amplitude collective response for
non-linearities to be important, and a noisy simulation of a linearly
stable disk quickly formed a strong bar, consistent with the results
reported by \cite{ELN82}.

Since density variations in the distribution of randomly distributed
particles are responsible for bar formation in this regime, the
reduced shot noise level from larger numbers of particles must result
in lower amplitude responses that ultimately should avoid non-linear
trapping.  The precise particle number required depends on the
responsiveness of the disk, which is weakened by random motion, lower
surface density, by increased disk thickness, or gravity softening.
\cite{ELN82} employed merely $20\,000$ particles, which was clearly
inadequate to capture the linear behavior.  However, robustly stable,
massive disks have been simulated by \cite{SM99} and \cite{SE01} that
employed only slightly larger particle numbers.  Note that these latter
models also benefitted from more careful set-up procedures to create
the initial equilibrium.

Thus the stabilizing mechanism proposed by \cite{Toom81} does indeed
work in simulations of high enough quality, and presumably therefore
also in real galaxy disks.  Indeed, \cite{BJM08} found a decreased
incidence of bars in galaxies having luminous bulges, and argued that
their result supports Toomre's stabilizing mechanism.

Thus the absence of bars in a significant fraction of high-mass disk
galaxies does not imply that the disk is sub-maximal.  The old
stability criteria proposed by \cite{OP73}, \cite{ELN82},
and \cite{CST95} apply only to disks that lack dense centers;
indeed \cite{ER98} showed explicitly that the power-law disks are
clear counter-examples to the simple bi-symmetric stability criterion
proposed by \cite{OP73}.

\subsection{Residual Concerns}
While all this represents real progress, a few puzzles remain.  The
most insistent is the 
absence of large, strong bars in galaxies like M33, which has a gently
rising rotation curve.  Although many spiral arms can be counted in
blue images \citep{SH80}, the near IR view \citep{Bloc04} manifests an
open two-arm spiral, suggesting that the disk cannot be far from
maximal, and also reveals a mild bar near the center of the galaxy.
\cite{CW07} also found kinematic evidence for a weak bar.  Perhaps the
stability of this galaxy can be explained by some slightly larger halo
fraction, or perhaps the weak bar has some unexpected effect, but
there is no convincing study to demonstrate that this galaxy, and
others like it, can support a 2-arm spiral without being disruptively
unstable.

The second concern is that lop-sided instabilities appear in extended
full-mass disks with flat or declining rotation curves, which is
discussed next.  A third concern, which is discussed
in \S\ref{barfreq}, is that the mechanism is unable to predict the
presence or absence of a bar in a real galaxy.

\ssection{Lop-sided Modes}
\label{lopsided}
Many galaxies have apparently lop-sided disks.  The treatment here
will not go into detail, since \cite{JC09} have recently reviewed both
the observational data and theory.

Both theoretical and simulation work on $m=1$ distortions to an
axisymmetric disk require special care, since the absence of
rotational symmetry can lead to artifacts unless special attention is
paid to linear momentum conservation.  Rigid mass components present
particular difficulties, since they should not be held fixed, and
extensive mass components are unlikely to respond as rigid objects.

As noted above, \cite{Zang76} found that the dominant instability of
the centrally cut-out Mestel disk was not the usual bar instability,
but a lop-sided mode, which persists in a full mass disk no matter how
large a degree of random motion or gentle the cutouts.  This
surprising finding was confirmed and extended to general power-law
disks by \cite{ER98}.  A lop-sided instability dominated simulations
\citep{Sell85} of a model having some resemblance to Zang's, in that
it had a dense massive bulge and no extended halo, while \cite{SCJ07}
reported similar behavior in simulations of a bare exponential disk.
\cite{LZKH} found pervasive lop-sided instabilities near the disk
center in a study of the collective modes of a set of mass rings.

Various mechanisms have been proposed to account for this instability.
\cite{BLBS} and \cite{ELB96} explored the idea that long-lived
lop-sidedness could be constructed from cooperative orbital responses
of the disk stars, along the lines discussed for bars by \cite{LB79}.
\cite{Trem05} discussed a self-gravitating secular instability in
near-Kep\-lerian potentials.\footnote{The ``sling amplification''
  mechanism proposed by \cite{Shu90} applies only to gaseous accretion
  disks, since it relies on sound waves propagating outside the OLR.}
A more promising mechanism is a cavity mode, similar to that for the
bar-forming instability \citep{DdRDD}: the mechanism again supposes
feedback to the swing-amplifier,\break which is still vigorous for
$m=1$ in a full-mass disk.

\begin{figure}[t]
\includegraphics[width=\hsize]{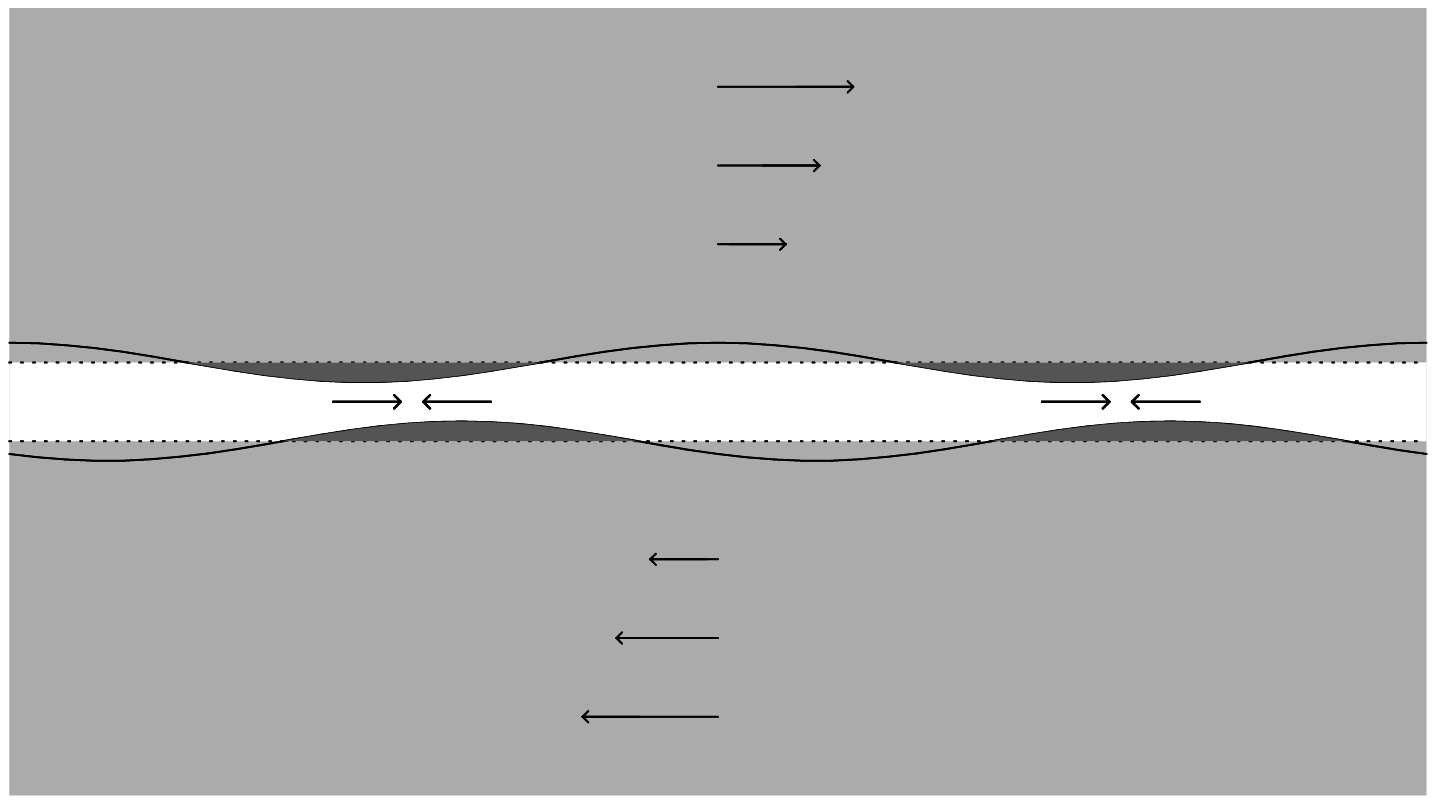}
\includegraphics[width=\hsize]{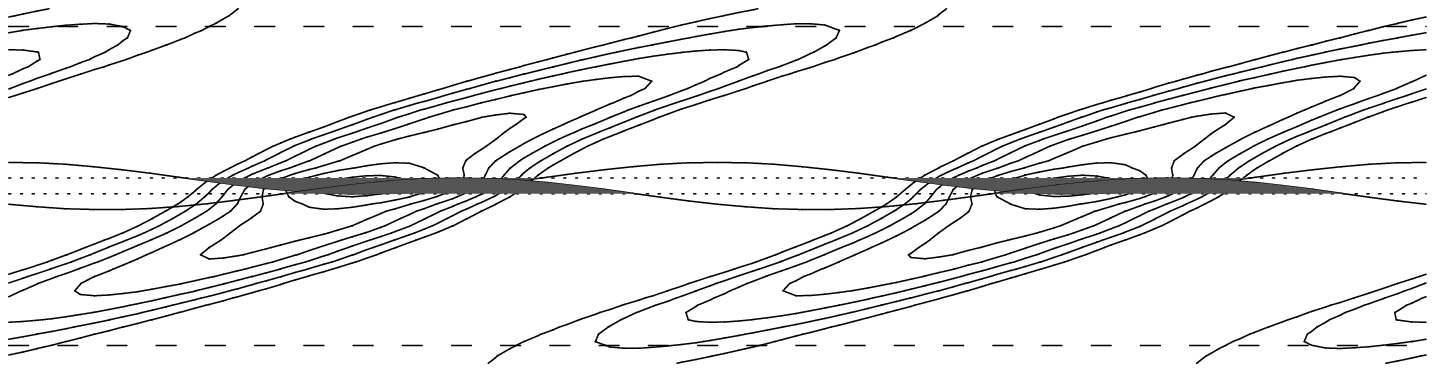}
\caption{The top panel illustrates a groove in the sheared sheet
  (\S\ref{swing}) model.  The light shaded region has the full
  undisturbed disk surface density, $\Sigma$, which in this frame has
  a linearly varying shear flow.  The white region is the groove,
  which has lower surface density, and the density excesses brought
  into the grove by small wave-like disturbances on either side of the
  groove are highlighted by the dark shading.  The width of the groove
  is exaggerated for clarity. \hfil\break The lower panel, which has
  the correct aspect ratio, shows the supporting responses to the
  density variations on the groove edges.  The dashed lines mark the
  locations of the Lindblad resonances.}
\label{groove}
\end{figure}

Feedback through the center cannot be prevented by an \ILR\ for $m=1$
waves, since the resonance condition $\Omega_p = \Omega_\phi -
\Omega_R$ (eq.~\ref{rescon}) can be satisfied only for retrograde waves.
But the lopsided mode can be stabilized by reducing the disk mass,
which reduces the $X$ parameter (eq.~\ref{Xdef}) until amplification
dies for $m=1$ \citep{Toom81}.  \cite{SE01} showed that, together with
a moderate bulge, the dark matter required for a globally stable disk
need not be much more than a constant density core to the minimum halo
needed for a flat outer rotation curve.

A qualitatively different lop-sided instability is driven by
counter-rotation.  This second kind of $m=1$ instability was first
reported by \cite{ZH78} in a series of $N$-body simulations designed
to explore the suppression of the bar instability by reversing the
angular momenta of a fraction of the stars; they found that a
lop-sided instability was aggravated as more retrograde stars were
included in their attempts to subdue the bar mode.  Analyses of
various disk models with retrograde stars \citep{Arak87,Sawa88,DdRDD}
have revealed that the growth rates of lop-sided instabilities
increase as the fraction of retrograde stars increases.  \cite{MS90}
and \cite{SV97} found lop-sided instabilities in simulations of oblate
spheroids with no net rotation.  Their flatter models had velocity
ellipsoids with a strong tangential bias, whereas \cite{SM94} found
that disks with half the stars retrograde, together with moderate
radial motion were surprisingly stable.

\cite{Wein94} pointed out that lop-sided distortions to near spherical
systems can decay very slowly, leading to a protracted period of
``sloshing''.  This {\bf seiche mode} in a halo, which decays
particularly slowly in mildly concentrated spherical systems, could
provoke lop-sidedness in an embedded disk \citep{Korn02,Idet02}.

\begin{figure*}[t]
\begin{center}
\includegraphics[width=.8\hsize,clip]{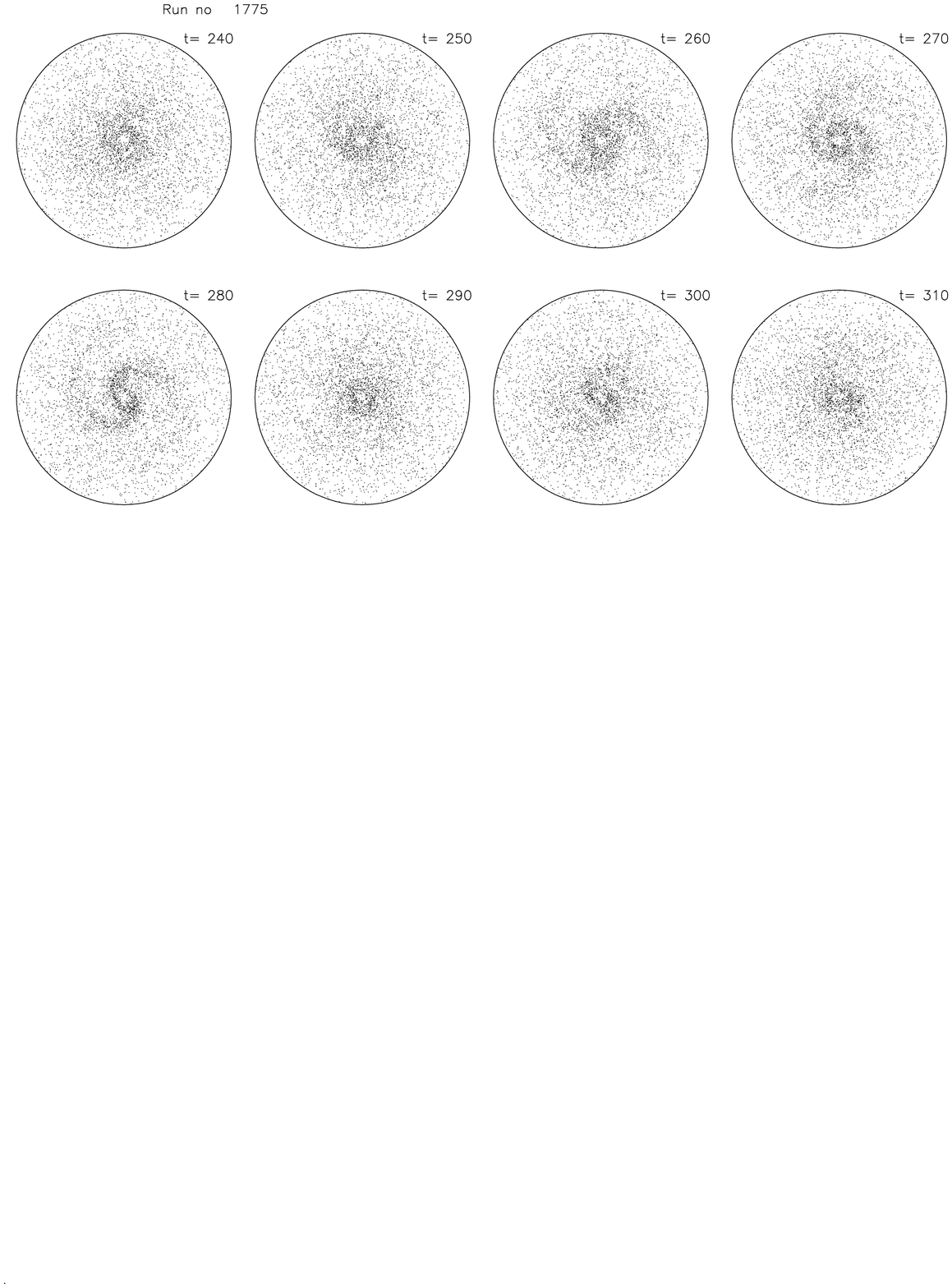}
\end{center}
\caption{The later part of the growth and subsequent decay of an
  isolated spiral mode in a disc that was seeded with a groove.  Disk
  rotation is counter-clockwise and disturbance forces in this
  simulation, taken from \cite{SB02}, were restricted to $m=2$ only.}
\label{SBsim}
\end{figure*}

If lop-sidedness is due to instability, then the limited work so far
suggests that it would imply a near-maximum disk.  But lop-sidedness
in the outer parts could also be caused by tidal interactions, or
simply by asymmetric disk growth, with the effects of differential
rotation being mitigated perhaps by the cooperative orbital responses
discussed by \cite{ELB96}.

\ssection{Groove and Edge Modes}
\label{gmodes}
Not all true global instabilities in disks require a feedback loop.
Another class of modes is simply driven from corotation, with trailing
waves propagating both inwards and outwards to be absorbed, in stellar
disks, at the Lindblad resonances on either side.

\cite{LH78} showed that disks are destabilized by a local extremum in
the radial variation of the ratio of surface density to vorticity,
$\Sigma / |{\bf \nabla}\cross\bv| \equiv \Sigma \Omega / (2\kappa^2)$,
or the reciprocal of potential vorticity, and proposed that the
instability created flat rotation curves for which the potential
vorticity is also flat.  \cite{SK91} demonstrated that the instability
caused by a rather\break insignificant, narrow, decrease in surface
density, \ie\ a ``groove'', is a global spiral mode.

The mechanism is easiest to visualize in a disk without random motion,
where small-scale surface density variations are not blurred by
epicyclic motions.  In this case, a deficiency of stars over a small
range of angular momentum creates a groove in the surface density, as
shown in the sheared sheet model in the top panel of
Fig.~\ref{groove}.  The groove itself is unstable because of the
gravitational coupling between disturbances on either side.  The dark
shaded areas in the Figure illustrate regions where small sinusoidal
radial displacements of material on each edge have created high
density regions where the density was previously low.  Disturbing
gravitational forces arise from the density excesses, as illustrated,
which are directed along the groove if the wavelength is long compared
with the groove width.  Material that is pulled back loses angular
momentum and sinks toward the center of the galaxy, while that which
is urged forward gains and rises outwards.  Thus each density excess
pulls on the other across the groove in such a way as to cause it to
grow exponentially, \ie\ the combined disturbance on the two sides is
unstable.  The groove edges need be only steep gradients, not
discontinuities.  Furthermore, the mechanism is the same, but harder
to visualize, in a disk with random motion where the density of stars
is depleted over a narrow range of angular momentum.

The growing disturbance in the groove creates wave-like mass
variations along the groove that are effectively growing co-orbiting
mass clumps of the type envisaged by \cite{JT66}.  Generalizing their
apparatus to allow for exponentially growing masses, \cite{SK91}
estimated the expected disk response, as shown in the lower panel of
Fig.~\ref{groove} for the parameters $Q=1.8$ and $X=2$, and low growth
rate.  The disk supporting response transforms the quite trivial
disturbance in the groove into an extensive spiral instability!

Unlike the sheared sheet, the azimuthal wavelength in a full disk can
take on only discrete values, $\lambda_y = 2\pi R/m$, which are all
unstable, but the one which grows most rapidly is that for which
swing-amplification (\S\ref{swing}) is the most vigorous, \ie\ $1 \la
X \la 2.5$ (eq.~\ref{Xdef}) in a flat rotation curve.  The simulation
shown in Fig.~\ref{SBsim} illustrates the scale and vigor of the mode,
which saturates at $\sim 20\%$ overdensity because of the onset of
horseshoe orbits at corotation \citep{SB02}.

\cite{LH78} and \cite{SK91} found that almost any narrow feature in
the angular momentum density is destabilizing, although the modes of a
simple ridge, for example, come in pairs with \CR\ some distance from
the ridge center.

\begin{figure}[t]
\includegraphics[width=\hsize]{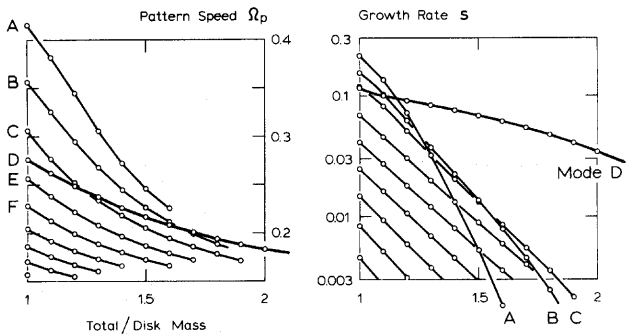}
\caption{The variation of pattern speeds and growth rates for the
  first few $m=2$ modes of the Gaussian disk as the active disk mass
  is decreased.  From \cite{Toom81}.}
\label{modeD}
\end{figure}

{\bf Edge modes} \citep{Toom81,PL89} are close cousins of groove
modes, and the mechanism can be understood in the same fashion.
Density variations on a single edge without a supporting disk response
are neutrally stable, but the necessarily one-sided wakes in the
massive interior disk add angular momentum to the density enhancements
on the edge, causing them to grow.

The eigenfrequencies of the low order bisymmetric\break modes of the
Gaussian disk vary with decreasing disk mass as shown in
Fig.~\ref{modeD}, taken from \cite{Toom81}.  Most are cavity modes,
but the edge-mode, labelled ``mode D'' stands out because its growth
rate in particular declines more slowly with decreasing disk mass.  If
the mechanism for both types of mode involves swing amplification, one
might expect their vigor to be similarly affected by the rise in the
$X$ parameter as the surface density decreases.  But the decreased
surface density also slows the group velocity while the rapidly
declining pattern speed of each cavity mode moves \CR\ farther out in
the disk.  Thus the growth rates of cavity modes drop more quickly
because of the increased travel time for the wave packet to complete
the feed-back loop, a factor that is absent for the edge mode.

The edge instability requires only a steep gradient in the surface
density, which need not drop to zero.  \cite{Toom89} gave the
condition for instability as ``the radial distance over which the disc
density undergoes most of its rapid change should be no larger than
about one quarter of the axisymmetric stability length $\lambda_{\rm
  crit}$'', assuming ``the disc is massive and cool enough for
vigorous swing-amplification.''

Curiously, global bar-forming modes were originally\break thought to
be related the rotational instabilities of uniform-density Maclaurin
spheroids of incompressible fluid\break \citep{OP73}, an idea
reinforced by the vigor of the bisymmetric mode of the sharp-edged
Maclaurin disk \citep{Kaln72}.  \cite{Toom81} noted that the
instability of such unrealistic galaxy models may be more closely
related to the edge mode than to the cavity mode described
in \S\ref{barmech}.

As the consequence of an edge instability in a realistic galaxy model
is to blur the edge, it seems unlikely that galaxy disks can retain
unstable density gradients for interestingly long periods.  Similarly,
the large-amplitude evolution of groove or ridge modes quickly erases
the feature that gave rise to them.  Nevertheless, the modes have
other possible consequences (see \S\ref{recrrnt}).

\ssection{Spiral Structure}
\label{spirals}
Despite many decades of effort, no theory to account for the graceful
spiral patterns in disk galaxies is widely accepted.  Most workers in
this field agree that spiral patterns are gravitationally driven
variations in surface density in the old stellar disk, as supported by
photometric data \citep[\eg][]{GPP04,ZCR9} and streaming motions in
high spatial resolution velocity maps \citep[\eg][]{SVOT}.

There seems little doubt that some spiral patterns are tidally driven
\cite[\eg][]{SL93,DTPB}, while others could be the driven responses to
bars \citep{Buta09}.  Although these two ideas may account for a large
fraction of the cases \citep{KN79}, especially if orbiting dark matter
clumps can excite patterns \citep{Dubi08}, spirals can still develop
in the absence of either trigger, as revealed in simulations.

The idea that spirals could be self-excited oscillations of the
stellar disk represents the greatest theoretical challenge.  While
there is general agreement that gas seems to be essential, no picture
seems complete, and current theories disagree on even the lifetimes of
the patterns.  One idea \citep[\eg][]{BL96}, is that spiral features
are manifestations of quasi-steady global modes of the underlying
disk.  Alternatively, they could be short-lived, recurrent transient
patterns that originate either from forcing by density
fluctuations \citep[\eg][]{Toom90}, or else from recurrent vigorous
instabilities \citep[\eg][]{Sell00}.

A serious obstacle to progress in this area has been the absence of
observational discriminants that would favor one of these radically
differing viewpoints over the other.  The predictions for density
variations or gas responses at a single instant are essentially
independent of the generating mechanism and do not depend strongly on
the lifetime of the pattern.  \cite{MRM09} attempted to measure radial
variations of pattern speeds using a generalization of the method
devised by \cite{TW84a} (see \S\ref{Rconst}).  They reported lower
pattern speeds at larger radii, but their measurements still tell us
little about spiral lifetimes.  However, the velocity distribution of
stars in the solar neighborhood \citep[][and
  Fig.~\ref{GCSsimp}]{Nord04} is most naturally accounted for in the
transient picture (see \S\ref{recrrnt}).

\subsection{Spirals as Global Modes of Smooth Disks}
Simple models of disk galaxies possess many linear instabilities
\citep[\eg][]{Jala07}.  The bar-forming mode (\S\ref{global}) is
usually the fastest growing, but it has almost no spirality.  These
studies are therefore important to understand stability, but even the
higher modes are not particularly promising for spiral
generation.\footnote{\cite{Korc05} calculated essentially
  gas-dynamical modes for models of specific galaxies, and argued that
  the shapes of one, or more, of the lower-order modes could be
  matched to the observed spiral pattern.  However, it is unclear that
  rapidly growing, linear modes can be seen for long at finite
  amplitude before non-linear effects will change their appearance,
  and it seems even less likely that two modes with different growth
  rates should have similar large amplitudes at the time a galaxy is
  observed.}

The ``density wave'' theory for spiral modes, described in detail by
\cite{BL96}, invokes a more specific galaxy model with a sub-maximal
disk that is dynamically cool in the outer parts and hot in the inner
disk.  The local stability parameter is postulated to be $1.0 \la Q
\la 1.2$ over most of the disk and to rise steeply to $Q > 2$ near the
center.  Bertin, Lin, and their co-workers perform a global analysis
using the hydrodynamic approximation (\BTii\ \S5.1), which reveals
slowly growing spiral modes under these specific conditions.

The mechanism \citep{Mark77} is a cavity mode, having qualitative
similarities to that for the bar mode (\S\ref{barmech}), but the
tightly-wrapped waves are trailing around the entire cycle.  The inner
turning point is at a $Q$-barrier: a steeply increasing $Q$ value
causes the forbidden zone (\S\ref{scwaves}) to broaden to the point
that in-going short waves get ``refracted'' into outgoing long waves,
which prevents the wave train from reaching the \ILR\ where it would
be damped.  The long waves then propagate out to near the \CR\ in the
$Q \sim 1$ part of the disk, where they switch back to the shortwave
branch with a small degree of amplification.  The \WASER\ mechanism
\citep{Mark76} at this turning point involves a third, transmitted
wave that is ``radiated'' outwards on the short-wave branch, carrying
away the angular momentum to excite the mode in the inner disk.  Thus
the amplifier involves a small ``swing'' from the long- to the
short-wave branch, whereas the bar instability uses a full swing from
leading to trailing.  This difference, together with their assumption
that the disk is sub-maximal (\ie\ $X > 3$ for $m=2$, see
eq.~\ref{Xdef}), allows the mode to be slowly growing.  \cite{Lowe94}
present a model of this kind to account for the spiral structure of
M81.

In order to justify the ``basic state'' of the disk they require,
\cite{BL96} argued heuristically that rapidly evolving features would
have disappeared long ago and that low-growth-rate instabilities in a
cool disk, created by gas dissipative processes and star formation,
will dominate at later times.  They invoked shocks in the gas to limit
the amplitude of the slowly growing mode, leading to a quasi-steady
global spiral pattern.

The main objection to their picture is that it is likely that such a
lively outer disk will suffer from other, more vigorous, collective
responses with $m>2$ that will quickly heat the disk, as described in
\S\ref{globsims}, and destroy their postulated background state.

\subsection{Recurrent Transients}
From the early work by \cite{MPQ70}, \cite{Hohl71}, \cite{HB74},
\cite{JS78}, and \cite{SC84}, $N$-body simulations of cool,
sub-maximal disks have exhibited recurrent transient spiral activity.
This basic result has not changed for several decades as numerical
quality has improved.

Claims of long-lived spiral waves \citep[\eg][]{TEDS} have mostly been
based on simulations of short duration.  For example, \cite{ET93}
presented a simulation that displayed spiral patterns for $\sim 10$
rotations, but the existence of some underlying long-lived wave is
unclear because the pattern changed from snapshot to snapshot.
\cite{DT94} ran their simulations for fewer than 7 disk rotations and
the bi-symmetric spiral they observed appeared to be an incipient bar
instability.  Zhang (1996,1998) adopted a similar mass distribution
and also reported long-lived patterns in her simulations.  The author
has attempted to reproduce her results, and indeed obtained similar
bi-symmetric features, but they appear to be the super-position of
several waves having differing pattern speeds.

\cite{SC84} stressed that patterns fade in simple simulations that do
not include the effects of gas dissipation; the reason is the disk
becomes less responsive as random motion rises due to particle
scattering by the spiral activity (\S\ref{scatt}), which is therefore
self-limiting.  They also demonstrated that mimicking the effects of
dissipative infall of gas, such as by adding fresh particles on
circular orbits, allowed patterns to recur ``indefinitely.''  Later
work \citep[\eg][]{CF85,Toom90} has shown that almost any method of
dynamical cooling can maintain spiral activity, as also happens in
modern galaxy formation simulations \citep[\eg][]{Ager10}.

Thus the transient spiral picture offers a natural explanation for the
absence of spiral patterns in S0 disk galaxies that have little or no
gas; maintenance of spiral activity requires a constant supply of new
stars on near-circular orbits.  Other pieces of indirect evidence that
also favor the transient spiral picture are reviewed in \S\ref{scatt}.

\subsection{Spirals as Responses to Density Fluctuations}
\cite{GLB2} and \cite{Toom90} suggested that a large part of the
spiral activity observed in disk galaxies is the collective response
of the disk to clumps in the density distribution.  As a spiral wake
is the collective response of a disk to an individual co-orbiting
perturber \citep{JT66}, multiple perturbers will create multiple
responses that all orbit at different rates.  The behavior of this
polarized disk reveals a changing pattern of trailing spirals, which
can equivalently be regarded as swing-amplified (\S\ref{swing})
noise.\footnote{\cite{Cuzz10} found evidence for similar behavior
  within Saturn's A ring.}

\cite{TK91} studied the amplified noise that arose in their $N$-body
simulations of the sheared sheet.  The massive particles themselves
provoke spiral responses with an amplitude proportional to the input
level of shot noise, caused by density variations in the distribution
of randomly distributed particles.  Comparison of their expectations
with linear theory predictions revealed that the simulations were
livelier than they expected, by a factor $\la 2$, apparently from a
gradual build-up of correlations between the mean orbital radii of the
particles.

This could be a mechanism for chaotic spirals in very gas rich discs,
where a high rate of dissipation may be able to maintain the
responsiveness of the disc \citep{Toom90} while the clumpiness of the
gas distribution may make the seed noise amplitude unusually high.
However, it seems likely that spiral amplitudes \citep[\eg][]{ZCR9}
are too large to be accounted for by this mechanism in most galaxies.
Also, the spiral structure should be chaotic, with little in the way
of clear symmetry expected.

\subsection{Non-linear Spiral Dynamics}
\label{nonlinear}
Tagger and his co-workers \citep{Tagg87,Sygn88,MT97} suggested that
global modes in stellar disks could be coupled through non-linear
interactions.  They proposed that wave 1 excites wave 2 through
second-order coupling terms that are large when \CR\ of wave 1 lies at
approximately the same radius as the \ILR\ of wave 2.  Conservation
rules require a third wave such that $m_3 = m_1 \pm m_2$; \ie\ most
likely an axisymmetric wave ($m_3=0$) if $m_1 = m_2$.  Many examples
of multiple waves in $N$-body simulations have been reported with
remarkable coincidences for the radii of the main resonances.

\cite{FDT5} developed a similar argument for waves in the sheared
sheet.  They found that amplified trailing waves could excite fresh
leading waves in their second-order theory.  They proposed this
mechanism as an alternative source of the amplitude excess noted above
in the simulations by \cite{TK91}, and they speculated that the same
mechanism may also account for the larger than expected amplitudes of
spirals in global $N$-body simulations (see \S\ref{globsims}).  As
both this mechanism, and that discussed in the previous paragraph,
employ terms that are second-order in the perturbation amplitude, they
become important only when features are strong.

\cite{PCG91} attempted to construct, by orbit superposition,
self-consistent steady spiral waves of finite amplitude to match the
observed non-axisymmetric patterns in specific galaxies.  They
experienced great difficulty in finding solutions near \CR, and suggested
that either this resonance or the 4:1 resonance\footnote{The resonance
  condition (eq.~\ref{rescon}) for a pure $\cos(m\theta)$ potential
  variation also implies frequency commensurabilities at multiples of
  $m$.  The small denominators that characterize the principal
  resonances (\BTii, \S3.3.3) arise at these ``ultra\-harmonic
  resonances'' only for non-circular orbits.  A new family of orbits
  appears at the 4:1 resonance that closes after 4 radial oscillations
  \citep[see \S\ref{barstr} and][]{SW93}.}  marks the outer radius of
the spiral.  Their finding is not unexpected for two reasons: (a) the
dispersion relation for steady, tightly-wrapped, small-amplitude waves
(Fig.~\ref{disprel}) predicts a forbidden region around \CR\ for all
$Q>1$, and (b) the non-linear dynamics of orbits in barred potentials
(see \S\ref{barstr}) finds only chaotic orbits near \CR, which are
unfavorable to self-consistency as \cite{PCG91} found.  It should be
noted that the difficulty of finding a self-consistent solution near
\CR\ is a direct consequence of their assumption of a steady wave
pattern; transient waves do not suffer from this problem
(Fig.~\ref{dusttoashes}) and rapidly growing groove modes
(\S\ref{gmodes}) have {\it peak\/} overdensities at \CR.

\cite{TKEC} and \cite{ARGM} suggested that spirals in barred galaxies
could be created by the slow migration of stars along Lyapunov
manifold tube orbits emanating from the unstable Lagrange points at
the ends of the bar.\footnote{Interestingly, this idea harks back to
  the old ``garden sprinkler'' model for spirals proposed by
  \cite{Jean23} \citep[see also][fig.\ 55 \& pp~357--360]{Jean29}.}
The ambitious hope of these preliminary papers is for an ultimate
unified picture for the co-existence of bars, spirals and rings, all
having the same pattern speed.

\subsection{Spirals in Global $N$-body Simulations}
\label{globsims}
\cite{SC84} and \cite{Sell89b} reported that their global simulations
manifested more vigorous spiral activity than was consistent with
amplified shot noise.  A brief summary of some further results to
support this claim is given here, and will be described more fully
elsewhere.

\begin{figure}[t]
\begin{center}
\includegraphics[width=.9\hsize]{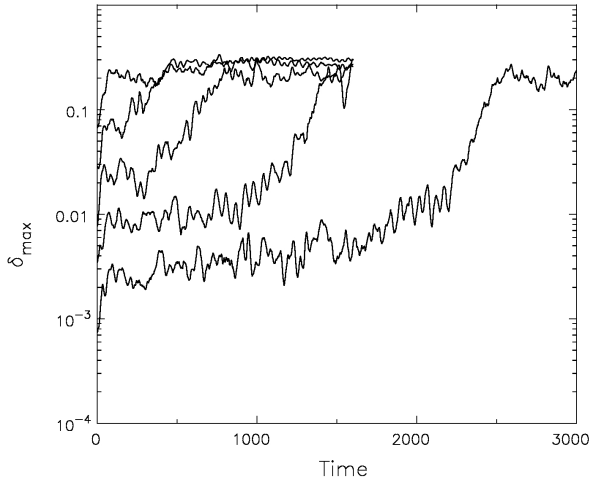}
\end{center}
\caption{The time evolution of the peak overdensity in a series of
  simulations of the half-mass Mestel disk with different numbers of
  particles.  The model was predicted by \cite{Toom81} to be globally
  stable.  The ordinate shows the maximum values of $\delta$ on grid
  rings.  The number of particles in each simulation rises by a factor
  10 from $N=5 \times 10^4$ to $N=5 \times 10^8$, and the initial
  amplitude is $\propto N^{-1/2}$.  Linear theory predicts the later
  amplitudes should have the same scaling.}
\label{cntrst}
\end{figure}

As noted in \S\ref{barmech}, \cite{Toom81} predicted the half-mass
Mestel disk to be globally stable to small-amplitude disturbances.
Thus $N$-body simulations of this disk should exhibit no activity in
excess of the inevitable swing-amp\-lified shot noise.
Fig.~\ref{cntrst} reveals that this is not the case.  The ordinate
shows the largest value of $\delta = \Delta \Sigma/ \Sigma$ from $m =
2$ disturbances in a sequence of simulations with increasing numbers
of particles.  The unit of time is $R_i/V_c$, where $V_cR_i$ is the
center of the inner angular momentum cut out.  Thus the orbit period
at this small radius is $2\pi$.

At $t=0$, $\delta \propto N^{-1/2}$, as appropriate for shot noise,
and swing-amplification causes an almost immediate jump by a factor of
a few for all $N$.  When $N = 5 \times 10^4$, amplified noise causes
$\sim 20\%$ overdensities almost immediately.  For larger $N$ the
amplitude eventually rises to similar values in later evolution, once
the inner disk has developed a pronounced bar.  But for the largest
two values of $N$ shown, a period of slow growth occurs after the
initial swing-amplified surge, offering tentative support for the
linear theory prediction of global stability, with the slow growth
perhaps arising from the gradual development of particle correlations
as described by \cite{TK91}.  Even in these cases, however, the
amplitude rises more rapidly once $\delta \ga 2\%$.

Thus even in these highly restricted simulations, spiral activity
always exhibits runaway growth -- albeit more and more delayed as
$N$ is increased -- behavior that is quite clearly not in accord with
linear theory predictions.  The rapid growth once $\delta \ga 2\%$
suggests that the behavior has already become non-linear in some
respect at this modest amplitude.  Note that the largest number of
particles, $N = 5\times 10^8$, is within a factor of 100 of the number
of stars in a real galaxy disk, where in reality the mass distribution
is far less smooth, owing to the existence of star clusters and giant
gas clouds.

\subsection{A Recurrent Instability Cycle?}
\label{recrrnt}
\cite{SL89} reported evidence for a recurrent instability cycle in
their simulations of a very low-mass disk.  They observed that each
spiral pattern created substantial changes to the distribution of
particles at the Lindblad resonances, which they suspected created
conditions for a new global instability of the groove mode kind (see
\S\ref{gmodes}).  They explicitly demonstrated that the later features
were indeed true instabilities, since when they continued a parallel
simulation after randomly shuffling the particles in azimuth at some
moment, the pattern speed of the next mode to appear was the same as
that in the original simulation that had not been
interrupted.\footnote{This behavior is inconsistent with the
  non-linear mode coupling ideas discussed in \S\ref{nonlinear}.}
Thus, each coherent wave leaves behind an altered \DF\ that apparently
provokes a new instability.

The runaway growth in the larger $N$ models shown in Fig.~\ref{cntrst}
is not caused by a single unstable mode, but appears to be a
succession of separate coherent waves, each having a lower rotation
rate and reaching a higher amplitude than the last.  \cite{Sell00}
reported similar behavior in lower-$N$ simulations, and demonstrated
that one of the waves did indeed cause strong scattering at the \ILR.
It should be noted that resonance scattering is a second-order effect,
but the evidence shown in Fig.~\ref{cntrst} suggests that it becomes
important at a relative overdensity of just $\Delta \Sigma / \Sigma
\sim 2$\%.  Exactly how the demise of one mode creates the conditions
for the next instability remains unclear, however.

Since the only evidence for this behavior came from a (well-tested)
$N$-body code, it seemed best not to pursue the idea further until
supporting evidence could be found.  \cite{Sell94} therefore expressed
the hope that evidence of resonance scattering could be found in
the \Hipp\ measurements of the local stellar kinematics.  The
publication of the \GCS\ with distances and full phase-space motions
of $\sim 14\,000$ F \& G dwarf stars (see Fig.~\ref{GCSsimp})
enabled \cite{Sell10} to show that the so-called Hyades stream is in
fact caused by scattering at a recent \ILR.

This very recent evidence supports the idea that spirals in the local
Milky Way, and presumably elsewhere, do in fact behave as the
simulations had indicated.  Further work is required to expose the
details of the recurrence mechanism.  However, it now seems misguided
to search for a spiral instability as some devious sort of cavity mode
in a smooth disk; \ie\ the assumption of a featureless \DF\ may have
thrown the spiral baby out with the bathwater!

\ssection{Buckling Instabilities and Warps}
\label{warps}
The optically visible parts of galactic disks are usually remarkably
thin and flat, whereas the more extended \ion{H}{1} disks of many
edge-on galaxies appear noticeably warped with an integral sign shape
\citep{Sanc76}.  Stellar warps \citep{RBCJ02,SdJH} are much less
pronounced than the warps in the extended gaseous disks.  The
long-known warp in the \ion{H}{1} layer of the Milky Way \citep{OKW58}
has been most recently analyzed by \cite{LBH06}, while the dust and
stars of the outer disk appear to be distorted in a similar, though
less extensive, shape \citep{Reyl09}.  Both gaseous and stellar warps
are frequently asymmetric, as appears to be the case for the Milky
Way.  The fact that stellar warps usually follow the same warped
surface as do the gaseous ones \citep[see also][]{CSvMS96}, is strong
evidence that warps are principally a gravitational phenomenon.

Warps are extremely common.  \ion{H}{1} observations of edge-on
galaxies \citep{Bosm91,GSK02} find very high fractions of warps, and
the true fraction must be even higher, since warps directed close to
our line of sight may be missed.  Warps can also be detected
kinematically even when the system is not edge-on.  Their ubiquity
suggests that warps are either repeatedly regenerated or long-lived.

\citet{Brig90} studied a sample of 12 warped galaxies with
high-quality 21-cm data, and found that galactic warps obey three
fairly simple rules:

\begin{enumerate}

\item The \ion{H}{1} layer typically is coplanar inside radius
  $R_{25}$, the radius where the B-band surface brightness is
  $25\,\hbox{mag arcsec}^{-2}=25\muB$, and the warp develops between
  $R_{25}$ and $R_{26.5}$ (the Holmberg radius).

\item The {\bf line of nodes} (\LON) is roughly straight inside $R_{26.5}$.

\item The \LON\ takes the form of a loosely-wound {\em leading spiral}
outside $R_{26.5}$.

\end{enumerate}

\cite{KW59} first drew attention to the winding problem presented by
warps (see also \BTii\ \S6.6.1), and argued that while self-gravity
would slow the differential precession, it could not be strong enough
to create a long-lived warp.  \cite{LB65}, on the other hand,
suggested that warps could result from a persisting misalignment
between the spin axis and the disk normal, \ie\ a long-lived mode.

This section describes the theory of bending waves in general before
addressing warps more directly.  Unstable bending modes will be
denoted {\bf buckling modes}, although other names that come from
plasma physics, such as hose, firehose, or hosepipe instabilities, are
commonly used.

\subsection{Local Theory of Bending Waves}
\label{bentsheet}
\cite{Toom66} considered the bending stability of an infinite, thin
slab of stars having a velocity dispersion $\sigma_x$ in the
$x$-direction and some characteristic thickness $z_0$.  The
self-gravity of the slab causes it to bend coherently provided the
vertical oscillations of stars are adiabatically invariant as they
move over the bend, which requires the slab to be thin compared with
the wavelength of the bend.  Furthermore, if the bending amplitude is
small, its effect on the horizontal motion is negligible.

Toomre derived the dispersion relation for small-amp\-litude,
long-wave ($kz_0 \ll 1$) distortions of the form\break $h(x,t)= H
e^{i(kx-\omega t)}$:
\be
\omega^2 = 2\pi G\Sigma|k| - \sigma_x^2k^2,
\label{sheet}
\ee
where $\Sigma$ is the vertically-integrated surface density of the
slab.  The first term on the {\small RHS} is the restoring force from
the perturbed gravity caused by the bend, while the second is the
inertia term due to the vertical acceleration needed for stars to
follow the corrugations.  The inertia term is destabilizing and
outweighs the gravitational restoring force when $\lambda<\lambda_{\rm
  J}=\sigma_x^2/G\Sigma$, causing the distortion to grow
exponentially.  The unstable range of the buckling instability is
precisely complementary to that of the Jeans instability in the plane
in the absence of rotation, which is unstable for wavelengths $\lambda
> \lambda_{\rm J}$ \citep{Toom66}.

The dispersion relation (\ref{sheet}) predicts a buckling instability,
at sufficiently short wavelengths, for any razor-thin system.
However, it does not hold for wavelengths shorter than or comparable
to the actual vertical thickness of the slab.  Since $z_0 \sim
\sigma_z^2 / (G\Sigma)$, where $\sigma_z$ is the vertical velocity
dispersion, one expects that bending at all wavelengths will be
suppressed when $\sigma_z/\sigma_x$ exceeds some critical value, which
\cite{Toom66} estimated to be 0.30.\footnote{\cite{KMC71} and
  \cite{FP84} considered the bending instability in a constant density
  slab of stars with sharp edges.}

\cite{Arak85} carried through a linear normal mode analysis of the
infinite, isothermal slab \citep{Spit42,Camm50}, which has the
properties $\rho = \rho_0 \, {\rm sech}^2(z/2z_0)$, $z_0 =
\sigma_z^2/(2\pi G\Sigma)$, and $\Sigma = 4\rho_0z_0$.  He assumed a
Gaussian distribution of $x$-velocities, with $\sigma_x \neq
\sigma_z$, and determined the range of unstable wavelengths as the
slab was made thicker.  He showed that the buckling instability could
be avoided at all wavelengths when $\sigma_z > 0.293\sigma_x$, in good
agreement with Toomre's earlier estimate.  At the marginal stability
threshold, the last unstable mode has a wavelength $\simeq 1.2
\lambda_{\rm J}$.

Galaxy disks are not, of course, infinite slabs subject to plane-wave
distortions, but the radial velocity dispersion, $\sigma_R$, which is
larger than the azimuthal dispersion, could perhaps be great enough to
drive a buckling instability.  As the observed ratio of velocity
dispersions for stars in the Solar neighborhood is $\sigma_z/\sigma_R
\sim 0.6$ \citep{Nord04}, \cite{Toom66} concluded that at least this
region of our Galaxy is ``apparently well clear of this stability
boundary.'' It should be noted that the approximate solution for the
potential, which requires $kR \gg 1$, is stretched in this case, since
$\lambda_{\rm J} \approx 7\,$kpc when $\sigma_R = 40\,$km~s$^{-1}$ and
$\Sigma = 50\,$M$_\odot$~pc$^{-2}$.

When the disk is embedded in some external potential, arising from a
halo or the distant bulge of the galaxy say, the dispersion relation
for short-wavelength waves in a thin slab (\ref{sheet})
acquires an additional stabilizing term
\be
\omega^2 = \nu_{\rm ext}^2 + 2\pi G\Sigma |k| - \sigma_R^2k^2,
\label{lbend}
\ee
where $\nu_{\rm ext}^2 = |\partial^2\Phi_{\rm ext} / \partial
z^2|_{z=0}$ (\BTii\ eq. 6.114).  Taking this additional factor into
account further reinforces local stability and global, axisymmetric
simulations \citep{Sell96a} confirmed that Toomre's conclusion holds
everywhere in an axisymmetric, but otherwise plausible, model of the
Milky Way disk.

As for WKB spiral waves (\S\ref{WKB}), the dispersion relation
(\ref{lbend}) can be generalized to {\it tightly-wrapped\/}
non-axisym\-metric bending waves simply by replacing $\omega$ with
$\omega - m\Omega$, with the angular rate of precession of the bending
wave that has $m$-fold rotational symmetry being $\Omega_p=\omega/m$.
It should be borne in mind that since observed warps in galaxies are
very far from being tightly wound, analyses that make this
approximation yield results that are at best only indicative of the
dynamical behavior.

Vertical resonances between the bending wave and the stars arise where
$m(\Omega_p - \Omega) = \pm\nu$ (\S\ref{vres}), although the meaning
of $\nu$ here depends on the context.  \BTii\ (\S6.6.1) considered
only razor thin disks, for which the internal oscillation frequency
$\nu_{\rm int}$ is infinite and the resonances occur where $m(\Omega_p
- \Omega) = \pm\nu_{\rm ext}$.  In disks of finite thickness, the
stars have a natural internal vertical frequency $\nu_{\rm
int}^2 \approx 4\pi G\rho_0$ (exact in the mid-plane of an infinite
slab), and for vertical resonances the appropriate value of $\nu =
(\nu_{\rm int}^2+\nu_{\rm ext}^2)^{1/2}$.  Henceforth, $\nu$ will be
used to mean either this total frequency in a thickened layer or
$\nu_{\rm ext}$ for a razor-thin sheet.

Eq.~(\ref{lbend}) is satisfied for both $\pm m(\Omega_p-\Omega)$,
implying two possible pattern speeds known as {\bf fast} and {\bf slow
waves}.  Because the gravity term raises the Doppler-shifted frequency
above $\nu$, waves in a cold disk ($\sigma_R=0$) extend away from the
vertical resonances, and do not exist in the region between them that
includes \CR.  The slow wave, which is a retrograde pattern for
$m=1$, is of most interest because the unforced precession rate,
$\Omega_p = \Omega - \nu$, has much the milder radial variation.

Much like density waves, neutrally stable bending waves propagate in a
disk in the radial direction at the {\bf group velocity} \citep{Toom83}
\be
v_g = {\partial\omega \over \partial k} = {{\rm sgn}(k) \pi G\Sigma -
  k\sigma_R^2 \over m(\Omega_p - \Omega)}.
\label{cgbend}
\ee
Since the denominator is negative for slow bending waves, trailing
waves ($k>0$) in a cold disk ($\sigma_R=0$) propagate inwards, while
leading waves ($k<0$) propagate outwards.  As for waves in non-uniform
whips, the bending amplitude rises when the wave propagates into a
region of lower surface density, and {\it vice versa}.

\begin{figure*}[t]
\begin{center}
\includegraphics[width=.9\hsize]{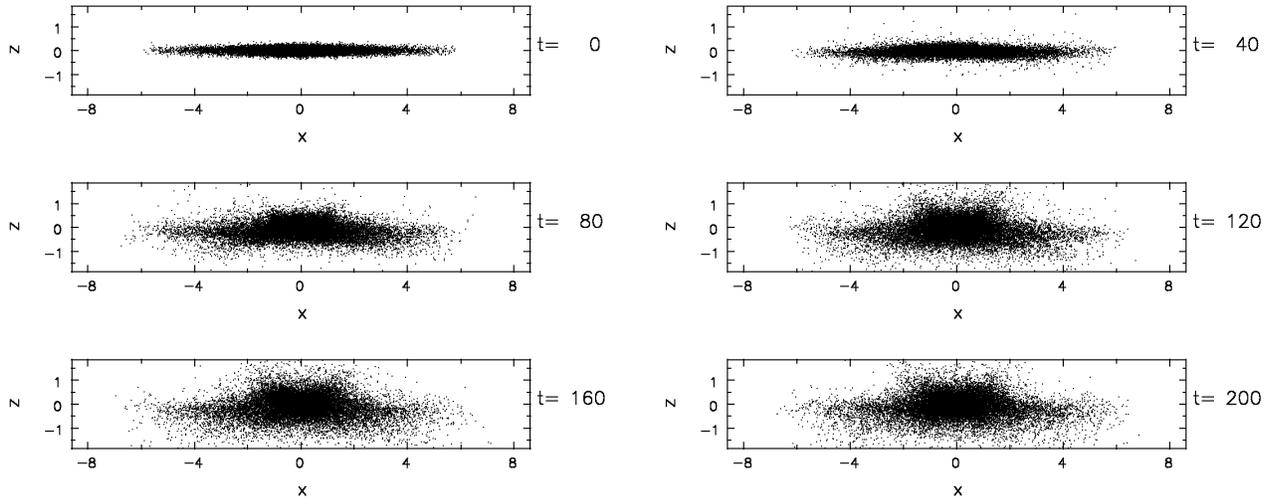}
\end{center}
\caption{The development of a buckling instability in a simple model
  of an isolated stellar disk with $Q\sim2$.  The disk is KT/4 model
  described by \cite{SM94}, with equal numbers of particles orbiting
  in each direction, but here the simulation uses a code of much
  higher spatial resolution.  The buckling mode is axisymmetric and,
  while non-axisymmetric features were permitted by the code, none
  developed.  The disk mass and radial scale are unity, and $G=1$; the
  orbit period at $R=2$ is 16 in these units.}
\label{buckle}
\end{figure*}

Unfortunately, a full description of wave propagation in a sheet of
finite thickness requires a solution of the linearized Boltzmann and
Poisson equations \citep{Toom66,Arak85,Wein91,Toom95}.  The results
are not analytic and surprisingly more complex than equation
(\ref{lbend}) for the razor-thin case.  Because the vertical potential
of the disk is anharmonic, stars whose vertical oscillation takes them
far from the mid-plane have lower vertical frequencies.  Thus, any
Doppler-shifted frequency $m(\Omega_p - \Omega)$ of the bending wave
will be in resonance with some stars that will damp the wave by
converting wave energy into increased random motion.  In general,
short-wavelength modes $kz_0 \ga 0.5$ damp in less than one
wavelength \citep{Wein91,Toom95}, while long-wavelength modes
$kz_0 \ll 1$ can propagate large distances.

\cite{SNT98} studied the propagation and\break damping of axisymmetric
bending waves in disks having both finite thickness and non-zero
random velocities.\break  Waves launched from the center of the disk at a
fixed frequency propagated outwards and were damped as they approached
the vertical resonance.  As a result the disk thickened over a small
radial range, with the peak occurring just interior to the resonance.

\subsection{Global Bending Modes}
Because of the complication caused by finite thickness, analytic work
almost always adopts the razor-thin approximation, which also requires
$\sigma_R=0$, since thin disks with velocity dispersion are buckling
unstable.\footnote{A razor thin disk is not destabilized by orbital
motions with no velocity spread.}  With these assumptions, the disk
can be approximated by a finite number of gravitationally coupled
circular rings of matter, each having the appropriate angular
momentum, which always admits a discrete spectrum of bending modes.
The real modes of the continuum disk can be distinguished by showing
they are independent of the number of rings employed.

The gravitational restoring forces are correctly captured in this
approach, but the lack of random motion omits the additional coupling
between adjacent mass elements caused by the epicyclic motions of the
stars.  This extra mechanism further stiffens the disk's resistance to
bending \citep{DS99}, especially in the high density inner regions
where radial epicycles are larger.

\cite{HT69} developed the coupled ring approach to study the
bending dynamics of rotationally-supported, razor-thin disks with no
random motion and no halo.  They were able to prove that general disks
of this kind have no axisymmetric ($m=0$) or warping ($m=1$)
instabilities, but they could not extend their proofs to higher
sectoral harmonics.  In fact they noted that were the disk composed of
two equal counter-rotating populations of stars (on circular orbits)
it would be buckling unstable for all $m \ge 2$.

\cite{HT69} also studied the particular case of the sharp-edged
Maclaurin disk in which all stars orbit at the same angular rate.
When the sense of rotation is the same for all stars, the disk is
stable to all bending waves, and there is a simple set of discrete
neutral modes for all $m$.  \cite{Poly77} extended their analysis to
Maclaurin disks with random motion (the Kalnajs disks) and was able to
solve for the complete spectrum of bending modes.  Needless to say,
the introduction of random velocities in this razor-thin system caused
buckling instabilities to appear for all $m$.

Since the Maclaurin disk bears little resemblance to real galaxies,
\cite{HT69} modified the disk to blur the sharp edge.  They
demonstrated that discrete warp modes in a cold, razor-thin disk can
exist only when the edge is unrealistically sharp.  Note that all
isolated disks admit two trivial zero-frequency modes: a vertical {\bf
  displacement} of the entire disk and a {\bf tilt} of the disk plane
to its original direction.  More interesting standing wave solutions
(modes) require traveling waves to reflect off the disk edge, but a
realistic disk with a fuzzy edge does not reflect bending waves,
because the group velocity decreases with the disk surface density
(eq.~\ref{cgbend}) and a wave packet in a cold disk will never reach
the edge \citep{Toom83}.

\subsection{Simulations of Buckling Modes}
\cite{SM94} used $N$-body simulations to study the global
instabilities of hot disks with no net rotation, \ie\ with half the
particles counter-rotating.  (See \S\ref{buckbar} for buckling modes
of more normal disks with large net rotation.)  The form and vigor of
the principal instabilities in any one of their models varied with the
balance between radial and azimuthal pressure and with disk thickness:
an in-plane lop-sided instability was the most disruptive for cool
models (\S\ref{lopsided}).  The radially hotter thin disks were
disrupted by axisymmetric bending instabilities (bell modes), as
illustrated in Fig.~\ref{buckle}.  The instability creates a thick
inner disk resembling a pseudo-bulge \citep{KK04}.  Very thin disks
also buckled in an $m=2$ ``saddle'' mode, and an $m=1$ warp
instability was detectable in some models, but never dominant.

Remarkably, instabilities in a counter-streaming model having
intermediate radial pressure caused rather mild changes and led to an
apparently stable, moderately thin, and almost axisymmetric disk.  The
in-plane velocities in this model resemble those reported by
\citep{RGK92} for the S0 galaxy NGC 4550, and indicate this galaxy
could be stable even without large quantities of dark matter.  The
stability of this end product also demonstrated that thin axisymmetric
systems with modest radial pressure are more stable to bending modes
than those having isotropic or radially biased \DF s.

\cite{MS94} proposed a criterion for the stability of a stellar system
to buckling modes that can be applied globally.  A particle moving at
speed $\dot x$ in a mildly bent sheet with characteristic wavenumber
$k$ experiences a periodic vertical forcing at frequency $k\dot x$
and, like any harmonic oscillator, the phase of its response depends
upon whether the forcing frequency is greater or less than its natural
vertical frequency.  If $k\dot x<\nu$, the density response of the
system to an imposed perturbation will be supportive, and the
disturbance can be sustained or even grow.  However, if $k\dot x>\nu$
for most particles, the overall density response to the perturbation
will produce a potential opposite to that imposed and the disturbance
will be damped.  Thus short waves are stable.  They showed that their
proposal successfully accounted for their $N$-body results, for the
behavior of the infinite slab at short wavelengths, and for the
apparent absence of elliptical galaxies more flattened than E7.

\cite{Sell96b} found a long-lived bending oscillation in an $N$-body
simulation of a warm disk that was constrained to be axi\-symmetric.
Apparently the system was able to support a standing wave, at most
mildly damped, between the center and the edge of the disk, even
though the stars had random motion in a disk of finite thickness.  The
frequency of the bending or flapping mode was low enough to avoid
vertical resonances with almost all disk particles, in agreement with
the requirement stated by \cite{Math90}.  While this result remains an
isolated curiosity, since the flapping mode would probably be quickly
damped in a halo, it is a clear counter-example to the argument by
\cite{HT69} that realistic disks do not possess global bending modes,
which apparently holds only for disks without random motion.

\subsection{Disks in Halos}
Since the paper by \cite{HT69}, ideas of warp formation have relied in
some way on the interaction between the disk and its dark matter halo.
\cite{DS83} and \cite{Toom83} suggested that a flattened halo
misaligned with the disk could form a long-lasting warp.  \cite{SC88}
and \cite{Kuij91} obtained long-lived warps (dubbed {\bf modified tilt
  modes}) of disks in rigid, misaligned halos, which were insensitive
to the details of the disk edge.  \cite{Love98} studied the tilting
dynamics of a set of rings also in a rigid halo, but assumed that the
inner disk lay in the symmetry plane of the spheroidal halo.

\cite{DC09} noted that the dark matter halos that result from cosmic
structure formation simulations are usually aspherical, with
frequent misalignments between the principal axes of the inner and
outer halo.  The disks in their simulations warped nicely when forced
with slowly rotating, but otherwise rigid, perturbing fields
representative of such halos.

However, dark matter halos are not rigid, and a responsive halo alters
the dynamics in several ways.  \cite{NT95} showed that were the inner
disk misaligned with the principal plane of the flattened halo, as
supposed by \cite{SC88}, its precession would be damped through
dynamical friction, bringing the disk into alignment with the halo on
time scales much shorter than a Hubble time.  But a more compelling
objection to the modified tilt mode emerged from $N$-body simulations
with live halos: \cite{DK95} found that the warp did not survive
while \cite{BJD98} showed that the inner halo quickly aligns itself
with the disk, not {\it vice versa}.  The large store of angular
momentum in the disk maintains its spin axis, but the pressure
supported inner halo can readily adjust its shape slightly to align
itself with the disk.

\subsection{Misaligned Infall}
The idea that galaxy warps are manifestations of {\it eternal\/} warp
modes seems doomed by the damping effect of a live halo.  But
slowly evolving warps remain viable, provided that suitable external
perturbations occur in enough cases.

In hierarchical galaxy formation scenarios, late infalling material
probably has an angular momentum axis misaligned with the disk spin
axis.  \cite{OB89} therefore proposed that warps arise due to the
slewing of the galactic potential as material with misaligned angular
momentum is accreted.  Structure formation simulations by \cite{QB92}
confirmed that the mean spin axis of a galaxy must slew as late
arriving material rains down on the early disk.  The
less-than-critical matter density in modern \LCDM\ universe
models implies that infall is less pervasive at later times, but it
manifestly continues to the present day in gravitationally bound
environments \citep[][and chapter by van Woerden \& Bakker]{Sanc08}.

\cite{JB99} and \cite{SS06} presented results of experiments in which a
disk was subjected to the torque from a misaligned, massive torus at a
large radius.  This well-defined perturbation is a very crude model of
an outer halo that is rotationally flattened, and having its spin axis
misaligned with that of the disk.  It is misaligned and farther out
because, in hierarchical scenarios, the mean angular momentum of the
later arriving outer halo is probably greater and misaligned from that
of the original inner halo and disk.  The accretion axis is, in
reality, likely to slew continuously over time, so a model with a
constant inclination is somewhat unrealistic.

Rather than striving for realism, \cite{SS06} used this simple forcing
to reach an understanding of how the warp develops and why the \LON\
usually forms a loosely-wound, leading spiral.  The inner disk
maintained a coherent plane because of both self-gravity and random
motion.  The torque arising from the misaligned outer torus caused the
inner disk, and the aligned inner halo, to precess rigidly even though
the torque increased with radius, but the outer disk beyond $\sim
4R_d$ started to warp.

As soon as the outer disk became misaligned with the inner disk, the
strongest torque on the outer parts of the disk arose from the inner
disk.  The torque from the interior mass was responsible for the
leading spiral of the line of nodes, even though the adopted external
field would have produced a trailing spiral.  The fact that the
\LON\ of most warps forms a leading spiral over an extended radial
range seems to imply massive disks.

Even though the disk precessed due to the external\break torque, its
motion was barely damped over many Gyr, in contrast to the
expectations from \cite{NT95}.  Damping was weak because the slow
precession rate allowed the inner halo to remain closely aligned with
the disk, which therefore caused little drag.  The weak damping seemed
to be caused more by the relative precession of the inner and outer
parts of the halo.  Also the warp evolved slowly as the layer settled
to the main plane at gradually increasing radii, in apparent agreement
with the decreasing outward group velocity.

\begin{figure}[t]
\centerline{\hspace{.05\hsize}\includegraphics[width=.87\hsize]{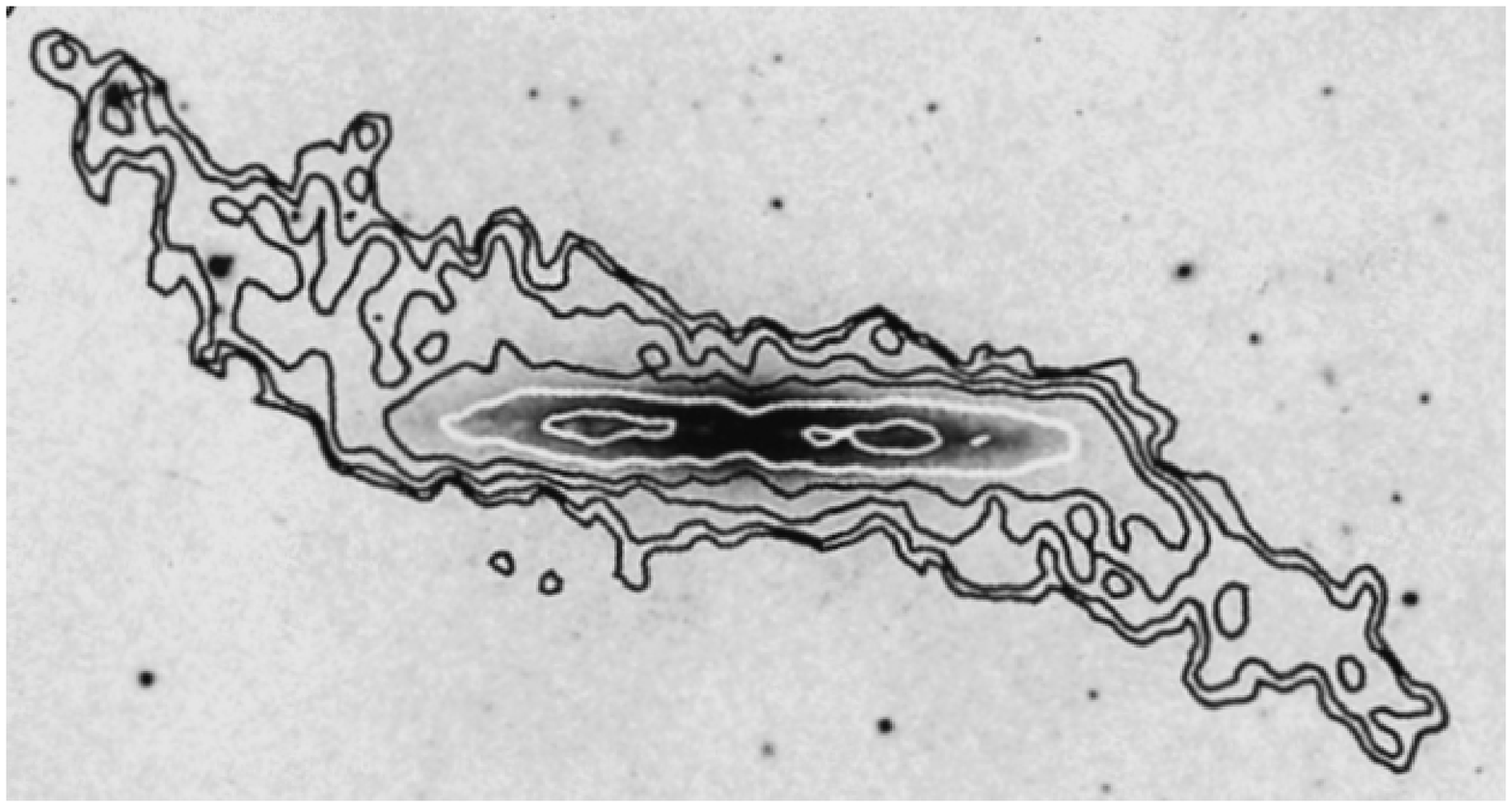}}
\centerline{\includegraphics[angle=-90,width=.91\hsize]{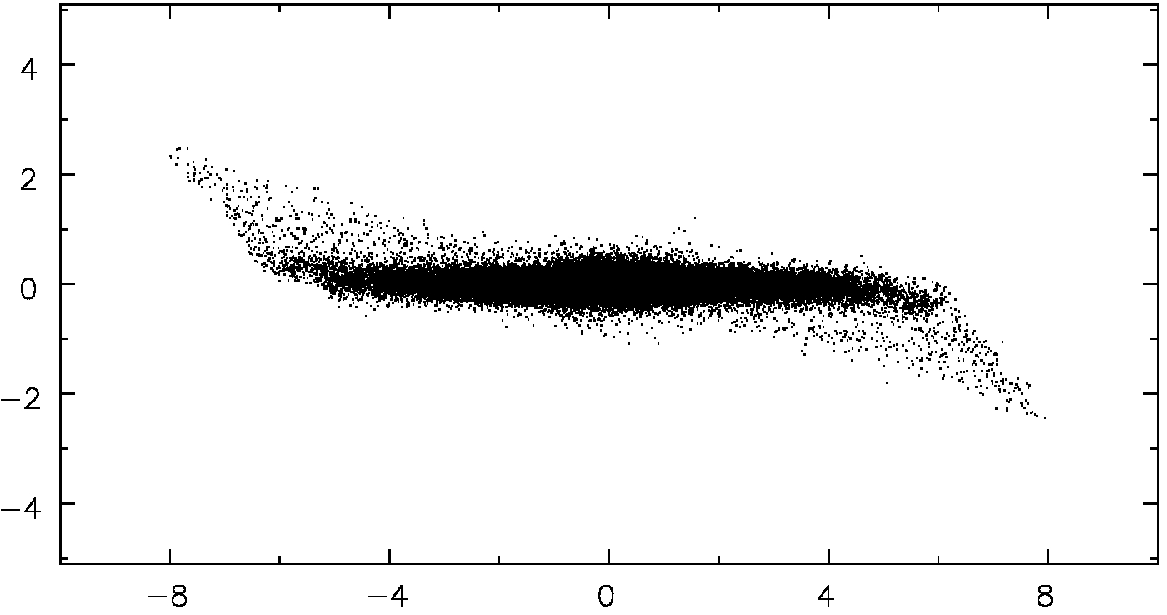}}
\caption{The upper panel shows the observed \ion{H}{1} warp of NGC
  4013 \citep{Bott96}, and the lower panel the warp in the simulation
  by \cite{SS06} that closely resembles it. The length unit on the
  axes is the scale length $\Rd$ of the exponential disc.}
\label{SSwarp}
\end{figure}

Fig.~\ref{SSwarp} shows the \ion{H}{1} observation of NGC~4013 by
\cite{Bott96}, together with the warp obtained in the simulation by
\cite{SS06}.  Their simulations revealed that the warp persisted for
cosmologically interesting times, even when the external forcing field
was removed.  Thus the persistence of warps is not nearly as
perplexing as previous studies had suggested.  Furthermore, the model
had a flat inner disk and the warp in the outer disk matched all of
Briggs's rules quite well.

A fixed outer torus is clearly unrealistic and the halo axis probably
shifts continuously or episodically, as argued by \cite{QB92}, making
warp lifetimes a side issue.  Warps formed this way can be repeatedly
regenerated when a new infall event happens.  Since cosmic infall and
mergers are more likely to happen in a denser environment, warps can
be induced more frequently in such an environment, which is consistent
with the statistics \citep{GSK02}.

\subsection{Warps Driven by Tides}
Tidal interactions between galaxies could be an additional cause of
warps in disks, and one that is quite likely to produce asymmetrical
warps.  This idea has been explored most fully to explain the warp of
the Milky Way's (\MW) disk that results from the proximity of the
Magellanic Clouds, especially the Large Cloud (\LMC).  The orientation
of the principal axes of the warp \citep{LBH06} at least seems
favorable to this hypothesis.  \cite{Bail03} suggested that the
Sagittarius dwarf galaxy is another possible culprit.

\cite{HT69}, in models that did not include dark matter, concluded
that the \LMC\ could be responsible for the warp of the \MW, but only
if it was a lot more massive than was then suspected and had recently
passed close to the edge of the disk.  \cite{GRKD} tested the
hypothesis in fully self-consistent simulations that included live
halos and used updated information about the distance and motion of
the \LMC.  They concluded that neither the amplitude, nor the
orientation, of the warp in the disk of the \MW\ was consistent with
the tidal hypothesis.  \cite{WB06}, on the other hand, suggested that
a co-operative response from the halo to the passage of the
\LMC\ could have generated the observed warp in the disk.  Thus there
is no consensus yet on the origin of the warp in the Milky Way.

\vfill\eject
\ssection{Bars}
\label{bars}
Bars are common features in disk galaxies.  An earlier review
\citep{SW93} of the vast topic of bars is now somewhat dated but, as
it provides a still useful summary of the basic dynamics, the present
article will update a few main points and the reader is referred to
the earlier review for a more detailed discussion.

\subsection{Origin of Bars}
\label{barorig}
For a long time, dynamicists were struggling to understand the absence
of bars in some disk galaxies \citep{OP73}, but since reasonable
models of luminous galaxies that include a dense bulge of moderate
mass are now known to be stable (\S\ref{global}), the problem has
become almost the opposite!  Strong bars are seen in many galaxies
whose mass distributions now appear unfavorable to the dynamical bar
instability, as evidenced by a nuclear gas ring (see \S\ref{gasflow}),
which can form only if the center is dense enough to have inhibited
bar formation by Toomre's mechanism.

However, the fact that the center is dense today does not require that
it was dense when the bar formed; secular inflow of gas
(\S\ref{gasflow}) can build up the central density after the bar has
formed.  Alternatively, the bar could have grown in size and slowed
trhough disk evolution \citep{Sell81,Bere07} or halo
friction \citep[][see \S\ref{dynfr}]{Atha02}, or have been triggered
by large density fluctuations in the disk \citep{Sell89a}, by tidal
interactions \citep{Nogu87,Bere04,CMM06}, halo
substructure \citep{Roma08b} or a non-axisym\-metric
halo \citep{DC09}.  Any of these considerations, or that described in
the next paragraph, could plausibly reconcile the existence of bars in
these galaxies with Toomre's stabilizing mechanism.

In an elegant piece of dynamics, \cite{LB79} demonstrated that the
inner parts of galaxies are regions where eccentric orbits have a
tendency to align themselves, which allows a bar to grow slowly
through orbit trapping.  The region where a cooperative response to a
mild perturbation occurs is where the overall radial density profile
of the galaxy flattens into a more uniform core.  In this region, the
radial variation of $\Omega - \kappa/2$ has a maximum at some non-zero
radius and an infinitesimal bar pattern can have a pattern speed that
allows two \ILR s.  Lynden-Bell's aligning mechanism, which operates
best on the more eccentric orbits, requires $\Omega_p \simeq
\Omega_\phi - \Omega_R/2$ of orbits in the aligning region.  As this
pattern speed is much lower than that expected from the global bar
instability (\S\ref{barmech}), the aligning mechanism offers an
additional route to bar formation in otherwise globally stable disks.
Although the cooperative region has a small radial extent, \cite{LB79}
suspected the bar could be much larger.

\cite{Erwi05} presented a useful study of bar properties and pointed
out that bars in late-type galaxies are often much smaller relative to
the disk size than are those formed in sumulations.

\subsection{Frequency of Bars}
\label{barfreq}
Strong bars are clearly visible in $25\% - 30\%$ of disk galaxies
\citep[\eg][]{Mast10}, and the fraction rises to $\ga 50\%$ when more
objective criteria are applied to red or near-IR images \citep[][and
  further references cited below]{Eskr00,MJ07,Rees07}.  \cite{BJM08}
found a higher bar fraction in later Hubble types, while \cite{MSA10}
found no bars in either very luminous, or very faint, galaxies.

Whatever may be the mechanism responsible for the formation of bars in
real galaxies, none of the above suggestions makes a clear prediction
for the frequency of bars.  \cite{Bosm96}, \cite{Cour03}, and others
have pointed out that barred galaxies seem little different from their
unbarred cousins in most respects, \eg\ they lie on the same
Tully-Fisher relation.  Minor systematic differences do exist:
\eg\ \cite{DC04} note that barred galaxies seem to have smaller mass
fractions of neutral \ion{H}{1} gas, but this seems more likely to be
the result, rather than the cause, of the bar.  The anti-correlation
of bar frequency with the bulge half-light \citep{BJM08} possibly
results from Toomre's stabilizing mechanism, but this cannot be the
whole story because some near-bulgeless disks are unbarred while
other barred disks have massive bulges.

If no dynamical property, other than their eponymous one, can be
identified that cleanly separates barred from unbarred galaxies, then
the existence of a bar in a galaxy may possibly be determined by
external factors, such as a chance encounter.  \cite{EEB90} reported
an increased fraction of bars in groups and clusters, but more recent
work \citep{Bara09,LGMK09,AMC09} has found little or no variation of
bar fraction with environment.  It is also possible that the bar
fraction could be changing with time; different groups disagree
\citep{Joge04,Shet08}, probably because observations of galaxies at
significant look-back times are subject to systematic difficulties due
to band-shifting and changing spatial resolution \citep[see
also][]{Elme07}.

A radical alternative is to regard bars as transient features that
form and decay, and that the current fraction of barred galaxies
represents the duty cycle \citep{BCS05}.  Bars in $N$-body simulations
are dynamically rugged objects that appear to last indefinitely.  Of
course, they could be destroyed by a merger event, for example,
although not much in the way of a cool disk would survive such an
event.  Norman and his co-workers \citep{PN90,NSH96} have argued that
bars can be destroyed by the accumulation of mass at their centers,
which may lead to a pseudo-bulge and/or a hot inner disk.  However,
\cite{SS04} and \cite{ALD05} found that unreasonably large and/or
dense mass concentrations were required to cause their bars to
dissolve.  The simulations by \cite{BCS05} uniquely show that gas
accretion may aid the dissolution of the bar and, with star formation,
could recreate a cool disk that would be needed to make a new bar.
Even if this behavior can be confirmed by others, the model requires
very substantial gas infall.  Moreover, the continued existence of the
hot old disk and the build-up of a dense center makes every cycle of
this speculative picture harder to achieve.

In the distant future, galaxy formation simulations may have the
quality and resolution perhaps to be able to predict the correct bar
fraction, and thereby reveal their cause.

\subsection{Structure of Bars}
\label{barstr}
This section gives a brief description of a few important aspects of
the orbital behavior in large-amplitude bars, and the reader is
referred to \cite{SW93} for a more comprehensive discussion.  Weak
bars can be treated using epicycle theory (\BTii\ \S3.3.3).  Most
early orbit studies in strongly barred potentials considered motion
confined to the plane perpendicular to the rotation axis.  Even though
3D motion is much richer, the fundamental structure of bars is most
easily understood from in-plane orbits.

Since bars are believed to be steadily-rotating, long-lived objects,
it makes sense to discuss their structure in a frame that co-rotates
with the potential well at the angular rate $\Omega_p$.  A rotating
frame has the {\bf effective potential}
\be
\Phi_{\rm eff} = \Phi - \textstyle{1\over2}\Omega_p^2R^2,
\label{peff}
\ee
where $\Phi$ is the potential in an inertial frame.

The effective potential surface in the disk plane (eq.~\ref{peff})
resembles a volcano, with a central crater, a rim, and a steeply
declining flank.  The crater is elongated in the direction of the bar,
and the rim has four {\bf Lagrange points}: two maxima, $L_4$ and
$L_5$, on the bar minor axis and two saddle points, $L_1$ and $L_2$ on
the bar major axis.  (The fifth Lagrange point, $L_3$, is the local
potential minimum at the bar center.)  Because of Poisson's equation,
the density contours of the bar must be more elongated than those of
the inner $\Phi_{\rm eff}$.

Neither $E$ nor $L_z$ is conserved in non-axisymmetric potentials, but
Jacobi's invariant $I_J$ (eq.~\ref{jacobi}) is conserved even for
strong bars that rotate steadily.  Since $I_J = {1\over2}|\bv|^2 +
\Phi_{\rm eff}$, where $\bv$ is the velocity in the rotating frame,
contours of $\Phi_{\rm eff}$ bound the possible trajectories of stars
having $I_J$ less than that contour value.  Stars that are confined to
the bar, which also have $I_J < \Phi_{\rm eff}(L_1)$, are of most
interested here.

A {\bf periodic orbit} is a possible path of a star in the rotating
frame that retraces itself, usually after a single turn around the
center, but always after a finite number of turns.  Because the orbits
close, the orbital period in the rotating frame is commensurable with
the radial period and these orbits are also described as resonant
orbits of the (strongly non-axisymmetric) potential.

Periodic orbits can be either {\bf stable}, in which case a star
nearby in phase space oscillates (librates) about its {\bf parent}
periodic orbit in an epicyclic fashion, or they are unstable, in which
case the trajectory of a star nearby in phase space diverges
exponentially (at first) from the periodic orbit.  The orbits of stars
that librate around a periodic orbit are known as {\bf regular}
orbits, those that do not librate about any periodic orbit are known
as {\bf irregular} or {\bf chaotic} orbits.  Chaotic orbits have only
a single integral, $I_J$, while regular orbits have an additional
integral (two more in 3D) that confines their motion to a hypersurface
of smaller dimension in phase space.  Regular orbits are the more
interesting because the star's orbit can be more elongated than the
potential surface that confines it, which is of great value when
building a self-consistent bar model.

\begin{figure}[t]
\includegraphics[width=\hsize]{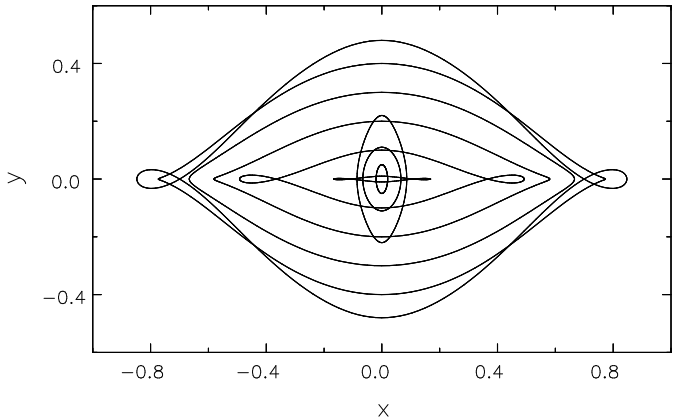}
\caption{Examples, in a rotating bar potential, of important periodic
  orbits that close after two radial oscillations for every turn about
  the center, the 2:1 resonant families.  Those orbits elongated
  parallel to the bar axis (horizontal) are members of the $x_1$
  family.  The $x_2$ orbits are elongated perpendicular to the bar.}
\label{barorbs}
\end{figure}

The main features can be illustrated in the simple potential
(\cf\ \BTii\ eq.~3.103)
\be
\Phi_{\rm eff}(x,y) = {\textstyle{1\over2}}v_0^2 \ln \left( 1 + {x^2 +
  y^2/q^2 \over R_c^2} \right) - {\textstyle{1\over2}}\Omega_p^2R^2,
\ee
where $R^2 = x^2 + y^2$, $R_c$ is a core radius inside of which the
potential is approximately harmonic, $q \leq 1$ is the flattening, and
$v_0$ is the circular speed at large $R$ when $q=1$.  As in \BTii, the
values are: $v_0 = 1$, $q=0.8$, $R_c=0.03$, and $\Omega_p=1$.  With
these parameters, the major-axis Lagrange points lie at a distance
$R_L \simeq 0.9996$ from the bar center.

The 2:1 resonant periodic orbits shown in Fig.~\ref{barorbs} have a
range of $I_J$ values, but all close in the rotating frame after two
radial oscillations for every turn about the center.  Orbits of the
main family, denoted $x_1$, are elongated parallel to the bar
(horizontal in the Figure), and are referred to as the {\bf backbone
of the bar}, since the majority are stable.  Stars on these orbits all
move in the same direction as the bar rotates, but have shorter
periods because they are interior to the Lagrange points (or more
loosely, they are inside \CR).  Some $x_1$ orbits are simple closed
figures, but others have loops near the outer ends where the star
progresses around the center of the galaxy more slowly than does the
rotating frame.  \cite{SS87} and \cite{VHC07} found that a large
fraction of particles in an $N$-body bar librate around them, and the
same behavior is expected for stars in real bars.

Another important family of 2:1 orbits appears near the center of the
bar, but is elongated perpendicular to the bar axis, as also shown in
Fig.~\ref{barorbs}.  This family, denoted $x_2$, is almost always
stable and is present in many realistic bar potentials.  Since
Lindblad resonances (\S\ref{resonances}) are defined for infinitesimal
perturbations to axisymmetric potentials, the label \ILR\ is very
loose usage in strong bars.  It is true that the $x_2$ family appears
in barred potentials only, but not always, when the axisymmetric mass
distribution and pattern speed admit one or more \ILR s.  Furthermore,
$x_2$ orbits orient themselves perpendicular, while the $x_1$ family
is parallel, to the bar axis, which is directly analogous to the
abrupt phase shift that occurs in the response of a forced harmonic
oscillator as the forcing frequency crosses its natural frequency
\citep[\cf][]{SH76}.  At finite amplitude, the near circular orbit
sequence acquires gaps at the 2:1 resonance bifurcations, which become
broader as the bar amplitude rises.  Thus the radial extent of the
$x_2$ family shrinks as the strength of the bar is increased, and it
may disappear entirely.  Even if one is careful to say that the
appearance of the $x_2$ family is the generalization of the \ILR\ to
strong bars, the radius of this resonance is still badly defined
because the orbits that appear inside it can be quite eccentric.

As $I_J$ approaches the value of $\Phi_{\rm eff}(L_1)$, the time to
complete a full turn in the rotating frame lengthens and additional
orbit families appear.  Orbits that close after any number of radial
oscillations can be found in principle, but of these only the 4:1
resonant orbits are of dynamical significance to bar structure.  As
the period lengthens, the proliferation of orbit families causes a
precipitous decrease in the stable regions around each parent and
chaotic behavior ensues \citep{Chir79}.  The onset of chaos near \CR\
led \cite{Cont80} to expect the density of a self-consistent bar to
drop steeply near the major axis Lagrange points, leading to the rule
that the length of a bar is limited by \CR.  This rule predicts that
the parameter \citep{Elme96}
\be
{\cal R} \equiv R_L/a_B > 1, 
\ee
where $a_B$ is the semi-major axis of the bar.  In principle,
self-consistent bar dynamics could allow bars with ${\cal R} \gg 1$,
although empirical bar pattern speed estimates (\S\ref{Rconst}) mostly
find that \CR\ is in fact only slightly beyond the end of the bar.

There are many more in-plane orbit families, but few are of dynamical
importance to the structure of the bar.  See \cite{SW93} for a fuller
account.

The extension to 3D allows for many more resonances between the
in-plane motion and the vertical oscillations.  While there is a rich
variety of behavior \citep{PF91,SPA02}, the backbone $x_1$ family from
2D continues to be the most important, but now with a ``tree'' of
orbits also librating vertically.  The new periodic orbits that appear
in 3D have a similar projected shapes as the in-plane $x_1$ family,
but they also librate vertically a small number of times over the same
period as the motion in the plane.  \cite{PSA02} highlighted the orbit
families that they found to be of importance for the ``boxy''
appearance of edge-on bars (see also \S\ref{buckbar}).

\subsection{Gas Flow}
\label{gasflow}
When pressure and magnetic forces can be neglected, any mild
dissipation will drive gas to move on stable periodic orbits.  An
organized streaming gas flow pattern is expected wherever the simplest
periodic orbits over a range of energies can be nested and intersect
neither with neighboring orbits, nor with themselves.  Shocks, where
pressure ceases to be negligible, must occur in flows either where
periodic orbits self-intersect, or where gas flows on two separate
orbits cross.  Fig.~\ref{barorbs} shows that were gas to flow in that
adopted bar potential, shocks would be inevitable because many orbits
self-intersect (the loops) and, in particular, $x_2$ orbits cross the
$x_1$ family.  Thus shocks are a general feature of cool (low
pressure) gas flows in bars.\footnote{Shocks may be avoided when
  pressure is significant \citep{EG97}.}

Full hydrodynamic simulations \citep[\eg][]{RHvA,Atha92,Fux99} are
needed to determine the flow pattern.  Shocks are offset to the
leading sides of the bar major-axis in models having reasonable
parameters.  \cite{Pren62} seems to have been the first to associate
the dust lanes in bars with the locations of shocks in the gas.

Shocks convert some kinetic energy of bulk motion in the gas into heat,
which is radiated efficiently.  Furthermore, the offset location of
the shock causes the gas to spend more than half its orbit on the
leading sides of the bar, where it is attracted backwards towards the
bar major axis, causing it to lose angular momentum.\footnote{The
  opposite happens in shocks outside \CR, where the gas gains angular
  momentum from the bar.}  Thus gas in the bar region must settle a
little deeper into the potential well on each passage through a shock,
\ie\ the bar drives gas inwards, the angular momentum it loses being
given to the bar.

The inflow stalls where gas settles onto the $x_2$ orbit family, which
is found in bar models that have dense centers, leading to a build up
of gas.  This behavior can be associated with nuclear rings of dense
gas \citep{Rega02}, which are often the sites of intense star
formation also \citep{Maoz01,Bene02}.  If this were the whole story,
the gas could not be driven any further inwards, but there is both
observational evidence, in the form of spiral dust lanes and star
formation \citep[\eg][]{CSM98}, and some theoretical work
\citep{Wada01} to suggest that self-gravity causes inflow to continue.
However, the existence of high gas density and rapid star formation in
the nuclear ring indicates that only a small fraction of the gas
continues inwards.

Modeling the gas flow in a specific galaxy allows one to determine two
properties of the bar that are hard to constrain otherwise.
Estimating the gravitational potential of the galaxy from a
photometric image plus a dark halo, \cite{WSW01} fitted for both the
disk mass-to-light (\ML) and $\Omega_p$ in the galaxy NGC~4123.  These
authors also described the procedure in detail.  Results for a number
of other galaxies were reported by \cite{PFF04}, \cite{Wein04}, and
\cite{ZSSWW}, although these last authors were unable to obtain an
entirely satisfactory fit.  \cite{LLA96} fixed the \ML\ and fitted
only for $\Omega_p$.  All these fits preferred rapidly rotating bars
in heavy disks.  \cite{Pere08} confirmed that the best fit parameters
of \ML\ and $\Omega_p$ were the same for both 2D Eulerian (Godunov)
and 3D Lagrangian ({\small SPH}) hydrodynamic methods.

Even though the quadrupole field of a bar decays quickly with radius,
it can be strong enough to drive a spiral shock in the gas of the
outer disk, as originally demonstrated by \cite{SH76}.  \cite{Schw81}
showed that when gas is modeled as inelastic particles, it is driven
outwards to form an outer ring \citep[see review by][]{BC96}.
However, it is unclear that spirals in the outer disks of real barred
galaxies are the responses to the bar, and they may owe more to
self-excited structures than to bar driving
\citep[\S\ref{spirals},][]{SS88,Buta09}.  In addition, the outer
spiral response to an imposed bar is not steady in modern simulations,
with the shapes of the driven arms cycling through a broad range.  For
these reasons, the gas flow models fitted to individual galaxies
should rely primarily on the fit within the bar and pay little
attention to the outer disk.

\cite{KSR03} tried a similar approach, but fitted a spiral pattern
instead of a bar, which may yield unreliable results for two reasons:
(i) The lifetimes of spiral arms are believed to be short
(\S\ref{spirals}) leading to broader resonances and stronger gas
responses than would arise in simulations that assume a slowly
evolving pattern, and (ii) the observed spirals could be the
superposition of several features with different angular rotation
rates.  As bars undoubtedly last for longer than do spirals and
dominate the non-axisymmetric potential, fits to bars in galaxies are
likely to yield better disk mass estimates.

\subsection{Bar Pattern Speeds}
\label{Rconst}
\cite{TW84a} devised a method to measure the pattern speed of a bar
directly from observations of a tracer component, which must obey the
equation of continuity.  Their original method assumes that the galaxy
has but a single pattern, and would yield a misleading result were
there more than one pattern, each rotating at a different angular
rate.

\begin{figure}[t]
\begin{center}
\includegraphics[width=.8\hsize,clip]{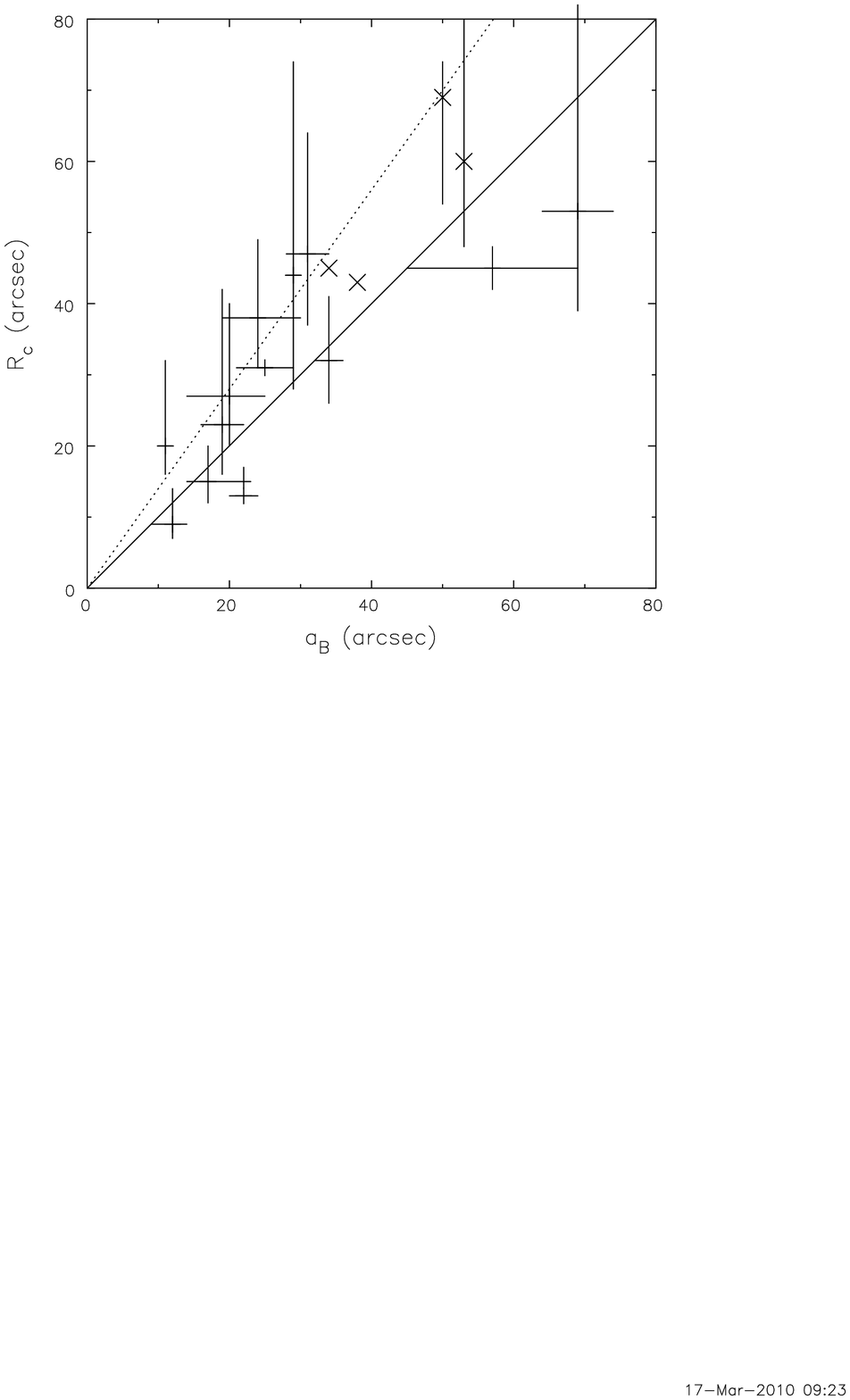}
\end{center}
\caption{Summary of direct pattern speed measurements for bars collected
  by \cite{Cors08}.  The diagonal line shows ${\cal R} =1$ and the
  dotted line ${\cal R} =1.4$.  The crosses mark values for which
  error bars are unavailable.}
\label{corsini}
\end{figure}

The stellar component of early-type barred galaxies is believed to
obey the equation of continuity because these galaxies have little
dust obscuration and no star formation.  They also rarely possess
prominent spirals in the outer disk.  Results of many studies using
this method for early-type barred galaxies were summarized by
\cite{Cors08} and are shown in Fig.~\ref{corsini}.  While some
individual measurements are quite uncertain, the data seem to favor $1
< {\cal R} \la 1.4$.  \cite{CH09} found a counter-example in a
low-luminosity galaxy.

\cite{Fath09} and \cite{MRM09} applied the method of \cite{TW84a}
to ionized and molecular gas, respectively.  Both groups argue that
this is valid, even though the separate gas components do not obey the
continuity equation that undelies the method.  \cite{Fath09} generally
find fast bars.  \cite{Meid08} generalized the method to attempt to
measure radial variations in the pattern speed and \cite{MRM09} found
suggestions of pattern speeds that are lower at large radii than those
near the center.

Other methods can yield indirect estimates of bar pattern speeds.
Fits of models of the gas flow (\S\ref{gasflow}) have been reported
for a few galaxies, finding ${\cal R} \sim 1.2$.  \cite{Atha92}
argued that the shapes and locations of dust lanes in bars also seem
to suggest that ${\cal R} \simeq 1.2$.  Associating a ring in a barred
galaxy with the location of a major resonance with the bar
\citep{BC96} yields, with kinematic information, an estimate of the
pattern speed.

\cite{RSL08} computed models of the stellar and gas (sticky particles)
responses to forcing by photometric models of 38 barred galaxies, in
which they assumed that the entire non-axisymmetric structure rotated at
the same pattern speed.  They attempted to match the model to the visual
morphology of the galaxy, and found a range of values for $\cal R$.
However, in most cases where ${\cal R} \gg 1$, the fit is dominated by
the outer spiral, which may have a lower angular speed than does the
bar.

\subsection{Bars Within Bars}
\cite{ES02} and others have found inner {\bf secondary bars} within
the inner parts of $>25\%$ of large-scale or {\bf primary bars}.  They
reported that the secondary bar has a length some $\sim 12\%$ of that
of the primary bar and the deprojected angles between the principal
axes of the two bars appeared to be randomly distributed, suggesting
that the two bars may tumble at differing rates.  This inference was
supported by \cite{CDA03}, who used the \cite{TW84a} method to show
that the two bars in NGC~2950 could not have the same rotation rates;
\cite{Maci06} used their data to argue that the secondary bar has a
large retrograde pattern speed.

The theoretical challenge presented by these facts is substantial, and
progress to understand the dynamics has been slow.  \cite{MS00}
studied the orbital structure in a potential containing two
nonaxisymmetric components rotating at differing rates.  However, it
is almost certain that the secondary bar can neither rotate at a
uniform rate \citep{LG88} nor can it maintain the same shape at all
relative phases to the primary.

\cite{FM93} argued that gas was essential to forming secondary
bars \citep[see also \eg][]{HSE01,ES04}.  However, some of the
collisionless simulations reported by \cite{RS99} and \cite{RSL02}
manifested dynamically decoupled inner structures when the inner disk
had high orbital frequencies due to a dense bulge.  The structure was
more spiral like in some models, but others appeared to show inner
bars that rotated more rapidly than the main bar.

\cite{DS07} created long-lived, double-barred galaxy models in
collisionless $N$-body simulations having dense inner disks, which
they described as pseudo-bulges.  They followed up with a more
detailed study \citep{ShD09} that also made some predictions for
observational tests.  The secondary bars in their models indeed
rotated at non-uniform rates, while their shape varied systematically
with phase relative to that of the primary.  These models prove that
collisionless dynamics can support this behavior, but it is unclear
that their initial conditions mimicked those that have given rise to
double barred galaxies in nature.

The possible consequence of gas inflow in these galaxies has attracted
a lot of attention.  \cite{SFB89} speculated that bars within bars
might lead to gas inflow over a wide dynamic range of scales, from
global to the parsec scale where accretion onto a black hole might
cause AGN activity.  While inflows may have been
observed \citep[\eg][]{Haan09}, understanding of gas flow in these
non-steady potentials remains rather preliminary \citep{Maci02,Hell07}.

\subsection{Buckling of Bars}
\label{buckbar}
\cite{CS81} first reported that the bars in their 3D simulations were
thicker than the disk from which they had formed, and had acquired a
pronounced ``boxy'' shape when viewed edge-on.  Boxy isophotes in
edge-on disk galaxies are now believed to be an indicator of a bar, as
is supported by kinematic evidence in the gas \citep{MK99,BA05}.

The reason the bar thickened was explained by \cite{RSJK}, who showed
that bars are subject to the buckling instability (\S\ref{bentsheet}).
The bar buckles because the formation of the bar created a structure
supported by elongated orbits that stream along the bar in the near
radial direction.  Even though the ingoing and outgoing stars stream
on different sides of the bar, the effective averaged $\sigma_R$ has
risen as a result without changing $\sigma_z$.  The simulation by
\cite{RSJK} revealed that the buckling instability produced a large
amplitude arch just before it saturated, after which the bar became
thicker.  The energy to increase vertical motion in the bar appeared
to have been released by the further concentration of mass towards the
bar center \citep[see also][]{MVS04}.  It is delightful that the
evolution of one instability, the bar-forming mode, should create a
new structure, the bar, that is itself unstable.

Bars still thicken in more recent simulations with grids having higher
spatial resolution (\eg\ Fig.~\ref{barform}), but do not seem to
exhibit the spectacular arch reported by \cite{RSJK} unless
bi-symmetry is enforced.  Low spatial resolution or significant
gravity softening (which are equivalent) weakens the restoring force
in eq.~(\ref{sheet}) and artificially increases $\lambda_J$, which is
the characteristic length for instability.  Stronger gravity causes
the preferred buckling modes to have shorter wavelength allowing, say,
an upward arch on one side of the center and downward arch on the
other.  Enforcing bi-symmetry prevents the bar from bending in this
anti-symmetric manner, and forces it to buckle through the single
arch mode.

\subsection{Dynamical Friction on Bars}
\label{dynfr}
Friction between a rotating bar and a massive halo was first reported
many years ago \citep{Sell80}, but the implications for dark matter
halos have fueled a renewed intense study of the topic.

\subsubsection{Theory}
Dynamical friction \citep{Chan43} is the retarding force experienced
by a massive perturber as it moves through a background sea of
low-mass particles.  It arises, even in a perfectly collisionless
system, from the vector sum of the impulses the perturber receives
from the particles as they are deflected by its gravitational field
(see Appendix).  Equivalently, friction can be viewed as the
gravitational attraction on the perturber of the density excess, or
wake, that develops behind it as it moves, as was nicely illustrated
by \cite{Muld83}.

\begin{figure}[t]
\begin{center}
\includegraphics[width=.8\hsize]{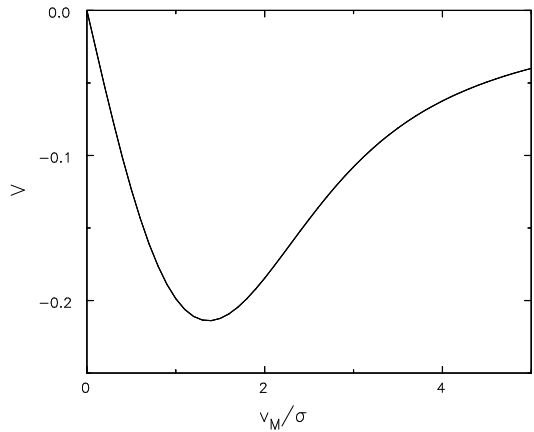}
\end{center}
\caption{The dimensionless acceleration function $V$ defined in
  eq.~(\ref{Chandra}) for the case of a Gaussian distribution of
  velocities among the background particles.  The function is negative
  because the acceleration is directed oppositely to the velocity.}
\label{dynv}
\end{figure}

Chandrasekhar's formula (\BTii\ eq.~8.6) for the acceleration of a
perturber of mass $M$ moving at speed $v_M$ through a uniform
background, density $\rho$, of non-inter\-acting particles having an
isotropic velocity distribution with a 1D rms velocity spread
$\sigma$, may be written as
\be
{dv_M \over dt} = 4\pi \ln\Lambda G^2 {M \rho \over \sigma^2} V\left(
{v_M \over \sigma} \right).
\label{Chandra}
\ee
The Coulomb logarithm is defined in the Appendix, and the
dimensionless function $V$ is drawn in Fig.~\ref{dynv} for a Gaussian
distribution of velocities; other velocity distributions would yield a
different functional form.  Physically, the retarding acceleration
must vanish when the perturber is at rest and it must also tend to
zero when the perturber moves so fast that the background particles
receive only small impulses and the feeble wake lies far downstream
from the perturber.  Friction is strongest when the speed of the
perturber is somewhat greater than the rms speeds of the background
particles.

The simplifying assumptions in its derivation invalidate application
of eq.~(\ref{Chandra}) to the physically more interesting problem of
friction in a non-uniform medium in which the background particles are
confined by a potential well and interact with the perturber
repeatedly.

Repeated encounters between the perturber and the\break background
particles require the more sophisticated treatment presented in
\cite{TW84b}, who adopted a rotating potential perturbation in a
gravitationally bound spherical halo of test particles.  They showed, as
did \cite{LBK} for spiral waves, that lasting changes to the orbits of
the halo particles appear to second order in the perturbing potential,
and can occur only at resonances.  They derived a daunting formula for
the torque on the halo caused by the perturbation that sums the
contributions from infinitely many resonances.  The contribution at
each resonance is proportional to the gradient of the \DF\ near the
phase-space location of the resonance, in a manner that is directly
analogous to Landau damping.  \cite{Wein85} computed the surprisingly
large torque expected on a rotating bar, and his conclusion was
confirmed in restricted tests \citep{LC91,HW92}.

\cite{WK07} pointed out that friction is dominated by a few important
resonances.  They estimated the widths of these resonances for a
perturbation having constant pattern speed and finite amplitude, and
argued that immense simulations would be needed to populate the
resonance with sufficient particles to capture the correct net
response.  However, the loss of angular momentum from the perturber
causes its pattern speed to change, and the resulting time-dependence
of the forcing frequency is a much more important factor in broadening
the resonances; thus friction can in fact be captured correctly in
simulations having moderate numbers of particles \citep{Sell08a}.
Note that the pattern speed of an orbiting satellite rises as it loses
angular momentum, while that of a bar usually decreases.

Despite the complicated language of resonant dynamics, the upshot is
simply that the perturber induces a wake-like response in the halo, as
was beautifully illustrated by \citet[][their Fig.~1]{WK07}.
As for the infinite medium, friction can be thought of more simply as
the torque between the perturber and the induced halo response.
\citet[][his Fig.~2b]{Sell06} shows the lag angle between the forcing
bar and the halo response, which is about 45\degr\ when friction is a
maximum and gradually decreases to zero as the bar slows, until
eventually friction ceases when the halo response co-rotates with the
bar.

\cite{LT83} for an orbiting satellite, and \cite{Sell06} for a
rotating bar, demonstrated that the frictional drag on the
perturbation scales with the mass of the perturber, $M$, the halo
density, $\rho$, and the halo velocity dispersion, $\sigma$ exactly as
in eq.~(\ref{Chandra}).  Furthermore, the dimensionless function that
describes the dependence on the angular speed of the perturber shares
the general properties with $V(x)$ that it is negative (for reasonable
non-rotating halos), and must $\rightarrow 0$ as $x \rightarrow
\infty$, and that it should be $\propto x$ as $x \rightarrow 0$.
Including self-gravity in the halo response causes a further slight
increase in friction, but does not otherwise change the behavior.

\subsubsection{Halo Density Constraint}
Fully self-consistent simulations of bar formation in a live halo by
\citet[][2000]{DS98} showed that strong bars are indeed slowed
rapidly.  The fact that observed bars appear not to have been slowed
(\S\ref{Rconst}) may imply an upper bound to the density of the dark
matter halo in barred disk galaxies.  \cite{VK03} claimed a
counter-example of a bar that does not experience much friction in a
``cosmo\-logically-motivated'' halo, even though their result
disagreed with all others for strong bars \citep{OD03,Atha03} and with
theory!

Investigation of their anomalous result by \cite{SD06} revealed that
friction can be avoided {\it temporarily\/} if the gradient of the
\DF\ at the most important resonance(s) has been flattened by earlier
evolution, which they described as a metastable state.  \cite{LT83}
reported similar behavior as a result of driving the perturber at
constant frequency for a protracted period.  They showed, as
did \cite{SD06} and \cite{VSH09}, that the full frictional drag
resumes after some delay, the duration of which seems to vary
stochastically \citep{SD09}.  Delayed friction can happen only in
simulations of disks in isolated, smooth halos, since any reasonable
amount of halo substructure, or a tidal encounter, disturbs the
delicate metastable state of the halo, causing friction to appear with
its full force.  Thus simulations that do not find strong friction
from moderately dense halos \citep[\eg][]{KVCQ} have simply not been
run for long enough.

While \cite{DS00} argued for near maximal disks, and their requirement
for a low halo density is in agreement with the disk masses derived
from fitting bar flow models (\S\ref{gasflow}), their constraint on
the halo density may be specific to their adopted halo models.  Thus
additional careful studies of other halo models seem warranted.

\subsubsection{Halo Density Reduction by Bars}
While a full discussion of processess that may lower the dark matter
density in the centers of halos is outside this review, a brief
mention of the effect of bar friction is appriate here.

\cite{WK02} argued that the transfer of angular momentum from the bar
to the halo could reduce the central density of the dark matter halo
by a substantial factor.  However, the possible density reduction is
quite modest \citep{HBWK,MD05,Sell08a} because the disk has only a
finite amount angular momentum to give to the halo.  Furthermore, as
the disk loses angular momentum, its mass distribution contracts, and
the deepening potential well further compresses the halo, which
actually overwhelms the slight density reduction \citep{Sell03,CVK06}.

\begin{figure}[t]
\begin{center}
\includegraphics[width=.7\hsize]{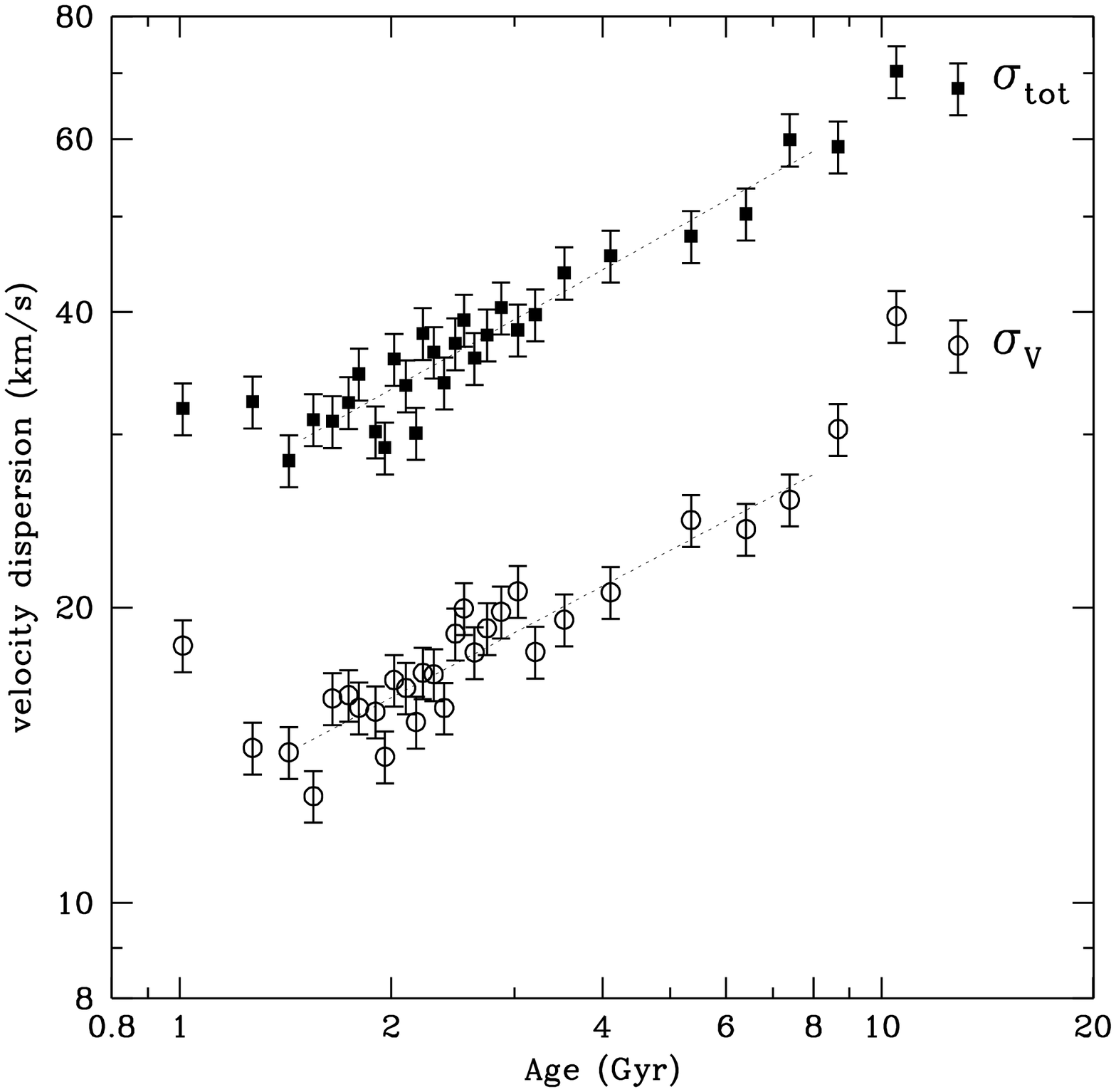}
\includegraphics[width=.7\hsize]{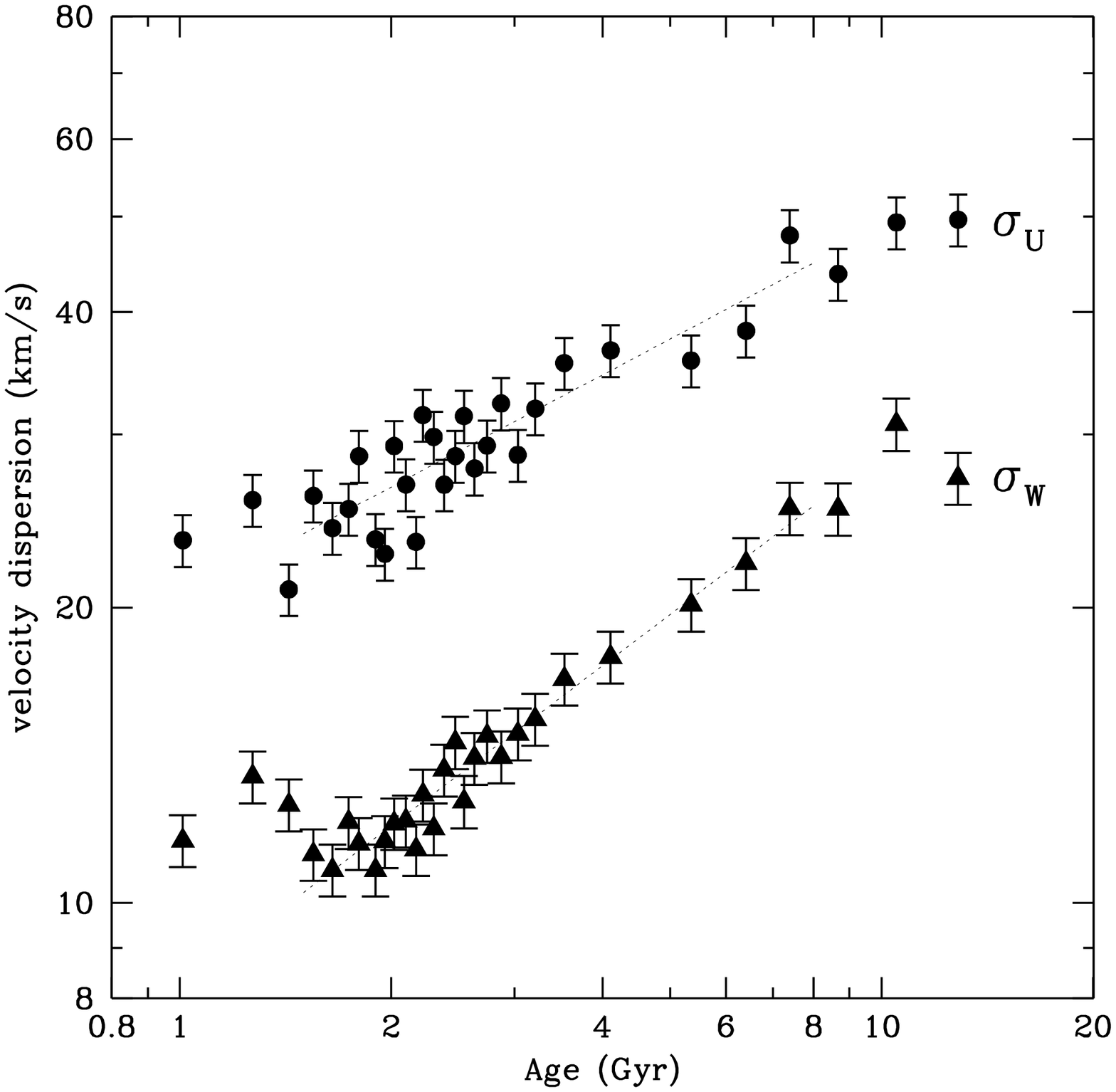}
\end{center}
\caption{The dispersion of stars in all three components, and the
  total dispersion of the \GCS\ sample of solar neighborhood stars
  using revised ages \citep{Holm07}.  The radial, azimuthal and
  vertical components are $\sigma_U$, $\sigma_V$, and $\sigma_W$
  respectively, and $\sigma_{\rm tot}^2 = \sigma_U^2 +\sigma_V^2 +
  \sigma_W^2$.  The fitted straight lines do not include the first
  three or last two points.}
\label{avr}
\end{figure}

\ssection{Secular Evolution within Disks}
\label{scatt}
Most of the behavior discussed so far, such as the non-linear
evolution of instabilities, causes changes to the host galaxy on a
dynamical time scale.  The broad topic of secular evolution in
galaxies describes changes that occur more gradually, such as the
secular formation of pseudo-bulges \citep[see][for excellent
reviews]{Korm93,KK04}, the formation of rings \citep[\eg][]{BC96}, or
dynamical friction between components (\S\ref{dynfr}).  The discussion
in this section concerns processes that scatter disk stars only.

It has long been realized that old stars in the solar neighborhood
have larger peculiar motions relative to the local standard of rest
(hereafter \LSR) than do young
stars \citep[\eg][]{Wiel77,Nord04,AB09}.  Postulating that older stars
were born with larger random speeds, say in a thicker disk, is
unattractive because it makes the present epoch of low velocity
dispersion special.  Some mechanism to scatter stars must therefore be
invoked to create the larger random speeds of older stars.

The trends presented in Fig.~\ref{avr} use the revised ages
\citep{Holm07} assigned to the \GCS\ stars \citep[see][for a
  review]{Sode10}.  Figs.~2, 3 \& 4 of \cite{AB09} show the
dispersions estimated as a function of color on the main sequence;
while blue stars are necessarily young, red stars are expected to have
a range of ages.  The dispersions of the supposed oldest stars
estimated by \cite{Holm07} are no greater than those estimated by
\cite{AB09} for their reddest stars, suggesting that the ``older''
bins in Fig.~\ref{avr} include stars having a wide range of ages, as
argued by \cite{Reid07}.  It should be noted that the reported
dispersions are simple second moments of the perhaps complex velocity
distributions (see Fig.~\ref{GCSsimp}).

A reliable determination of the variation of dispersion as a function
of time could provide another useful constraint on the scattering
mechanism (\eg\ \BTii\ \S8.4).  \cite{QG00} and \cite{SG07} argued that
the dispersion may saturate for stars above a certain age; with a much
older surge to account for the highest velocities.  However,
\cite{AB09} found better fits to the data with continuous acceleration,
and deduced $\sigma \sim t^{0.35}$, with $t$ being the current age of
the stellar generation, in tolerable agreement, in fact, with 0.38 for
the logarithmic slope of $\sigma_{\rm tot}$ in Fig.~\ref{avr}.

Three principal scattering agents have been discussed: dense gas
clumps in the disk, massive black holes in the halo, and recurrent
short-lived spirals.  Note that the first two are essentially
collisional processes that accelerate the relaxation rate (see the
Appendix), while the changes\break caused by spirals can increase
random motion without leading to a more relaxed \DF.  As none in
isolation fits the data, a combination of spirals and gas clumps seems
to be favored.  Minor mergers and the effects of halo substructure are
discussed in \S\ref{survival}.

An orbiting mass clump induces a collective spiral wake in the
surrounding disk that enhances its mass and size by substantial
factors \citep{JT66}, a complication that is ignored in many studies
of cloud scattering.  Since molecular gas is mostly concentrated in
spiral arms \citep{Niet06,Grat10,Efre10}, it is probably futile to
draw a sharp distinction between spiral arms and the wakes of dense
gas clumps, and a correct treatment would be to calculate the effects
of the combined star and gas disk.  \cite{BL88} took a step in this
direction, but a full calculation may remain unreachable for some time
if one tries to include a self-consistent treatment of the formation
and dispersal of the gas clumps: molecular gas concentrations probably
grow in the converging gas flow into a spiral arm, and are
subsequently dispersed by star formation.

Treating spirals and mass clumps in the disk as distinct scattering
agents may be justified, therefore, if the wakes of cloud complexes
can be lumped with spirals into a single scattering agent that is
distinct from the clouds that caused them.  At the very least, this
simplifying assumption separates the problem into tractable pieces.

\subsection{Heating by Spirals}
\cite{LBK} showed that stars are scattered by a slowly changing
potential perturbation only near resonances.  More precisely, a spiral
potential that grows and decays adiabatically, \ie\ on a time-scale
long compared with the orbital and epicyclic periods, will not cause a
lasting change to a star's $E$ and $L_z$.  Wave-particle interactions
become important near the resonances, where stars experience secular
changes through ``surfing'' on the potential variations at \CR\ or
through a periodic forcing close to their epicyclic frequency at the
Lindblad resonances.  Either case produces a lasting change to a
star's orbit.

The width of a resonance, \ie\ the range of orbit frequencies of stars
that are strongly affected, depends only on the amplitude of the
potential when the perturbation is long-lived.  But perturbations of
shorter lifetimes have a broader range of frequencies and more stars
experience lasting changes.

The discussion in \S\ref{resonances} and Fig.~\ref{lindblad} indicates
that stars that lose (gain) $L_z$ near the \ILR\ (\OLR) move onto more
eccentric orbits, which is the root cause of heating by spirals.
Exchanges at \CR\ move stars to new orbits also, but with no change to
the energy of non-circular motion, as discussed in \S\ref{churning}.

Significant heating by spiral waves over a large part of a disk
requires them to be transient; a quasi-steady pattern, of the type
envisaged by \cite{BL96} say, will cause localized heating at an
exposed resonance, while stars elsewhere will move through the pattern
without otherwise being affected.  \cite{BW67}, \cite{CS85} and
\cite{BL88} calculated the heating caused by transient spirals.
\cite{JB90}, \cite{DeSi04}, and \cite{MQ06} presented numerical
studies of the consequences for a disk of test particles subject to
some assumed set of spiral wave perturbations.

It is important to realize that the vertical oscillations of stars are
little affected by spiral potential variations \citep[\S\ref{vres}
  and][]{Carl87}.  In the absence of heavy clumps that can redirect
disk velocities (\S\ref{clouds}), the increasing in-plane motions in
simulations of initially cool, thin disks may ultimately cause the
velocity ellipsoid to become sufficiently anisotropic as to cause it
to thicken through mild buckling instabilities (\S\ref{bentsheet}).
This may account for claims \citep[\eg][]{MD07} that disks thicken due
to spiral heating.

\subsection{Churning by Transient Spirals}
\label{churning}
Studies of the metallicities and ages of nearby stars
\citep{Edva93,Nord04} found that older stars tend to have lower
metallicities on average.  As the ages of individual stars are
disputed \citep{Reid07,Holm07}, the precise form of the relation is
unclear.  However, there seems to be general agreement that there is a
spread of metallicities at each age, which is also supported by other
studies \citep{CHW03,Hayw08,SH10}.  The spread seems to be more than
twice that expected from simple blurring of the gradient by stellar
epicyclic excursions.  In the absence of radial mixing, a metallicity
spread amongst coeval stars is inconsistent with a simple chemical
evolution model in which the metallicity of the disk rises
monotonically in each annular bin.

\cite{SB02} showed that scattering at \CR\ causes very effective
mixing.  In a few Gyr, multiple transient spirals caused stars to
diffuse in radius.  Churning of the stellar disk occurs at corotation
of the spirals with no associated heating, and is able to account for
the apparent metallicity spread with age.  Ro\u skar \etal\ (2008ab)
presented more detailed simulations that included infall, star
formation and feedback that confirmed this behavior.  \cite{SB09}
developed the first chemical evolution model for the \MW\ disk to
include radial churning.

\subsection{Cloud Scattering}
\label{clouds}
Many years before the discovery of giant molecular gas clouds,
\cite{SS53} postulated their existence to account for the secular
heating of disk stars.  \cite{Lace84} extended their calculation to 3D
and concluded that cloud scattering should cause the vertical
dispersion, $\sigma_z$, to be intermediate between the radial,
$\sigma_R$, and azimuthal, $\sigma_\phi$, components.\footnote{The
ratio $\sigma_R/\sigma_\phi \approx 2\Omega/\kappa$ (\BTii, eq.~8.117)
is forced by epicyclic motions of disk stars.}  This result seems
physically plausible on energy equipartition grounds: scattering by
massive clouds redirects the peculiar motions of stars through random
angles, and therefore isotropizes the motions as far as the epicycle
gyrations allow.

Despite the fact that redirecting peculiar motions happens much more
rapidly than they can be increased by the same scatterers, the data do
not reveal Lacey's expected axis ratio.  The second moments of the
velocity distribution of solar neighborhood stars in the three
orthogonal directions \citep[Fig.~\ref{avr} and][]{Wiel77,AB09}
satisfy the inequality $\sigma_z < \sigma_\phi < \sigma_R$.  The ratio
of the two in-plane components is in reasonable agreement with
expectations from epicyclic motions, but the vertical component is the
smallest, and this remains true for all groups when the stars are
subdivided according to color or estimates of their ages.
\cite{Gers00} also observed a flattened ellipsoid in the disk of
NGC~2895.

\cite{Carl87} and \cite{JB90} therefore developed the plausible
argument that spirals drive up the in-plane components more rapidly
than scattering is able to redirect those motions into the vertical
direction, thereby accounting for the observed axis ratios of the
velocity ellipsoid.  \cite{Sell00} cited their argument as offering
strong support for the transient spiral picture, but it now seems to
be incorrect.

\cite{IKM93} claimed that cloud scattering alone would lead to the
vertical component being the smallest, with the precise axis ratio
depending on the local slope of the rotation curve.  Their simulations
\citep{SI99}, and others \citep[\eg][]{Vill85,HF02}, confirmed their
expectation.

\begin{figure}[t]
\begin{center}
\includegraphics[width=.9\hsize,clip=]{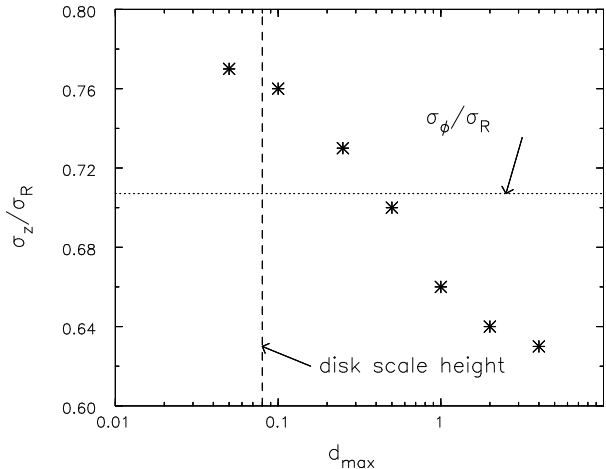}
\end{center}
\caption{The equilibrium axis ratio of the velocity ellipsoid of
  particles plotted as a function of the limiting range of the
  perturbation forces from the heavy particles. See \cite{Sell08b} for
  a description of the calculations.}
\label{axratio}
\end{figure}

\cite{Sell08b} resolved this disagreement using simulations of test
particles in the sheared sheet (see \S\ref{swing}).  Scattering by
randomly distributed co-orbiting mass\break clumps confirmed the
flattened velocity ellipsoid predicted by \cite{IKM93}.
Fig.~\ref{axratio} reveals why Ida's prediction differs from Lacey's:
Lacey, as others \citep{SS53,BL88}, assumed that cloud scattering is
local, but the $\ln\Lambda$ term in the formulae in the Appendix
implies that distant encounters dominate any scattering process in 3D,
at least to distances a few times the disk thickness.  Distant
scatterers in the flattened geometry of a disk must predominantly
affect the in-plane star velocities, and couple much less strongly to
the vertical component.  Fig.~\ref{axratio} shows the equilibrium
ratio $\sigma_z/\sigma_R$ when scatterers beyond the finite range
$d_{\rm max}$ artificially exert no forces.  The ratio settles to
something close to Lacey's energy equipartition prediction when none
but the closest heavy clumps perturb the stars, but the equilibrium
ellipsoid flattens in separate experiments as more distant clouds are
included, tending towards Ida's result with no artificial cut off.

Thus the {\it shape\/} of the local velocity ellipsoid,
Fig.~\ref{avr}, is apparently consistent with cloud scattering and
does not, as seemed attractive, require concurrent spiral arm
scattering.  However, the data do not imply that spirals are
unimportant: cloud scattering seems unable to generate the random
speeds of the oldest stars
\citep[\eg][]{Lace91,HF02}, and there are hints in Fig.~\ref{avr} of
some evolution of the velocity ellipsoid shape that may demand a
compound origin.

\subsection{Black Holes in the Halo}
The possibility that the dark matter halos of galaxies are made up of
massive black holes (\BH s) is not yet excluded. \cite{LO85} calculated
the consequences for the stars in the disk of the \MW, assuming
the \BH s to have orbits characteristic of a pressure-supported halo
and to impart impulses to disk stars as they pass through the disk.
The high speeds of their encounters with disk stars would cause the
velocity dispersion to rise as $t^{1/2}$, while the predicted shape of
the velocity ellipsoid is in reasonable agreement with that observed.

\cite{LO85} addressed a number of issues with their model, such as the
X-ray accretion luminosity as the \BH s pass through the gas disk, the
accumulation of \BH s in the galactic center through dynamical
friction, and the survival of dwarf galaxies.  They also acknowledged
that it does not predict the correct variation of $\sigma_{\rm tot}$
with Galactic radius.  \BTii\ (\S7.4.4) added that the idea could be
ruled out on the grounds that wide binary star systems would be
disrupted too quickly.

If disk scattering is dominated by spirals, as argued below, then
scattering by \BH s would be needed merely to redirect the peculiar
motions into the vertical direction.  This reviewer is not aware of
any such calculation, but since passing \BH s scatter stars in the
direction perpendicular to their orbits, it seems unlikely that they
could redirect peculiar motion without also increasing in-plane
motions.  However, if this expectation is too pessimistic and the
desired axis ratio could be achieved with lower \BH\ masses, many of
the other objections are weakened.

\subsection{Discussion}
The extraordinary phase-space structure of the solar neighborhood
\citep[][and Fig.~\ref{GCSsimp}]{Dehn98,Nord04} indicates that there
is little in the way of an underlying smooth component and the stellar
distribution is broken into several ``streams'' \citep{BHR9}.  The
features are too substantial to have simply arisen from groups of
stars that were born with similar kinematics \citep[\eg][]{Egge96}, as
confirmed in detailed studies \citep{Fama07,Bens07,BH09}.

Thus it is clear that the entire \DF\ has been sculptured by dynamical
processes.  Were the large part of the spread in velocities caused by
multiple scatterings off molecular clouds in the disk, or off black
holes in the halo, the distribution should approximate the simple
double Gaussian proposed by K. Schwarzschild (see \BTii\ \S4.4.3).
The vertical velocity distribution \citep{Nord04}, on the other hand,
does have a relaxed appearance, as noted by \cite{SG07}.

Various dynamical agents have been proposed to account for kinematic
features in the solar neighborhood.  \cite{Kaln91} argued that the
\OLR\ of the bar in the \MW\ might be close to the solar circle.
Features in the subsequently-released \Hipp\ data have also been
attributed to the \OLR\ of the bar \citep{Rabo98,Dehn00,Fux01}, while
\cite{Sell10} attributed another to a recent \ILR.

\cite{DeSi04} were able to produce distributions of stars with a
similar degree of substructure in simulations of test particles moving
in the adopted potential perturbations representing a succession of
short-lived spiral transients; see also \cite{MQ06}.  Other models
that included both bars and spirals were presented by \cite{Quil03},
\cite{Chak07}, and \cite{Anto09}, while \cite{Helm06} suggested that
some of the substructure may also be caused by satellite infall (see
also \S\ref{survival}).

Since spirals (and bars) are inefficient at exciting vertical motions,
another scattering process needs to be invoked to redirect in-plane
random motion into vertical motion.  The influence of clumps in the
disk, which are known to exist, seems consistent with the relaxed
appearance of the vertical velocity distribution.  Furthermore, the
observed axes of the velocity ellipsoid, except possibly for the
youngest stars (Fig.~\ref{avr}), are consistent with the prediction of
cloud scattering.  Thus cloud scattering seems to be just sufficient
to redirect velocities, which they are good at, but not to contribute
significantly to heating.

\ssection{Fragility of Disks}
\label{survival}
Most of the stars in spiral galaxies reside in remarkably thin disks.
In a photometric study of edge-on galaxies, \cite{YD06} found that the
fraction of the baryons in the thin disk component is in the range
70\% to 90\%, with higher fractions being characteristic of more
massive galaxies.  The so-called superthin disks are even more
extreme; the disk of the low-luminosity galaxy UGC~7321, for example,
has a radial scale-length some 14 times that of its characteristic
thickness \citep{Matt00}.

Hierarchical galaxy formation scenarios predict that galaxy formation
is far from monolithic, with occasional major mergers with other
halos, and frequent minor mergers.  Stellar disks that have formed in
the centers of the halos are torn apart in major mergers \citep{BH92},
but the consequences of minor mergers are harder to determine from
simulations \citep[\eg][]{Walk96}.  \cite{TO92} argued that the
existence of thin disks in galaxies can be used to constrain the rate
of minor mergers.  Their numerical estimates have been criticized on
various grounds \citep[\eg][]{HC97,SNT98,VW99}, but it is clear that a
tight constraint remains.

Wyse (this volume) stresses that the thick disk of the \MW\ contains
only very old stars, and that the ages of thin disk stars stretch back
10~Gyr.  This fact would seem to imply that the last galactic merger
to have stirred the \MW\ disk occurred some 10~Gyr ago, and that no
comparable disturbance could have occurred since.  The \MW\ may not be
unique in this respect, as \cite{Moul05} found that thick disks in
four nearby galaxies also appear to be old.

\cite{Stew08} estimated the rate of mergers in the current
\LCDM\ cosmology.  They concluded that 95\% of parent halos of some
$10^{12}h^{-1}\;M_\odot$ will have merged with another halo of at
least 20\% of its mass in the last 10~Gyr, and consequently the disk
hosted by the parent must somehow survive in most cases.

Possibly only a small fraction of infalling satellites pose a real
threat to a disk.  All satellites will be tidally stripped as they
fall into the main halo, and some may even dissolve completely before
they can affect the disk -- the Sagittarius dwarf galaxy appears to be
a good example \citep{LJM05}.  A massive satellite loses orbital
energy through dynamical friction (\S\ref{dynfr}) causing it to settle
deeper into the main halo.  Its survival depends on its density (see
\BTii\ \S8.3); it will be stripped of its loosely bound envelope until
its mean density, $\bar \rho$, is about one third the mean density of
the halo interior to its orbital radius (\BTii\ eq.~8.92), and
disrupted entirely as it reaches a radius in the main halo where even
its central density falls below the mean interior halo density.  This
process has been studied in more detail by \cite{BM07} and by
\cite{Choi09}.  Thus the vulnerability of disks depends on the inner
densities, in both dark matter and baryons, of the accreted sub-halos.

If even a single, moderately massive core survives to the inner halo,
it could cause an unacceptable increase in the disk thickness.
Vertical heating of the disk occurs when a passing or penetrating
satellite is able to increase the vertical motions of the disk stars.
High speed passages therefore deposit little energy into disk motions,
but if the satellite's orbit remains close to the disk mid-plane, then
its vertical frequency will couple strongly to that of some of the
disk stars and heating will be rapid.  Indeed, \cite{Read08} argued
that the accretion event(s) that created the thick disk of the
\MW\ most probably resulted from the infall of a subhalo whose orbit
plane was inclined at 10\degr\ -- 20\degr\ to that of the disk.  The
energy deposited could take the form of exciting bending waves in the
disk that can propagate radially until they are damped at vertical
resonances \citep{SNT98}.

\cite{Kaza09} simulated several minor mergers with sub-halos in
detailed models with plausible parameters.  They found that the disk
is substantially thickened and heated by the mergers, although they
noted that their simulations lacked a gaseous component.  Simulations
that include a gas component cannot resolve the small-scale physical
processes of gas fragmentation, star formation, feedback, \etc, and
therefore the behavior of the dissipative gas component necessarily
includes somewhat {\it ad hoc\/} prescriptions for these aspects of
sub-grid physics.  \cite{Hopk09} stressed that gas in mergers settles
quickly to begin to form a new disk; however, it is the fate of the
stars that had been formed prior to the merger that is the principal
concern.  \cite{Kaza09} argued it is possible that a dissipative
component in the disk could absorb some of the orbital energy of the
satellite which would reduce the heating of the stellar disk, and such
an effect appears to have occured in the simulations by \cite{Most10}.
A clear conclusion has yet to emerge from this on-going research
effort, but exclusively old thick disks together with the prevalence
of thin disks poses a substantial, though perhaps surmountable,
challenge to the \LCDM\ model.

It should be noted that ideas to reduce the density of the inner main
halo through frictional energy loss to halo
\citep[\eg][]{ElZa01,MCW06,Roma08a} require the kinds of dense massive
fragments that are themselves a danger to the disk.  This will be of
less importance in the early stages of galaxy assembly, but disk
survival adds the requirement that any such process be completed
quickly.

\ssection{Conclusions}
This review has focused on understanding the mechanisms that underlie
the various instabilities and processes that affect the structure of
galaxy disks.  The discussion has referred extensively to simulations
that are sufficiently simple that they clearly illustrate each
particular aspect of the behavior.  A complementary, and powerful,
approach is to add additional physical processes to simulations, with
the aim of improving realism to address otherwise inaccessible
questions.  However, the increased complexity of the behavior that
comes with increasing realism, makes a deep understanding of the
results harder to achieve.  Hopefully, the behavior described here
will form a solid basis on which to build as the challenges presented
by more realistic systems are addressed.

A contrast with accretion disk theory seems appropriate here.
\cite{SS73} proposed a scaling relation for turbulent viscosity that
led to rapid progress in modeling accretion disks some 25 years before
the likely origin of the viscosity was identified \citep{BH98}.  A
theory for the structural evolution of galaxies seems much farther
away.  While most of the important physical processes may have been
identified, exactly how they drive evolution is still not fully
understood, and even the evolutionary path remains vague.  In the
absence of this understanding, simplifying scaling laws cannot be
identified with confidence and it seems best to work at improved
understanding of the mechanisms at play.

Galaxy dynamics has made immense strides in the 45 years since the
chapter by \cite{Oort65} in the corresponding volume of the preceding
series.  Understanding of local wave mechanics (\S\ref{local}) and
the mechanisms for the principal global gravitational instabilities
(\S\S\ref{global} \& \ref{gmodes}) is well advanced.  Bending waves
and global buckling modes (\S\ref{warps}) are mostly understood, but
not at quite the same level, while lop-sided modes (\S\ref{lopsided})
perhaps still require some more concerted effort.  Understanding
of bar structure (\S\ref{bars}) and the role of bars in galaxy
evolution has developed beyond recognition, and disk heating
(\S\ref{scatt}) seems more solidly understood, with our confidence
being strongly boosted by the extremely valuable data from
\cite{Nord04}.

But answers to a number of major questions of galaxy dynamics are
still incomplete.  Although the bar-forming instability is now
understood (\S\ref{barmech}), it does not provide a clear picture of
why only a little over half of all bright disk galaxies are barred
(\S\ref{barfreq}).  After bars, spirals arms are the second most
prominent feature of disk galaxies and are probably responsible for
the most dynamical evolution (\S\ref{scatt}), yet a deep understanding
of their origin (\S\ref{spirals}) remains elusive.  The good progress
made in recent years to understand the warps of galaxy disks
(\S\ref{warps}) still has not supplied a satisfactory account of their
incidence.

Many of these outstanding issues may be bound up with how galaxies
form, our understanding of which is currently making particularly
rapid progress.

\section*{Acknowledgments}
The author is indebted to James Binney, Victor Debattista, Agris
Kalnajs, Juntai Shen, Alar Toomre, and Scott Tremaine for numerous
valuable comments on a draft of this review.

\vfill\eject
\appendix

\section{Relaxation Time in Spheroids and Disks}
A test particle moving at velocity $\bv$ along a trajectory that
passes a stationary field star of mass $m$ with impact parameter $b$
is deflected by the attraction of the field star.  For a distant
passage, it acquires a transverse velocity component $|\bv_\perp|
\simeq 2Gm/(bv)$ to first order (\BTii\ eq.\ 1.30).  Encounters at
impact parameters small enough to produce deflections where this
approximation fails badly are negligibly rare and relaxation is driven
by the cumulative effect of many small deflections.

If the density of field stars is $n$ per unit volume, the test
particle will encounter $\delta n = 2\pi b\delta b\,nv$ stars per unit
time with impact parameters between $b$ and $b+\delta b$.  Assuming
stars to have equal masses, each encounter at this impact parameter
produces a randomly directed $\bv_\perp$ that will cause a mean square
net deflection per unit time of
\be
\delta v_\perp^2 \simeq \left( {2Gm \over bv }\right)^2 \times
2\pi b\delta b\,nv = {8\pi G^2m^2n \over v}{\delta b \over b}.
\label{elerate}
\ee
The total rate of deflection from all encounters is the integral over
impact parameters, yielding
\be
v_\perp^2 = {8\pi G^2m^2n \over v}\int_{b_{\rm min}}^{b_{\rm max}}{db \over b}
= {8\pi G^2m^2n \over v} \ln\Lambda,
\label{trate}
\ee
where $\ln\Lambda \equiv \ln(b_{\rm max}/b_{\rm min})$ is the Coulomb
logarithm.  Typically one chooses the lower limit to be the impact
parameter of a close encounter, $b_{\rm min} \simeq 2Gm/v^2$, for
which $|\bv_\perp|$ is overestimated by the linear formula, while the
upper limit is, say, the half-mass (or effective) radius, $R$, of the
stellar distribution beyond which the density decreases rapidly.  The
vagueness of these definitions is not of great significance to an
estimate of the overall rate because we need only the logarithm of
their ratio.  The Coulomb logarithm implies equal contributions to the
integrated deflection rate from every decade in impact parameter
simply because the diminishing gravitational influence of more distant
stars is exactly balanced by their increasing numbers.

Note that the first order deflections that give rise to this steadily
increasing random energy come at the expense of second order
reductions in the forward motion of the same particles that we have
neglected \citep{Heno73}.  Thus the system does indeed conserve
energy, as it must.

We define the {\bf relaxation time} to be the time needed for $v_\perp^2
\simeq v^2$, where $v$ is the typical velocity of a star.  Thus
\be
\tau_{\rm relax} = {v^3 \over 8\pi G^2m^2n \ln\Lambda}.
\ee
To order of magnitude, a typical velocity $v^2 \approx GNm/R$, where
$N$ is the number of stars each of mean mass $m$, yielding $\Lambda
\approx N$. Defining the dynamical time to be $\tau_{\rm dyn} = R/v$
and setting $N \sim R^3n$, we have
\be
\tau_{\rm relax} \approx {N \over 6 \ln N}\tau_{\rm dyn},
\label{trelax}
\ee
which shows that the collisionless approximation is well satisfied in
galaxies, which have $10^8 \la N \la 10^{11}$ stars.  Including the
effect of a smooth dark matter component in this estimate would
increase the typical velocity, $v$, thereby further lengthening the
relaxation time.

This standard argument, however, assumed a pressure-supported
quasi-spherical system in several places.  \cite{Rybi72} pointed out
that the flattened geometry and organized streaming motion within
disks affects the relaxation time in two important ways.  First, the
assumption that the typical encounter velocity is comparable to the
orbital speed $v=(GNm/R)^{1/2}$ is clearly wrong; stars move past each
other at the typical random speeds in the disk, say $\beta v$ with
$\beta \sim 0.1$, causing larger deflections and decreasing the
relaxation time by a factor $\beta^3$.

Second, the distribution of scatterers is not uniform in 3D, as was
implicitly assumed in eq.~(\ref{elerate}).  Assuming a razor thin
disk, changes the volume element from $2\pi v\,b\delta b$ for 3D to
$2v\,\delta b$ in 2D, which changes the integrand in eq.~(\ref{trate})
to $b^{-2}$ and replaces the Coulomb logarithm by the factor $(b_{\rm
  min}^{-1} - b_{\rm max}^{-1})$.  In 2D therefore, relaxation is
dominated by close encounters.  Real galaxy disks are neither razor
thin, nor spherical, so the spherical dependence applies at ranges up
to the typical disk thickness, $z_0$, beyond which the density of
stars drops too quickly to make a significant further contribution to
the relaxation rate.  Thus we should use $\Lambda \simeq z_0/b_{\rm
  min}$ for disks.  More significantly, the local mass density is also
higher, so that $N \sim R^2z_0n$.  These considerations shorten the
relaxation time by the factor $(z_0/R)\ln(R/z_0)$.  An additional
effect of flattened distribution of scatterers is to determine the
shape of the equilibrium velocity ellipsoid, as discussed in
\S\ref{clouds}.

A third consideration for disks is that the mass distribution is much
less smooth than is the case in the bulk of pressure supported
galaxies.  A galaxy disk generally contains massive star clusters and
giant molecular clouds whose influence on the relaxation rate turns
out to be non-negligible (see \S\ref{scatt}).


\begin{thebibliography}{99}

\def\bold#1{{\bf #1}}

\def\aap{A\&A}
\def\aapr{A\&A Rev.}
\def\aj{AJ}
\def\apj{ApJ}
\def\apjl{ApJL}
\def\apjs{ApJS}
\def\apss{Ap.\ Sp. Sci.}
\def\araa{ARAA}
\def\jcoph{J. Comp.\ Phys.}
\def\mnras{MNRAS}
\def\newa{NewA}
\def\nat{Nature}
\def\phr{Phys.\ Rep.}
\def\pnas{Proc.\ Nat.\ Acad.\ Sci.}
\def\PhD{PhD thesis}
\def\rmp{Rev.\ Mod.\ Phys.}
\def\rpp{Rep.\ Prog.\ Phys.}
\def\sci{Sience}
\def\sva{Sov.\ Ast.}
\def\sval{Sov.\ Ast.\ Lett.}

\bibitem[\protect\citeauthoryear{Agertz \etal}{2010}]{Ager10}
Agertz, O., Teyssier, R. \& Moore, B. 2010, arXiv:1004.0005

\bibitem[\protect\citeauthoryear{Aguerri \etal}{2009}]{AMC09}
Aguerri, J. A. L., M\'endez-Abreu, J. \& Corsini, E. M. 2009, \aap, {\bf 495}, 491

\bibitem[\protect\citeauthoryear{Antoja \etal}{2009}]{Anto09}
Antoja, T., Valenzuela, O., Pichardo, B., Moreno, E., Figueras, F. \& Fernández, D. 2009, \apjl, {\bf 700}, L78

\bibitem[\protect\citeauthoryear{Aoki \etal}{1979}]{ANI79}
Aoki, S., Noguchi, M. \& Iye, M. 1979, \pasj, {\bf 31}, 737

\bibitem[\protect\citeauthoryear{Araki}{1985}]{Arak85}
Araki, S. 1985, \PhD, MIT.

\bibitem[\protect\citeauthoryear{Araki}{1987}]{Arak87}
Araki, S. 1987, \aj, {\bf 94}, 99

\bibitem[\protect\citeauthoryear{Athanassoula}{1992}]{Atha92}
Athanassoula, E. 1992, \mnras, {\bf 259}, 345

\bibitem[\protect\citeauthoryear{Athanassoula}{2002}]{Atha02}
Athanassoula, E. 2002, \apjl, {\bf 569}, L83

\bibitem[\protect\citeauthoryear{Athanassoula}{2003}]{Atha03}
Athanassoula, E. 2003, \mnras, {\bf 341}, 1179

\bibitem[\protect\citeauthoryear{Athanassoula \etal}{1987}]{ABP87}
Athanassoula, E., Bosma, A. \& Papaioannou, S. 1987, \aap, {\bf 179}, 23

\bibitem[\protect\citeauthoryear{Athanassoula \etal}{2005}]{ALD05}
Athanassoula, E., Lambert, J. C. \& Dehnen, W. 	2005, \mnras, {\bf 363}, 496

\bibitem[\protect\citeauthoryear{Athanassoula \etal}{2009}]{ARGM}
Athanassoula, E., Romero-G\'omez, M. \& Masdemont, J. J. 2009, \mnras, {\bf 394}, 67

\bibitem[\protect\citeauthoryear{Athanassoula \& Sellwood}{1986}]{AS86}
Athanassoula, E. \& Sellwood, J. A. 1986, \mnras, {\bf 221}, 213

\bibitem[\protect\citeauthoryear{Aumer \& Binney}{2009}]{AB09}
Aumer, M. \& Binney, J. J. 2009, \mnras, {\bf 397}, 1286

\bibitem[\protect\citeauthoryear{Bailin}{2003}]{Bail03}
Bailin, J. 2003, \apjl, {\bf 583}, L79

\bibitem[\protect\citeauthoryear{Balbus \& Hawley}{1998}]{BH98}
Balbus, S. A. \& Hawley, J. F. 1998, \rmp, {\bf 70}, 1

\bibitem[\protect\citeauthoryear{Baldwin \etal}{1980}]{BLBS}
Baldwin J. E., Lynden-Bell D., Sancisi R. 1980, \mnras, {\bf 193}, 313

\bibitem[\protect\citeauthoryear{Barazza \etal}{2008}]{BJM08}
Barazza, F. D., Jogee, S. \& Marinova, I. 2008, \apj, {\bf 675}, 194

\bibitem[\protect\citeauthoryear{Barazza \etal}{2009}]{Bara09}
Barazza, F. D. \etal\ 2009, \aap, {\bf 497}, 713

\bibitem[\protect\citeauthoryear{Barbanis \& Woltjer}{1967}]{BW67}
Barbanis, B. \& Woltjer, L. 1967, \apj, {\bf 150}, 461

\bibitem[\protect\citeauthoryear{Barnes \& Hernquist}{1992}]{BH92}
Barnes, J. E. \& Hernquist, L. 1992, \araa, {\bf 30}, 705

\bibitem[\protect\citeauthoryear{Benedict \etal}{2002}]{Bene02}
Benedict, G. F., Howell, A., Jorgensen, I., Kenney, J. \& Smith, B. J. 2002, \aj, {\bf 123}, 1411

\bibitem[\protect\citeauthoryear{Bensby \etal}{2007}]{Bens07}
Bensby, T., Oey, M. S., Feltzing, S. \& Gustafsson, B. 2007, \apjl, {\bf 655}, L89

\bibitem[\protect\citeauthoryear{Berentzen \etal}{2004}]{Bere04}
Berentzen, I., Athanassoula, E., Heller, C. H. \& Fricke, K. J. 2004, \mnras, {\bf 347}, 220

\bibitem[\protect\citeauthoryear{Berentzen \etal}{2007}]{Bere07}
Berentzen, I., Shlosman, I., Martinez-Valpuesta, I. \& Heller, C. H. 2007, \apj, {\bf 666}, 189

\bibitem[\protect\citeauthoryear{Bertin \& Lin}{1996}]{BL96}
Bertin, G. \& Lin, C. C. 1996, {\it Spiral Structure in Galaxies\/} (Cambridge, MA: The MIT Press)

\bibitem[\protect\citeauthoryear{Binney, Jiang \& Dutta}{Binney \etal}{1998}]{BJD98}
Binney, J., Jiang, I. \& Dutta, S. 1998, \mnras, {\bf 297}, 1237

\bibitem[\protect\citeauthoryear{Binney \& Lacey}{1988}]{BL88}
Binney, J. J. \& Lacey, C. G. 1988, \mnras, {\bf 230}, 597

\bibitem[\protect\citeauthoryear{Binney \& Tremaine}{2008}]{BT08}
Binney, J. \& Tremaine, S. 2008, {\it Galactic Dynamics\/} (2nd ed.; Princeton: Princeton University Press) (BT08)

\bibitem[\protect\citeauthoryear{Block \etal}{2004}]{Bloc04}
Block, D. L., Freeman, K. C., Jarrett, T. H., Puerari, I., Worthey, G., Combes, F. \& Groess, R. 2004, \aap, {\bf 425}, L37

\bibitem[\protect\citeauthoryear{Bosma}{1991}]{Bosm91}
Bosma, A. 1991, in {\it Warped Disks and Inclined Rings Around Galaxies}, ed.\ S. Casertano, P. D. Sackett \& F. H. Briggs (Cambridge: Cambridge University Press), 181

\bibitem[\protect\citeauthoryear{Bosma}{1996}]{Bosm96}
Bosma, A. 1996, in IAU Colloq.\ {\bf 157}, {\it Barred Galaxies}, ed.\ R. Buta, D. A. Crocker \& B. G. Elmegreen (San Francisco: ASP Conf series {\bf 91}), 132

\bibitem[\protect\citeauthoryear{Bottema}{1996}]{Bott96}
Bottema, R. 1996, \aap, {\bf 306}, 345

\bibitem[\protect\citeauthoryear{Bournaud \etal}{2005}]{BCS05}
Bournaud, F., Combes, F. \& Semelin, B. 2005, \mnras, {\bf 364}, L18

\bibitem[\protect\citeauthoryear{Bovy \& Hogg}{2009}]{BH09}
Bovy, J. \& Hogg, D. W. 2009, arXiv:0912.3262

\bibitem[\protect\citeauthoryear{Bovy, Hogg \& Roweis}{Bovy \etal}{2009}]{BHR9}
Bovy, J., Hogg, D. W. \& Roweis, S. T. 2009, \apj, {\bf 700}, 1794

\bibitem[\protect\citeauthoryear{Boylan-Kolchin \& Ma}{2007}]{BM07}
Boylan-Kolchin, M. \& Ma, C-P. 2007, \mnras, {\bf 374}, 1227

\bibitem[\protect\citeauthoryear{Briggs}{1990}]{Brig90}
Briggs, F. H. 1990, \apj, {\bf 352}, 15

\bibitem[\protect\citeauthoryear{Bureau \& Athanassoula}{2005}]{BA05}
Bureau, M. \& Athanassoula, E., 2005, \apj, {\bf 626}, 159

\bibitem[\protect\citeauthoryear{Buta \& Combes}{1996}]{BC96}
Buta, R. \& Combes, F. 1996, \fcp, {\bf 17}, 95 

\bibitem[\protect\citeauthoryear{Buta \etal}{2009}]{Buta09}
Buta, R. J., Knapen, J. H., Elmegreen, B. G., Salo, H., Laurikainen, E., Elmegreen, D. M., Puerari, I. \& Block, D. L. 2009, \aj, {\bf 137}, 4487

\bibitem[\protect\citeauthoryear{Camm}{1950}]{Camm50}
Camm, G. L, 1950, \mnras, {\bf 110}, 305

\bibitem[\protect\citeauthoryear{Carlberg}{1987}]{Carl87}
Carlberg, R. G. 1987, \apj, {\bf 322}, 59

\bibitem[\protect\citeauthoryear{Carlberg \& Freedman}{1985}]{CF85}
Carlberg, R. G. \& Freedman, W. L. 1985, \apj, {\bf 298}, 486

\bibitem[\protect\citeauthoryear{Carlberg \& Sellwood}{1985}]{CS85}
Carlberg, R. G. \& Sellwood, J. A. 1985, \apj, {\bf 292}, 79

\bibitem[\protect\citeauthoryear{Carollo \etal}{1998}]{CSM98}
Carollo, C. M., Stiavelli, M. \& Mack, J. 1998, \aj, {\bf 116}, 68

\bibitem[\protect\citeauthoryear{Chakrabarty}{2007}]{Chak07}
Chakrabarty, D. 2007, \aap, {\bf 467}, 145

\bibitem[\protect\citeauthoryear{Chandrasekhar}{1943}]{Chan43}
Chandrasekhar, S. 1943, \apj, {\bf 97}, 255 

\bibitem[\protect\citeauthoryear{Chemenin \& Hernandez}{2009}]{CH09}
Chemin, L. \& Hernandez, O. 2009, \aap, {\bf 499}, L25

\bibitem[\protect\citeauthoryear{Chen, Hou \& Wang}{Chen \etal}{2003}]{CHW03}
Chen, L., Hou, J. L. \& Wang, J. J. 2003, \aj, {\bf 125}, 1397

\bibitem[\protect\citeauthoryear{Chirikov}{1979}]{Chir79}
Chirikov, B. V. 1979, \phr, {\bf 52}, 265-379

\bibitem[\protect\citeauthoryear{Christodoulou \etal}{1995}]{CST95}
Christodoulou, D. M., Shlosman, I. \& Tohline, J. E. 1995, \apj, {\bf 443}, 551

\bibitem[\protect\citeauthoryear{Choi \etal}{2009}]{Choi09}
Choi, J.-H., Weinberg, M. D. \& Katz, N. 2009, \mnras, {\bf 400}, 1247

\bibitem[\protect\citeauthoryear{Col\'\i n \etal}{2006}]{CVK06}
Col\'\i n, P., Valenzuela, O. \& Klypin, A. 2006, \apj, {\bf 644}, 687

\bibitem[\protect\citeauthoryear{Combes \& Sanders}{1981}]{CS81}
Combes, F. \& Sanders, R. H. 1981, \aap, {\bf 96}, 164

\bibitem[\protect\citeauthoryear{Contopoulos}{1980}]{Cont80}
Contopoulos, G. 1980, \aap, {\bf 81}, 198

\bibitem[\protect\citeauthoryear{Corbelli \& Walterbos}{2008}]{CW07}
Corbelli, E. \& Walterbos, R. A. M. 2007, \apj, {\bf 669}, 315

\bibitem[\protect\citeauthoryear{Corsini}{2008}]{Cors08}
Corsini, E. M. 2008, in {\it Formation and Evolution of Galaxy Bulges}, IAU Symp.~{\bf 245} (Dordrecht: Kluwer) p.~125

\bibitem[\protect\citeauthoryear{Corsini \etal}{2003}]{CDA03}
Corsini, E. M., Debattista, V. P. \& Aguerri, J. A. L. 2003, \apjl, {\bf 599}, L29

\bibitem[\protect\citeauthoryear{Courteau \etal}{2003}]{Cour03}
Courteau, S., Andersen, D. R., Bershady, M. A., MacArthur, L. A. \& Rix, H-W. 2003, \apj, {\bf 594}, 208

\bibitem[\protect\citeauthoryear{Cox \etal}{1996}]{CSvMS96}
Cox, A. L., Sparke, L. S., van Moorsel, G. \& Shaw, M. 1996, \aj, {\bf 111}, 1505

\bibitem[\protect\citeauthoryear{Curir \etal}{2006}]{CMM06}
Curir, A., Mazzei, P. \& Murante, G. 2006, \aap, {\bf 447}, 453

\bibitem[\protect\citeauthoryear{Cuzzi \etal}{2010}]{Cuzz10}
Cuzzi, J. N., \etal\ 2010, \sci, {\bf 327} 1470

\bibitem[\protect\citeauthoryear{Davoust \& Contini}{2004}]{DC04}
Davoust, E. \& Contini, T. 2004, \aap, {\bf 416}, 515

\bibitem[\protect\citeauthoryear{Debattista \& Sellwood}{1998}]{DS98}
Debattista, V. P. \& Sellwood, J. A. 1998, \apjl, {\bf 493}, L5

\bibitem[\protect\citeauthoryear{Debattista \& Sellwood}{1999}]{DS99}
Debattista, V. P. \& Sellwood, J. A. 1999, \apjl, {\bf 513}, L107

\bibitem[\protect\citeauthoryear{Debattista \& Sellwood}{2000}]{DS00}
Debattista, V. P. \& Sellwood, J. A. 2000, \apj, {\bf 543}, 704

\bibitem[\protect\citeauthoryear{Debattista \& Shen}{2007}]{DS07}
Debattista, V. P. \& Shen, J. A. 2007, \apjl, {\bf 654}, L127

\bibitem[\protect\citeauthoryear{Dehnen}{1998}]{Dehn98}
Dehnen, W. 1998, \aj, {\bf 115}, 2384

\bibitem[\protect\citeauthoryear{Dehnen}{2000}]{Dehn00}
Dehnen, W. 2000, \aj, {\bf 119}, 800

\bibitem[\protect\citeauthoryear{Dekel \& Shlosman}{1983}]{DS83}
Dekel, A. \& Shlosman, I. 1983, in IAU Symposium {\bf 100}, {\it Internal Kinematics and Dynamics of Galaxies}, ed.\ E. Athanassoula (Dordrecht: Reidel) p~187

\bibitem[\protect\citeauthoryear{De Simone, Wu \& Tremaine}{De Simone \etal}{2004}]{DeSi04}
De Simone, R. S., Wu, X. \& Tremaine, S. 2004, \mnras, {\bf 350}, 627

\bibitem[\protect\citeauthoryear{Dobbs \etal}{2010}]{DTPB}
Dobbs, C. L., Theis, C., Pringle, J. E. \& Bate, M. R. 2010, \mnras, (in press)	

\bibitem[\protect\citeauthoryear{Donner \& Thomasson}{1994}]{DT94}
Donner, K. J. \& Thomasson, M. 1994, \aap, {\bf 290}, 475

\bibitem[\protect\citeauthoryear{Dubinski \etal}{2009}]{DBS09}
Dubinski, J., Berentzen, I. \& Shlosman, I. 2009, \apj, {\bf 697}, 293

\bibitem[\protect\citeauthoryear{Dubinski \& Chakrabarty}{2009}]{DC09}
Dubinski, J. \& Chakrabarty, D. 2009, \apj, {\bf 703}, 2068

\bibitem[\protect\citeauthoryear{Dubinski \etal}{2008}]{Dubi08}
Dubinski, J., Gauthier, J.-R., Widrow, L. \& Nickerson, S. 2008, in {\it Formation and Evolution of Galaxy Disks}, ed.\ J. G. Funes SJ \&  E. M. Corsini (San Francisco: ASP {\bf 396}), p.~321

\bibitem[\protect\citeauthoryear{Dubinski \& Kuijken}{1995}]{DK95}
Dubinski, J. \& Kuijken, K. 1995, \apj, {\bf 442}, 492

\bibitem[\protect\citeauthoryear{Dury \etal}{2008}]{DdRDD}
Dury, V., de Rijcke, S., Debattista, V. P. \& Dejonghe, H. 2008, \mnras, {\bf 387}, 2 

\bibitem[\protect\citeauthoryear{Earn \& Lynden-Bell}{1996}]{ELB96}
Earn, D. J. D. \& Lynden-Bell, D. 1996, \mnras, {\bf 278}, 395

\bibitem[\protect\citeauthoryear{Edvardsson \etal}{1993}]{Edva93}
Edvardsson, B., Andersen, B., Gustafsson, B., Lambert, D. L., Nissen, P. E. \& Tomkin, J. 1993, \aap, {\bf 275}, 101

\bibitem[\protect\citeauthoryear{Efremov}{2010}]{Efre10}
Efremov, Yu. N. 2010, \mnras, to appear (arXiv:1002.4555)

\bibitem[\protect\citeauthoryear{Efstathiou \etal}{1982}]{ELN82}
Efstathiou, G., Lake, G. \& Negroponte, J. 1982, \mnras, {\bf 199}, 1069

\bibitem[\protect\citeauthoryear{Eggen}{1996}]{Egge96}
Eggen, O. J. 1996, \aj, {\bf 112}, 1595

\bibitem[\protect\citeauthoryear{Elmegreen}{1996}]{Elme96}
Elmegreen, B. 1996, in IAU Colloq.\ {\bf 157}, {\it Barred Galaxies}, ed.\ R. Buta, D. A. Crocker \& B. G. Elmegreen (San Francisco: ASP Conf series {\bf 91}), 197

\bibitem[\protect\citeauthoryear{Elmegreen \& Thomasson}{1993}]{ET93}
Elmegreen, B. G. \& Thomasson, M. 1993, \aap, {\bf 272}, 37

\bibitem[\protect\citeauthoryear{Elmegreen \etal}{2007}]{Elme07}
Elmegreen, B. G., Elmegreen, D. M., Knapen, J. H., Buta, R. J., Block, D. L. \& Puerari, I. 2007, \apjl, {\bf 670}, L97

\bibitem[\protect\citeauthoryear{Elmegreen \etal}{1990}]{EEB90}
Elmegreen, D. M., Elmegreen, B. G. \& Bellin, A. D. 1990, \apj, {\bf 364}, 415

\bibitem[\protect\citeauthoryear{El-Zant \etal}{2001}]{ElZa01}
El-Zant, A., Shlosman, I. \& Hoffman, Y. 2001, \apj, {\bf 560}, 636

\bibitem[\protect\citeauthoryear{Englmaier \& Gerhard}{1997}]{EG97}
Englmaier, P. \& Gerhard, O. 1997, \mnras, {\bf 287}, 57

\bibitem[\protect\citeauthoryear{Englmaier \& Shlosman}{2004}]{ES04}
Englmaier, P. \& Shlosman, I. 2004, \apjl, {\bf 617}, L115

\bibitem[\protect\citeauthoryear{Erwin}{2005}]{Erwi05}
Erwin, P. 2005, \mnras, {\bf 364}, 283

\bibitem[\protect\citeauthoryear{Erwin \& Sparke}{2002}]{ES02}
Erwin, P. \& Sparke, L. S. 2002, \aj, {\bf 124}, 65

\bibitem[\protect\citeauthoryear{Eskridge \etal}{2000}]{Eskr00}
Eskridge, P. B., \etal\ 2000, \aj, {\bf 119}, 536

\bibitem[\protect\citeauthoryear{Evans \& Read}{1998}]{ER98}
Evans, N. W. \& Read, J. C. A. 1998, \mnras, {\bf 300}, 106

\bibitem[\protect\citeauthoryear{Famaey \etal}{2007}]{Fama07}
Famaey, B., Pont, F., Luri, X., Udry, S., Mayor, M. \& Jorrissen, A. 2007, \aap, {\bf 461}, 957

\bibitem[\protect\citeauthoryear{Fathi \etal}{2009}]{Fath09}
Fathi, K., Beckman, J. E., Pi\~nol-Ferrer, N., Hernandez, O., Mart\'\i nez-Valpuesta, I. \& Carignan, C. 2009, \apj, {\bf 704}, 1657

\bibitem[\protect\citeauthoryear{Fridman \& Polyachenko}{1984}]{FP84}
Fridman, A. M. \& Polyachenko, V. L. 1984. {\it Physics of Gravitating Systems} (New York: Springer-Verlag)

\bibitem[\protect\citeauthoryear{Friedli \& Martinet}{1993}]{FM93}
Friedli, D. \& Martinet, L. 1993, \aap, {\bf 277}, 27

\bibitem[\protect\citeauthoryear{Fuchs \etal}{2005}]{FDT5}
Fuchs, B., Dettbarn, C. \& Tsuchiya, T. 2005, \aap, {\bf 444}, 1

\bibitem[\protect\citeauthoryear{Fux}{1999}]{Fux99}
Fux, R. 1999, \aap, {\bf 345}, 787

\bibitem[\protect\citeauthoryear{Fux}{2001}]{Fux01}
Fux, R. 2001, \aap, {\bf 373}, 511

\bibitem[\protect\citeauthoryear{Garc\'\i a-Ruiz \etal}{Garc\'\i a-Ruiz \etal}{2002a}]{GRKD}
Garc\'\i a-Ruiz, I., Kuijken, K. \& Dubinski. J. 2002a, \mnras, {\bf 337}, 459

\bibitem[\protect\citeauthoryear{Garc\'\i a-Ruiz, Sancisi \& Kuijken}{Garc\'\i a-Ruiz \etal}{2002b}]{GSK02}
Garc\'\i a-Ruiz, I., Sancisi, R. \& Kuijken, K. 2002b, \aap, {\bf 394}, 769

\bibitem[\protect\citeauthoryear{Gerssen \etal}{2000}]{Gers00}
Gerssen, J., Kuijken, K. \& Merrifield, M. R. 2000, \mnras, {\bf 317}, 545

\bibitem[\protect\citeauthoryear{Goldreich \& Lynden-Bell}{1965a}]{GLB1}
Goldreich, P. \& Lynden-Bell, D. 1965a, \mnras, {\bf 130}, 97

\bibitem[\protect\citeauthoryear{Goldreich \& Lynden-Bell}{1965b}]{GLB2}
Goldreich, P. \& Lynden-Bell, D. 1965b, \mnras, {\bf 130}, 125

\bibitem[\protect\citeauthoryear{Goldreich \& Tremaine}{1978}]{GT78}
Goldreich, P. \& Tremaine, S. 1978, \apj, {\bf 222}, 850

\bibitem[\protect\citeauthoryear{Gratier \etal}{2010}]{Grat10}
Gratier, P. \etal\ 2010, \aap, to appear (arXiv:1003.3222)

\bibitem[\protect\citeauthoryear{Grosb\o l \etal}{2004}]{GPP04}
Grosb\o l, P., Patsis, P. A. \& Pompei, E. 2004, \aap, {\bf 423}, 849

\bibitem[\protect\citeauthoryear{Haan \etal}{2009}]{Haan09}
Haan, S., Schinnerer, E., Emsellem, E., García-Burillo, S., Combes, F., Mundell, C. G. \& Rix, H.-W. 2009, \apj, {\bf 692}, 1623

\bibitem[\protect\citeauthoryear{H\"anninen \& Flynn}{2002}]{HF02}
H\"anninen, J. \& Flynn, C. 2002, \mnras, {\bf 337}, 731

\bibitem[\protect\citeauthoryear{Haywood}{2008}]{Hayw08}
Haywood, M. 2008, \mnras, {\bf 388}, 1175

\bibitem[\protect\citeauthoryear{Heller \etal}{2007}]{Hell07}
Heller, C. H., Shlosman, I. \& Athanassoula, E. 2007, \apj, {\bf 657}, L65

\bibitem[\protect\citeauthoryear{Heller \etal}{2001}]{HSE01}
Heller, C., Shlosman, I. \& Englmaier, P. 2001, \apj, {\bf 553}, 661

\bibitem[\protect\citeauthoryear{Helmi \etal}{2006}]{Helm06}
Helmi, A., Navarro, J. F., Nordstr\"om, B., Holmberg, J., Abadi, M. G. \& Steinmetz, M. 2006, \mnras, {\bf 365}, 1309

\bibitem[\protect\citeauthoryear{H\'enon}{1973}]{Heno73}
H\'enon, M. 1973, in {\it Dynamical Structure and Evolution of Stellar Systems}, ed.\ L. Martinet \& M. Mayor (Sauverny: Geneva Observatory) p.~182

\bibitem[\protect\citeauthoryear{Hernquist \& Weinberg}{1992}]{HW92}
Hernquist, L. \& Weinberg, M. D. 1992, \apj, {\bf 400}, 80

\bibitem[\protect\citeauthoryear{Hill}{1878}]{Hill}
Hill, G. W., 1878 {\it Am. J. Math.}, {\bf 1}, 5

\bibitem[\protect\citeauthoryear{Hockney \& Brownrigg}{1974}]{HB74}
Hockney, R. W. \& Brownrigg, D. R. K. 1974, \mnras, {\bf 167}, 351

\bibitem[\protect\citeauthoryear{Hohl}{1971}]{Hohl71}
Hohl, F. 1971, \apj, {\bf 168}, 343

\bibitem[\protect\citeauthoryear{Holley-Bockelmann \etal}{2005}]{HBWK}
Holley-Bockelmann, K., Weinberg, M. \& Katz, N. 2005, \mnras, {\bf 363}, 991

\bibitem[\protect\citeauthoryear{Holmberg \& Flynn}{2004}]{HF04}
Holmberg, J. \& Flynn, C. 2004, \mnras, {\bf 352}, 440

\bibitem[\protect\citeauthoryear{Holmberg, Nordstr\"om \& Andersen}{Holmberg \etal}{2007}]{Holm07}
Holmberg, J., Nordstr\"om, B. \& Andersen, J. 2007, \aap, {\bf 475}, 519

\bibitem[\protect\citeauthoryear{Holmberg, Nordstr\"om \& Andersen}{Holmberg \etal}{2009}]{HNA9}
Holmberg, J., Nordstr\"om, B. \& Andersen, J. 2009, \aap, {\bf 501}, 941
	
\bibitem[\protect\citeauthoryear{Hopkins \etal}{2009}]{Hopk09}
Hopkins, P. F., Cox, T. J., Younger, J. D. \&  Hernquist, L. 2009, \apj, {\bf 691}, 1168

\bibitem[\protect\citeauthoryear{Huang \& Carlberg}{1997}]{HC97}
Huang, S. \& Carlberg, R. G. 1997, \apj, 480, 503

\bibitem[\protect\citeauthoryear{Hunter \& Toomre}{1969}]{HT69}
Hunter, C. \& Toomre, A. 1969, \apj, {\bf 155}, 747

\bibitem[\protect\citeauthoryear{Ida \etal}{1993}]{IKM93}
Ida, S., Kokuba, E. \& Makino, J. 1993, \mnras, {\bf 263}, 875

\bibitem[\protect\citeauthoryear{Ideta}{2002}]{Idet02}
Ideta, M. 2002, \apj, {\bf 568}, 190

\bibitem[\protect\citeauthoryear{Jalali}{2007}]{Jala07}
Jalali, M. A. 2007, \apj, {\bf 669}, 218

\bibitem[\protect\citeauthoryear{James \& Sellwood}{1978}]{JS78}
James, R. A. \& Sellwood, J. A. 1978, \mnras, {\bf 182}, 331

\bibitem[\protect\citeauthoryear{Jeans}{1923}]{Jean23}
Jeans, J. H. 1923, \mnras, {\bf 84}, 60

\bibitem[\protect\citeauthoryear{Jeans}{1929}]{Jean29}
Jeans, J. H. 1929, {\it Astronomy and Cosmogony\/} (Cambridge: Cambridge University Press)

\bibitem[\protect\citeauthoryear{Jenkins \& Binney}{1990}]{JB90}
Jenkins, A. \& Binney, J. J. 1990, \mnras, {\bf 245}, 305

\bibitem[\protect\citeauthoryear{Jiang \& Binney}{1999}]{JB99}
Jiang, I. \& Binney, J. 1999, \mnras, {\bf 303}, L7

\bibitem[\protect\citeauthoryear{Jog \& Combes}{2009}]{JC09}
Jog, C. J. \& Combes, F. 2009, \phr, {\bf 471}, 75

\bibitem[\protect\citeauthoryear{Jog \& Solomon}{1992}]{JS92}
Jog, C. J. \& Solomon, P. M. 1992, \apj, {\bf 387}, 152

\bibitem[\protect\citeauthoryear{Jogee \etal}{2004}]{Joge04}
Jogee, S. \etal\ 2004, \apjl, {\bf 615}, L105

\bibitem[\protect\citeauthoryear{Julian \& Toomre}{1966}]{JT66}
Julian, W. H. \& Toomre, A. 1966, \apj, {\bf 146}, 810

\bibitem[\protect\citeauthoryear{Kahn \& Woltjer}{1959}]{KW59}
Kahn, F. D. \& Woltjer, L. 1959, \apj, {\bf 130}, 705

\bibitem[\protect\citeauthoryear{Kalnajs}{1965}]{Kaln65}
Kalnajs, A. J. 1965, \PhD, Harvard University

\bibitem[\protect\citeauthoryear{Kalnajs}{1972}]{Kaln72}
Kalnajs, A. J. 1972, \apj, {\bf 175}, 63

\bibitem[\protect\citeauthoryear{Kalnajs}{1976}]{Kaln76}
Kalnajs, A. J. 1976, \apj, {\bf 205}, 751

\bibitem[\protect\citeauthoryear{Kalnajs}{1978}]{Kaln78}
Kalnajs, A. J. 1978, in IAU Symposium {\bf 77} {\it Structure and Properties of Nearby Galaxies} eds.\ E. M. Berkhuisjen \& R. Wielebinski (Dordrecht:Reidel) p.~113

\bibitem[\protect\citeauthoryear{Kalnajs}{1991}]{Kaln91}
Kalnajs, A. J. 1991, in {\it Dynamics of Disc Galaxies}, ed.\ B. Sundelius ( Gothenburg: G\"oteborgs University) p.~323

\bibitem[\protect\citeauthoryear{Kazantzidis \etal}{2009}]{Kaza09}
Kazantzidis, S., Zentner, A. R., Kravtsov, A. V., Bullock, J. S. \& Debattista, V. P. 2009, \apj, {\bf 700}, 1896

\bibitem[\protect\citeauthoryear{Khoperskov \etal}{2007}]{KJKJ}
Khoperskov, A. V., Just, A., Korchagin, V. I. \& Jalali, M. A.	2007, \aap, {\bf 473}, 31

\bibitem[\protect\citeauthoryear{Klypin \etal}{2009}]{KVCQ}
Klypin, A., Valenzuela, O., Col\'\i n, P. \& Quinn, T. 2009, \mnras, {\bf 398}, 1027

\bibitem[\protect\citeauthoryear{Korchagin \etal}{2005}]{Korc05}
Korchagin, V., Orlova, N., Kikuchi, N., Miyama, S. M. \& Moiseev, A. V. 2005, arXiv:astro-ph/0509708

\bibitem[\protect\citeauthoryear{Kornreich \etal}{2002}]{Korn02}
Kornreich, D. A., Lovelace, R. V. E. \& Haynes, M. P. 2002, \apj, {bf 580}, 705

\bibitem[\protect\citeauthoryear{Kormendy}{1993}]{Korm93}
Kormendy, J. 1993, in IAU Symposium {\bf 153}, {\it Galactic Bulges}, eds. H. Dejonghe \& H. Habing (Dordrecht: Kluwer) p~209

\bibitem[\protect\citeauthoryear{Kormendy \& Kennicutt}{2004}]{KK04}
Kormendy, J. \& Kennicutt, R. C. 2004, \araa, {\bf 42}, 603

\bibitem[\protect\citeauthoryear{Kormendy \& Norman}{1979}]{KN79}
Kormendy, J. \& Norman, C. A. 1979, \apj, {\bf 233}, 539

\bibitem[\protect\citeauthoryear{Kranz \etal}{2003}]{KSR03}
Kranz, T., Slyz, A. D. \& Rix, H.-W. 2003, \apj, {\bf 586}, 143

\bibitem[\protect\citeauthoryear{Kuijken}{1991}]{Kuij91}
Kuijken, K. 1991, \apj, {\bf 376}, 467

\bibitem[\protect\citeauthoryear{Kulsrud \etal}{1971}]{KMC71}
Kulsrud, R. M., Mark, J. W-K. \& Caruso, A. 1971, \apss, {\bf 14}, 52

\bibitem[\protect\citeauthoryear{Lacey}{1984}]{Lace84}
Lacey, C. G. 1984, \mnras, {\bf 208}, 687

\bibitem[\protect\citeauthoryear{Lacey}{1991}]{Lace91}
Lacey, C. G. 1991, in {\it Dynamics of Disc Galaxies}, ed.\ B. Sundelius (Gothenburg: G\"oteborgs University) p.~257

\bibitem[\protect\citeauthoryear{Lacey \& Ostriker}{1985}]{LO85}
Lacey, C. G. \& Ostriker, J. P. 1985, \apj, {\bf 299}, 633

\bibitem[\protect\citeauthoryear{Law \etal}{2005}]{LJM05}
Law, D. R., Johnston, K. V. \& Majewski, S. R. 2005, \apj, {\bf 619}, 807

\bibitem[\protect\citeauthoryear{Levine \etal}{2006}]{LBH06}
Levine, E. S., Blitz, L. \& Heiles, C. 2006, \apj, {\bf 643}, 881

\bibitem[\protect\citeauthoryear{Li \etal}{2009}]{LGMK09}
Li, C., Gadotti, D. A., Mao, S. \& Kauffmann, G. 2009, \mnras, {\bf 397}, 726

\bibitem[\protect\citeauthoryear{Lin \& Shu}{1966}]{LS66}
Lin, C. C. \& Shu, F. H. 1966, \pnas, {\bf 55}, 229

\bibitem[\protect\citeauthoryear{Lin \& Tremaine}{1983}]{LT83}
Lin, D. N. C. \& Tremaine, S. 1983, \apj, {\bf 264}, 364

\bibitem[\protect\citeauthoryear{Lindblad \etal}{1996}]{LLA96}
Lindblad, P. A. B., Lindblad, P. O. \& Athanassoula, E. 1996, \aap, {\bf 313}, 65

\bibitem[\protect\citeauthoryear{Little \& Carlberg}{1991}]{LC91}
Little, B. \& Carlberg, R. G. 1991, \mnras, {\bf 250}, 161

\bibitem[\protect\citeauthoryear{Louis \& Gerhard}{1988}]{LG88}
Louis, P. D. \& Gerhard, O. E. 1988, \mnras, {\bf 233}, 337

\bibitem[\protect\citeauthoryear{Lovelace}{1998}]{Love98}
Lovelace, R. V. E. 1998, \aap, {\bf 338}, 819

\bibitem[\protect\citeauthoryear{Lovelace \& Hohlfeld}{1978}]{LH78}
Lovelace, R. V. E. \& Hohlfeld, R. G. 1978, \apj, {\bf 221}, 51

\bibitem[\protect\citeauthoryear{Lovelace \etal}{1999}]{LZKH}
Lovelace, R. V. E., Zhang, L., Kornreich, D. A. \& Haynes, M. P. 1999, \apj, {\bf 524}, 634

\bibitem[\protect\citeauthoryear{Lowe \etal}{1994}]{Lowe94}
Lowe, S. A., Roberts, W. W., Yang, J., Bertin, G. \& Lin, C. C. 1994, \apj, {\bf 427}, 184

\bibitem[\protect\citeauthoryear{Lynden-Bell}{1965}]{LB65}
Lynden-Bell, D. 1965, \mnras, {\bf 129}, 299

\bibitem[\protect\citeauthoryear{Lynden-Bell}{1979}]{LB79}
Lynden-Bell, D. 1979, \mnras, {\bf 187}, 101

\bibitem[\protect\citeauthoryear{Lynden-Bell \& Kalnajs}{1972}]{LBK}
Lynden-Bell, D. \& Kalnajs, A. J. 1972, \mnras, {\bf 157}, 1

\bibitem[\protect\citeauthoryear{Maciejewski}{2006}]{Maci06}
Maciejewski, W. 2006, \mnras, {\bf 371}, 451

\bibitem[\protect\citeauthoryear{Maciejewski \& Sparke}{2000}]{MS00}
Maciejewski, W. \& Sparke, L. S. 2000, \mnras, {\bf 313}, 745

\bibitem[\protect\citeauthoryear{Maciejewski \etal}{2002}]{Maci02}
Maciejewski, W., Teuben, P. J., Sparke, L. S. \&  Stone, J. M. 2002, \mnras, {\bf 329}, 502

\bibitem[\protect\citeauthoryear{Maoz \etal}{2001}]{Maoz01}
Maoz, D., Barth, A. J., Ho, L. C., Sternberg, A. \& Filippenko, A. V. 2001, \aj, {\bf 121}, 3048

\bibitem[\protect\citeauthoryear{Marinova \& Jogee}{2007}]{MJ07}
Marinova, I. \& Jogee, S. 2007, \apj, {\bf 659}, 1176

\bibitem[\protect\citeauthoryear{Mark}{1974}]{Mark74}
Mark, J. W-K. 1974, \apj, {\bf 193}, 539

\bibitem[\protect\citeauthoryear{Mark}{1976}]{Mark76}
Mark, J. W-K. 1976, \apj, {\bf 203}, 81

\bibitem[\protect\citeauthoryear{Mark}{1977}]{Mark77}
Mark, J. W-K. 1977, \apj, {\bf 212}, 645

\bibitem[\protect\citeauthoryear{}{2004}]{MVS04}
Martinez-Valpuesta, I. \& Shlosman, I. 2004, \apjl, {\bf 613}, L29

\bibitem[\protect\citeauthoryear{Mashchenko \etal}{2006}]{MCW06}
Mashchenko, S., Couchman, H. M. P. \& Wadsley, J. 2006, \nat, {\bf 442}, 539

\bibitem[\protect\citeauthoryear{Masset \& Tagger}{1997}]{MT97}
Masset, F. \& Tagger, M. 1997, \aap, {\bf 322}, 442

\bibitem[\protect\citeauthoryear{Masters \etal}{2010}]{Mast10}
Masters, K. L., Nichol, R. C., Hoyle, B., Lintott, C., Bamford, S., Edmondson, E. M., Fortson, L., Keel, W. C., Schawinski, K., Smith, A. \& Thomas, D. 2010, arXiv:1003.0449

\bibitem[\protect\citeauthoryear{Mathur}{1990}]{Math90}
Mathur, S. D. 1990, \mnras, {\bf 243}, 529

\bibitem[\protect\citeauthoryear{Matthews}{2000}]{Matt00}
Matthews, L. D. 2000, \aj, {\bf 120}, 1764

\bibitem[\protect\citeauthoryear{McMillan \& Dehnen}{2005}]{MD05}
McMillan, P. J. \& Dehnen, W. 2005, \mnras, {\bf 363}, 1205

\bibitem[\protect\citeauthoryear{McMillan \& Dehnen}{2007}]{MD07}
McMillan, P. J. \& Dehnen, W. 2007, \mnras, {\bf 378}, 541

\bibitem[\protect\citeauthoryear{Meidt \etal}{2008}]{Meid08}
Meidt, S. E., Rand, R. J., Merrifield, M. R., Debattista, V. P. \& Shen, J. 2008, \apj, {\bf 676}, 899

\bibitem[\protect\citeauthoryear{Meidt \etal}{2009}]{MRM09}
Meidt, S. E., Rand, R. J. \& Merrifield, M. R. 2009, \apj, {\bf 702}, 277

\bibitem[\protect\citeauthoryear{M\'endez-Abreu \etal}{2010}]{MSA10}
M\'endez-Abreu, J., S\'anchez-Janssen, R. \& Aguerri, J. A. L. 2010, \apjl, {\bf 711}, L61

\bibitem[\protect\citeauthoryear{Merrifield \& Kuijken}{1999}]{MK99}
Merrifield, M. R. \& Kuijken, K. 1999, \aap, {\bf 345}, L47

\bibitem[\protect\citeauthoryear{Merritt \& Sellwood}{1994}]{MS94}
Merritt, D. \& Sellwood, J. A. 1994, \apj, {\bf 425}, 551

\bibitem[\protect\citeauthoryear{Merritt \& Stiavelli}{1990}]{MS90}
Merritt, D. \& Stiavelli, M. 1990, \apj, {\bf 358}, 399-417

\bibitem[\protect\citeauthoryear{Mestel}{1963}]{Mest63}
Mestel, L. 1963, \mnras, {\bf 126}, 553

\bibitem[\protect\citeauthoryear{Miller, Prendergast \& Quirk}{Miller \etal}{1970}]{MPQ70}
Miller, R. H., Prendergast, K. H. \& Quirk, W. J. 1970, \apj, {\bf 161}, 903

\bibitem[\protect\citeauthoryear{Minchev \& Quillen}{2006}]{MQ06}
Minchev, I. \& Quillen, A. C. 2006, \mnras, {\bf 368}, 623

\bibitem[\protect\citeauthoryear{Moster \etal}{2010}]{Most10}
Moster, B. P., Macci\`o, A. V., Somerville, R. S., Johansson, P. H. \& Naab, T. 2010, \mnras, {\bf 403}, 1009

\bibitem[\protect\citeauthoryear{Mould}{2005}]{Moul05}
Mould, J. 2005, \aj, {\bf 129}, 698

\bibitem[\protect\citeauthoryear{Mulder}{1983}]{Muld83}
Mulder, W. A. 1983, \aap, {\bf 117}, 9

\bibitem[\protect\citeauthoryear{Nelson \& Tremaine}{1995}]{NT95}
Nelson, R. W. \& Tremaine, S. 1995, \mnras, {\bf 275}, 897

\bibitem[\protect\citeauthoryear{Nieten \etal}{2006}]{Niet06}
Nieten, Ch., Neininger, N., Gu\'elin, M., Ungerechts, H., Lucas, R., Berkhuijsen, E. M., Beck, R. \& Wielebinski, R. 2006, \aap, {\bf 453}, 459

\bibitem[\protect\citeauthoryear{Noguchi}{1987}]{Nogu87}
Noguchi, M. 1987, \mnras, {\bf 228}, 635

\bibitem[\protect\citeauthoryear{Nordstr\"om \etal}{2004}]{Nord04}
Nordstr\"om, B., Mayor, M., Andersen, J., Holmberg, J., Pont, F., J\o rgensen, B. R., Olsen, E. H., Udry, S. \& Mowlavi, N.  2004, \aap, {\bf 418}, 989

\bibitem[\protect\citeauthoryear{Norman \etal}{1996}]{NSH96}
Norman, C. A., Sellwood, J. A. \& Hasan, H. 1996, \apj, {\bf 462}, 114

\bibitem[\protect\citeauthoryear{O'Neill \& Dubinski}{2003}]{OD03}
O'Neill, J. K. \& Dubinski, J. 2003, \mnras, {\bf 346}, 251

\bibitem[\protect\citeauthoryear{Oort}{1965}]{Oort65}
Oort, J. H. 1965, in Stars and Stellar Systems, {\bf 5} {\it Galactic Structure}, ed.~A. Blaauw \& M. Schmidt (Chicago: University of Chicago Press), p.~455

\bibitem[\protect\citeauthoryear{Oort \etal}{1958}]{OKW58}
Oort, J. H., Kerr, F. J. \&  Westerhout, G. 1958, \mnras, {\bf 118}, 379

\bibitem[\protect\citeauthoryear{Ostriker \& Binney}{1989}]{OB89}
Ostriker, E. C. \& Binney, J. J. 1989, \mnras, {\bf 237}, 785

\bibitem[\protect\citeauthoryear{Ostriker \& Peebles}{1973}]{OP73}
Ostriker, J. P. \& Peebles, P. J. E. 1973, \apj, {\bf 186}, 467

\bibitem[\protect\citeauthoryear{Papaloizou \& Lin}{1989}]{PL89}
Papaloizou, J. C. B. \& Lin, D. N. C. 1989, \apj, {\bf 344}, 645

\bibitem[\protect\citeauthoryear{Patsis \etal}{1991}]{PCG91}
Patsis, P. A., Contopoulos, G. \& Grosbol, P. 1991, \aap, {\bf 243}, 373

\bibitem[\protect\citeauthoryear{Patsis \etal}{2002}]{PSA02}
Patsis, P. A., Skokos, Ch. \& Athanassoula, E. 2002, \mnras, {\bf 337}, 578

\bibitem[\protect\citeauthoryear{P\'erez}{2008}]{Pere08}
P\'erez, I. 2008, \aap, {\bf 478}, 717

\bibitem[\protect\citeauthoryear{P\'erez \etal}{2004}]{PFF04}
P\'erez, I., Fux, R. \& Freeman, K. 2004, \aap, {\bf 424}, 799

\bibitem[\protect\citeauthoryear{Pfenniger \& Friedli}{1991}]{PF91}
Pfenniger, D. \& Friedli, D. 1991, \aap, {\bf 252}, 75

\bibitem[\protect\citeauthoryear{Pfenniger \& Norman}{1990}]{PN90}
Pfenniger, D. \& Norman, C. 1990, \apj, {\bf 363}, 391

\bibitem[\protect\citeauthoryear{Pichon \& Cannon}{1997}]{PC97}
Pichon, C. \& Cannon, R. C 1997, \mnras, {\bf 291}, 616

\bibitem[\protect\citeauthoryear{Polyachenko}{2004}]{Poly04}
Polyachenko, E. V. 2004, \mnras, {\bf 348}, 345

\bibitem[\protect\citeauthoryear{Polyachenko}{2005}]{Poly05}
Polyachenko, E. V. 2005, \mnras, {\bf 357}, 559

\bibitem[\protect\citeauthoryear{Polyachenko}{1977}]{Poly77}
Polyachenko, V. L. 1977, \sval, {\bf 3}, 51

\bibitem[\protect\citeauthoryear{Prendergast}{1962}]{Pren62}
Prendergast, K. H. 1962, in {\it Interstellar Matter in Galaxies}, ed.\ L. Woltjer (New York: Benjamin), p.~217

\bibitem[\protect\citeauthoryear{Quillen}{2003}]{Quil03}
Quillen, A. C. 2003, \aj, {\bf 125}, 785

\bibitem[\protect\citeauthoryear{Quillen \& Garnett}{2000}]{QG00}
Quillen, A. C. \& Garnett, D. R. 2000, arXiv:astro-ph/0004210

\bibitem[\protect\citeauthoryear{Quinn \& Binney}{1992}]{QB92}
Quinn, T. \& Binney, J. 1992, \mnras, {\bf 255}, 729

\bibitem[\protect\citeauthoryear{Raboud \etal}{1998}]{Rabo98}
Raboud, D., Grenon, M., Martinet, L., Fux, R. \& Udry, S. 1998, \aap, {\bf 335}, L61

\bibitem[\protect\citeauthoryear{Rafikov}{2001}]{Rafi01}
Rafikov, R. R. 2001, \mnras, {\bf 323}, 445

\bibitem[\protect\citeauthoryear{Raha \etal}{1991}]{RSJK}
Raha, N., Sellwood, J. A., James, R. A. \& Kahn, F. D. 1991, \nat, {\bf 352}, 411

\bibitem[\protect\citeauthoryear{Rautiainen \& Salo}{1999}]{RS99}
Rautiainen, P. \& Salo, H. 1999, \aap, {\bf 348}, 737

\bibitem[\protect\citeauthoryear{Rautiainen \etal}{2002}]{RSL02}
Rautiainen, P., Salo, H. \& Laurikainen, E. 2002, \mnras, {\bf 337}, 1233

\bibitem[\protect\citeauthoryear{Rautiainen \etal}{2008}]{RSL08}
Rautiainen, P., Salo, H. \& Laurikainen, E. 2008, \mnras, {\bf 388}, 1803

\bibitem[\protect\citeauthoryear{Read \etal}{2008}]{Read08}
Read, J. I., Lake, G., Agertz, O. \& Debattista, V. P. 2008, \mnras, {\bf 389}, 1041

\bibitem[\protect\citeauthoryear{Reese \etal}{2007}]{Rees07}
Reese, A., Williams, T. B., Sellwood, J. A., Barnes, E. I. \& Powell, B. A. 2007, \aj, {\bf 133}, 2846

\bibitem[\protect\citeauthoryear{Regan \etal}{2002}]{Rega02}
Regan, M. W., Sheth, K., Teuben, P. J. \& Vogel, S. N. 2002, \apj, {\bf 574}, 126

\bibitem[\protect\citeauthoryear{Reid \etal}{2007}]{Reid07}
Reid, I. N., Turner, E. L., Turnbull, M. C., Mountain, M. \& Valenti, J. A. 2007, \apj, {\bf 665}, 767

\bibitem[\protect\citeauthoryear{Reshetnikov \etal}{2002}]{RBCJ02}
Reshetnikov, V., Battaner, E., Combes, F. \& Jim\'enez-Vicente, J. 2002, \aap, {\bf 382}, 513

\bibitem[\protect\citeauthoryear{Reyl\'e \etal}{2009}]{Reyl09}
Reyl\'e, C., Marshall, D. J., Robin, A. C. \& Schultheis\'e, M. 2009, \aap, {\bf 495}, 819

\bibitem[\protect\citeauthoryear{Roberts \etal}{1979}]{RHvA}
Roberts, W. W., Huntley, J. M. \& van Albada, G. D. 1979, \apj, {\bf 233}, 67 

\bibitem[\protect\citeauthoryear{Romano-D\'\i az \etal}{2008a}]{Roma08a}
Romano-D\'\i az, E., Shlosman, I., Hoffman, Y. \& Heller, C. 2008a, \apjl, {\bf 685}, L105

\bibitem[\protect\citeauthoryear{Romano-D\'\i az \etal}{2008b}]{Roma08b}
Romano-D\'\i az, E., Shlosman, I., Heller, C. \& Hoffman, Y. 2008b, \apjl, {\bf 687}, L13

\bibitem[\protect\citeauthoryear{Romeo}{1992}]{Rome92}
Romeo, A. B. 1992, \mnras, {\bf 256}, 307

\bibitem[\protect\citeauthoryear{Ro\u skar \etal}{2008a}]{Rosk8a}
Ro\u skar, R., Debattista, V. P., Quinn, T. R., Stinson, G. S. \& Wadsley, J. 2008, \apjl, {\bf 684}, L79

\bibitem[\protect\citeauthoryear{Ro\u skar \etal}{2008b}]{Rosk8b}
Ro\u skar, R., Debattista, V. P., Stinson, G. S., Quinn, T. R., Kaufmann, T. \& Wadsley, J. 2008, \apjl, {\bf 675}, L65

\bibitem[\protect\citeauthoryear{Rubin \etal}{1992}]{RGK92}
Rubin, V. C., Graham, J. A. \& Kenney, J. D. P. 1992, \apjl, {\bf 394} L9

\bibitem[\protect\citeauthoryear{Rybicki}{1972}]{Rybi72}
Rybicki, G. B. 1972, in IAU Colloq.\ {\bf 10}, {\it Gravitational $N$-body Problem}, ed.\ M. Lecar (Dordrecht: Reidel), 22

\bibitem[\protect\citeauthoryear{Saha \etal}{2007}]{SCJ07}
Saha, K., Combes, J. \& Jog, C. 2007, \mnras, {\bf 382}, 419

\bibitem[\protect\citeauthoryear{Saha \etal}{2009}]{SdJH}
Saha, K., de Jong, R. \& Holwerda, B. 2009, \mnras, {\bf 396}, 409

\bibitem[\protect\citeauthoryear{Salo \& Laurikainen}{1993}]{SL93}
Salo, H. \& Laurikainen, E. 1993, \apj, {\bf 410}, 586

\bibitem[\protect\citeauthoryear{Sancisi}{1976}]{Sanc76}
Sancisi, R. 1976, \aap, {\bf 53}, 159

\bibitem[\protect\citeauthoryear{Sancisi \etal}{2008}]{Sanc08}
Sancisi, R., Fraternali, F., Oosterloo, T. \& van der Hulst, T. 2008, \aapr, {\bf 15}, 189

\bibitem[\protect\citeauthoryear{Sandage \& Humphreys}{1980}]{SH80}
Sandage, A. \& Humphreys, R. M. 1980, \apjl, {\bf 236}, L1

\bibitem[\protect\citeauthoryear{Sanders \& Huntley}{1976}]{SH76}
Sanders, R. H. \& Huntley, J. M. 1976, \apj, {\bf 209}, 53

\bibitem[\protect\citeauthoryear{Sawamura}{1988}]{Sawa88}
Sawamura, M. 1988, \pasj, {\bf 40}, 279

\bibitem[\protect\citeauthoryear{Sch\"on\-rich \& Binney}{2009}]{SB09}
Sch\"onrich, R. \& Binney, J. 2009, \mnras, {\bf 396}, 203

\bibitem[\protect\citeauthoryear{Schwarz}{1981}]{Schw81}
Schwarz, M. P. 1981, \apj, {\bf 247}, 77

\bibitem[\protect\citeauthoryear{Seabroke \& Gilmore}{2007}]{SG07}
Seabroke, G. M. \& Gilmore, G. 2007, \mnras, {\bf 380}, 1348

\bibitem[\protect\citeauthoryear{Sellwood}{1980}]{Sell80}
Sellwood, J. A. 1980, \aap, {\bf 89}, 296

\bibitem[\protect\citeauthoryear{Sellwood}{1981}]{Sell81}
Sellwood, J. A. 1981, \aap, {\bf 99}, 362

\bibitem[\protect\citeauthoryear{Sellwood}{1985}]{Sell85}
Sellwood, J. A. 1985, \mnras, {\bf 217}, 127

\bibitem[\protect\citeauthoryear{Sellwood}{1989a}]{Sell89a}
Sellwood, J. A. 1989a, \mnras, {\bf 238}, 115

\bibitem[\protect\citeauthoryear{Sellwood}{1989b}]{Sell89b}
Sellwood, J. A. 1989b, in {\it Dynamics of Astrophysical Discs}, ed.\ J. A. Sellwood (Cambridge: Cambridge University Press) p.~155

\bibitem[\protect\citeauthoryear{Sellwood}{1994}]{Sell94}
Sellwood, J. A. 1994, in {\it Galactic and Solar System Optical Astrometry} ed L Morrison (Cambridge: Cambridge University Press) p.~156

\bibitem[\protect\citeauthoryear{Sellwood}{1996a}]{Sell96a}
Sellwood, J. A. 1996a, in IAU Symp.\ {\bf 169}, {\it Unsolved Problems of the Milky Way\/}, ed.\ L. Blitz \& P. Teuben (Dordrecht: Kluwer) p.~31

\bibitem[\protect\citeauthoryear{Sellwood}{1996b}]{Sell96b}
Sellwood, J. A. 1996b, \apj, {\bf 473}, 733

\bibitem[\protect\citeauthoryear{Sellwood}{2000}]{Sell00}
Sellwood, J. A. 2000, in {\it Astrophysical Dynamics -- in Commemoration of F. D. Kahn}, ed.\ D. Berry, D. Breitschwerdt, A. da Costa \& J. E. Dyson, \apss, {\bf 272}, p.~31 (astro-ph/9909093)

\bibitem[\protect\citeauthoryear{Sellwood}{2003}]{Sell03}
Sellwood, J. A. 2003, \apj, {\bf 587}, 638

\bibitem[\protect\citeauthoryear{Sellwood}{2006}]{Sell06}
Sellwood, J. A. 2006, \apj, {\bf 637}, 567

\bibitem[\protect\citeauthoryear{Sellwood}{2008a}]{Sell08a}
Sellwood, J. A. 2008a, \apj, {\bf 679}, 379

\bibitem[\protect\citeauthoryear{Sellwood}{2008b}]{Sell08b}
Sellwood, J. A. 2008b, in {\it Formation and Evolution of Galaxy Disks}, ed.\ J. G. Funes SJ \&  E. M. Corsini (San Francisco: ASP {\bf 396}), p.~341 (arXiv:0803.1574)

\bibitem[\protect\citeauthoryear{Sellwood}{2010}]{Sell10}
Sellwood, J. A. 2010, \mnras, submitted (arXiv:1001.5197)

\bibitem[\protect\citeauthoryear{Sellwood \& Binney}{2002}]{SB02}
Sellwood, J. A. \& Binney, J. J. 2002, \mnras, {\bf 336}, 785

\bibitem[\protect\citeauthoryear{Sellwood \& Carlberg}{1984}]{SC84}
Sellwood, J. A. \& Carlberg, R. G. 1984, \apj, {\bf 282}, 61

\bibitem[\protect\citeauthoryear{Sellwood \& Debattista}{2006}]{SD06}
Sellwood, J. A. \& Debattista, V. P. 2006, \apj, {\bf 639}, 868

\bibitem[\protect\citeauthoryear{Sellwood \& Debattista}{2009}]{SD09}
Sellwood, J. A. \& Debattista, V. P. 2009, \mnras, {\bf 398}, 1279

\bibitem[\protect\citeauthoryear{Sellwood \& Evans}{2001}]{SE01}
Sellwood, J. A. \& Evans, N. W. 2001, \apj, {\bf 546}, 176

\bibitem[\protect\citeauthoryear{Sellwood \& Kahn}{1991}]{SK91}
Sellwood, J. A. \& Kahn, F. D. 1991, \mnras, {\bf 250}, 278

\bibitem[\protect\citeauthoryear{Sellwood \& Lin}{1989}]{SL89}
Sellwood, J. A. \& Lin, D. N. C. 1989, \mnras, {\bf 240}, 991

\bibitem[\protect\citeauthoryear{Sellwood \& Merritt}{1994}]{SM94}
Sellwood, J. A. \& Merritt, D. 1994, \apj, {\bf 425}, 530

\bibitem[\protect\citeauthoryear{Sellwood \& Moore}{1999}]{SM99}
Sellwood, J. A. \& Moore, E. M. 1999, \apj, {\bf 510}, 125

\bibitem[\protect\citeauthoryear{Sellwood \etal}{1998}]{SNT98}
Sellwood, J. A., Nelson, R. D. \& Tremaine, S. 1998, \apj, {\bf 506}, 590

\bibitem[\protect\citeauthoryear{Sellwood \& Sparke}{1988}]{SS88}
Sellwood, J. A. \& Sparke, L. S. 1988, \mnras, {\bf 231}, 25P

\bibitem[\protect\citeauthoryear{Sellwood \& Valluri}{1997}]{SV97}
Sellwood, J. A. \& Valluri, M. 1997, \mnras, {\bf 287}, 124

\bibitem[\protect\citeauthoryear{Sellwood \& Wilkinson}{1993}]{SW93}
Sellwood, J. A. \& Wilkinson, A. 1993, \rpp, {\bf 56}, 173

\bibitem[\protect\citeauthoryear{Shakura \& Sunyaev}{1973}]{SS73}
Shakura, N. I. \& Sunyaev, R. A. 1973, \aap, {\bf 24}, 337

\bibitem[\protect\citeauthoryear{Shen \& Debattista}{2009}]{ShD09}
Shen, J. \& Debattista, V. P. 2009, \apj, {\bf 690}, 758

\bibitem[\protect\citeauthoryear{Shen \& Sellwood}{2004}]{SS04}
Shen, J. \& Sellwood, J. A. 2004, \apj, {\bf 604}, 614

\bibitem[\protect\citeauthoryear{Shen \& Sellwood}{2006}]{SS06}
Shen, J. \& Sellwood, J. A. 2006, \mnras, {\bf 370}, 2

\bibitem[\protect\citeauthoryear{Sheth \etal}{2008}]{Shet08}
Sheth, K. \etal\ 2008, \apj, {\bf 675}, 1141

\bibitem[\protect\citeauthoryear{Shetty \etal}{2007}]{SVOT}
Shetty, R., Vogel, S. N., Ostriker, E. C. \& Teuben, P. J. 2007, \apj, {\bf 665}, 1138

\bibitem[\protect\citeauthoryear{Shiidsuke \& Ida}{1999}]{SI99}
Shiidsuke, K. \& Ida, S. 1999, \mnras, {\bf 307}, 737

\bibitem[\protect\citeauthoryear{Shlosman \etal}{1989}]{SFB89}
Shlosman, I., Frank, J. \& Begelman, M. C. 1989, \nat, {\bf 338}, 45

\bibitem[\protect\citeauthoryear{Shu \etal}{1990}]{Shu90}
Shu, F. H., Tremaine, S., Adams, F. C. \& Ruden, S. P. 1990, \apj, {\bf 358}, 495

\bibitem[\protect\citeauthoryear{Skokos \etal}{2002}]{SPA02}
Skokos, Ch., Patsis, P. A. \& Athanassoula, E. 2002, \mnras, {\bf 333}, 847

\bibitem[\protect\citeauthoryear{Soderblom}{2010}]{Sode10}
Soderblom, D. R. 2010, \araa, to appear (arXiv:1003.6074)

\bibitem[\protect\citeauthoryear{Sparke \& Casertano}{1988}]{SC88}
Sparke, L. S. \& Casertano, S. 1988, \mnras, {\bf 234}, 873

\bibitem[\protect\citeauthoryear{Sparke \& Sellwood}{1987}]{SS87}
Sparke, L. S. \& Sellwood, J. A. 1987, \mnras, {\bf 225}, 653

\bibitem[\protect\citeauthoryear{Spitzer}{1942}]{Spit42}
Spitzer, L. 1942, \apj, {\bf 95}, 329

\bibitem[\protect\citeauthoryear{Spitzer \& Schwarz\-schild}{1953}]{SS53}
Spitzer, L. \& Schwarzschild, M. 1953, \apj, {\bf 118}, 106

\bibitem[\protect\citeauthoryear{Stanghellini \& Haywood}{2010}]{SH10}
Stanghellini, L. \& Haywood, M. 2010, \apj, {\bf 714}, 1096

\bibitem[\protect\citeauthoryear{Stewart \etal}{2008}]{Stew08}
Stewart, K. R., Bullock, J. S., Wechsler, R. H., Maller, A. H. \& Zentner, A. R. 2008, \apj, {\bf 683}, 597

\bibitem[\protect\citeauthoryear{Sygnet \etal}{1987}]{Sygn88}
Sygnet, J. F., Tagger, M., Athanassoula, E. \& Pellat, R. 1988, \mnras, {\bf 232}, 733 

\bibitem[\protect\citeauthoryear{Tagger \etal}{1987}]{Tagg87}
Tagger, M., Sygnet, J. F., Athanassoula, E. \& Pellat, R. 1987, \apjl, {\bf 318}, L43

\bibitem[\protect\citeauthoryear{Thomasson \etal}{1990}]{TEDS}
Thomasson, M., Elmegreen, B. G., Donner, K. J. \& Sundelius, B. 1990, \apjl, {\bf 356}, L9

\bibitem[\protect\citeauthoryear{Toomre}{1964}]{Toom64}
Toomre, A. 1964, \apj, {\bf 139}, 1217

\bibitem[\protect\citeauthoryear{Toomre}{1966}]{Toom66}
Toomre, A. 1966, in {\it Geophysical Fluid Dynamics}, notes on the 1966 Summer Study Program at the Woods Hole Oceanographic Institution, ref. no. 66-46

\bibitem[\protect\citeauthoryear{Toomre}{1969}]{Toom69}
Toomre, A. 1969, \apj, {\bf 158}, 899

\bibitem[\protect\citeauthoryear{Toomre}{1981}]{Toom81}
Toomre, A. 1981, in {\it The Structure and Evolution of Normal Galaxies}, ed.\ S. M. Fall \& D. Lynden-Bell (Cambridge: Cambridge University Press), p.~111

\bibitem[\protect\citeauthoryear{Toomre}{1983}]{Toom83}
Toomre, A. 1983, in IAU Symposium {\bf 100}, {\it Internal Kinematics and Dynamics of Galaxies}, ed.\ E. Athanassoula (Dordrecht: Reidel) p~177

\bibitem[\protect\citeauthoryear{Toomre}{1989}]{Toom89}
Toomre, A. 1989, in {\it Dynamics of Astrophysical Discs}, ed.\ J. A. Sellwood (Cambridge: Cambridge University Press) p.~153

\bibitem[\protect\citeauthoryear{Toomre}{1990}]{Toom90}
Toomre, A. 1990, in {\it Dynamics \& Interactions of Galaxies}, ed.\ R. Wielen (Berlin, Heidelberg: Springer-Verlag), p.~292

\bibitem[\protect\citeauthoryear{Toomre}{1995}]{Toom95}
Toomre, A. 1995, unpublished notes

\bibitem[\protect\citeauthoryear{Toomre \& Kalnajs}{1991}]{TK91}
Toomre, A. \& Kalnajs, A. J. 1991, in {\it Dynamics of Disc Galaxies}, ed.\ B. Sundelius (Gothenburg: G\"oteborgs University) p.~341

\bibitem[\protect\citeauthoryear{T\'oth \& Ostriker}{1992}]{TO92}
T\'oth, G. \& Ostriker, J. P. 1992, \apj, {\bf 389}, 5

\bibitem[\protect\citeauthoryear{Tremaine}{2005}]{Trem05}
Tremaine, S. 2005, \apj, {\bf 625}, 143

\bibitem[\protect\citeauthoryear{Tremaine \& Weinberg}{1984a}]{TW84a}
Tremaine, S. \& Weinberg, M. D. 1984a, \apjl, {\bf 282}, L5

\bibitem[\protect\citeauthoryear{Tremaine \& Weinberg}{1984b}]{TW84b}
Tremaine, S. \& Weinberg, M. D. 1984b, \mnras, {\bf 209}, 729

\bibitem[\protect\citeauthoryear{Tsoutsis \etal}{2009}]{TKEC}
Tsoutsis, P., Kalapotharakos, C., Efthymiopoulos, C. \& Contopoulos, G. 2009, \aap, {\bf 495}, 743

\bibitem[\protect\citeauthoryear{Valenzuela \& Klypin}{2003}]{VK03}
Valenzuela, O. \& Klypin, A. 2003, \mnras, {\bf 345}, 406

\bibitem[\protect\citeauthoryear{Vandervoort}{1970}]{Vand70}
Vandervoort, P. O. 1970, \apj, {\bf 161}, 87

\bibitem[\protect\citeauthoryear{Vauterin \& Dejonghe}{1996}]{VD96}
Vauterin, P. \& Dejonghe, H. 1996, \aap, {\bf 313}, 465

\bibitem[\protect\citeauthoryear{Vel\'azquez \& White}{1999}]{VW99}
Vel\'azquez, H. \& White, S. D. M.. 1999, \mnras, {\bf 304}, 254

\bibitem[\protect\citeauthoryear{Villa-Vargas \etal}{2009}]{VSH09}
Villa-Vargas, J., Shlosman, I. \& Heller, C. 2009, \apj, {\bf 707}, 218

\bibitem[\protect\citeauthoryear{Villumsen}{1985}]{Vill85}
Villumsen, J. V. 1985, \apj, {\bf 290}, 75

\bibitem[\protect\citeauthoryear{Voglis \etal}{2007}]{VHC07}
Voglis, N., Harsoula, M. \& Contopoulos, G. 2007, \mnras, {\bf 381}, 757

\bibitem[\protect\citeauthoryear{Wada}{2001}]{Wada01}
Wada, K. 2001, \apjl, {\bf 559}, L41

\bibitem[\protect\citeauthoryear{Walker \etal}{1996}]{Walk96}
Walker, I. R., Mihos, J. C. \& Hernquist, L. 1996, \apj, {\bf 460}, 121

\bibitem[\protect\citeauthoryear{Weinberg}{1985}]{Wein85}
Weinberg, M. D. 1985, \mnras, {\bf 213}, 451

\bibitem[\protect\citeauthoryear{Weinberg}{1991}]{Wein91}
Weinberg, M. D. 1991, \apj, {\bf 373}, 391

\bibitem[\protect\citeauthoryear{Weinberg}{1994}]{Wein94}
Weinberg, M. D. 1994, \apj, {\bf 421}, 481

\bibitem[\protect\citeauthoryear{Weinberg \& Blitz}{2006}]{WB06}
Weinberg, M. D. \& Blitz, L. 2006, \apjl, {\bf 641}, L33

\bibitem[\protect\citeauthoryear{Weinberg \& Katz}{2002}]{WK02}
Weinberg, M. D. \& Katz, N. 2002, \apj, {\bf 580}, 627

\bibitem[\protect\citeauthoryear{Weinberg \& Katz}{2007}]{WK07}
Weinberg, M. D. \& Katz, N. 2007, \mnras, {\bf 375}, 425

\bibitem[\protect\citeauthoryear{Weiner}{2004}]{Wein04}
Weiner, B. J. 2004, in IAU Symp.\ {\bf 220}, {\it Dark Matter in Galaxies}, ed.\ S. Ryder, D. J. Pisano, M. Walker \& K. C. Freeman (Dordrecht: Reidel), p.~35

\bibitem[\protect\citeauthoryear{Weiner \etal}{2001}]{WSW01}
Weiner, B. J., Sellwood, J. A. \& Williams, T. B. 2001, \apj, {\bf 546}, 931

\bibitem[\protect\citeauthoryear{Wielen}{1977}]{Wiel77}
Wielen, R. 1977, \aap, {\bf 60}, 263

\bibitem[\protect\citeauthoryear{Yoachim \& Dalcanton}{2006}]{YD06}
Yoachim, P. \& Dalcanton, J. J. 2006, \aj, {\bf 131}, 226

\bibitem[\protect\citeauthoryear{Zang}{1976}]{Zang76}
Zang, T. A. 1976, \PhD, MIT

\bibitem[\protect\citeauthoryear{Zang \& Hohl}{1978}]{ZH78}
Zang, T. A. \& Hohl, F. 1978, \apj, {\bf 226}, 521

\bibitem[\protect\citeauthoryear{Z\'anmar S\'anchez \etal}{2008}]{ZSSWW}
Z\'anmar S\'anchez, R., Sellwood, J. A., Weiner B. J. \& Williams, T. B. 2008, \apj, {\bf 674}, 797

\bibitem[\protect\citeauthoryear{Zhang}{1996}]{Zhan96}
Zhang, X. 1996, \apj, {\bf 457}, 125

\bibitem[\protect\citeauthoryear{Zhang}{1998}]{Zhan98}
Zhang, X. 1998, \apj, {\bf 499}, 93

\bibitem[\protect\citeauthoryear{Zibetti, Charlot \& Rix}{Zibetti \etal}{2009}]{ZCR9}
Zibetti, S., Charlot, S. \& Rix, H.-W. 2009, \mnras, {\bf 400}, 1181

\end{thebibliography}
\end{document}